\documentclass[journal ]{new-aiaa}
\usepackage[utf8]{inputenc}
\usepackage{textcomp}

\usepackage{graphicx}
\usepackage{subcaption}
\usepackage{amsmath}
\usepackage{commath}
\usepackage[version=4]{mhchem}
\usepackage{siunitx}
\usepackage{longtable,tabularx}

\usepackage{algorithm}
\usepackage{algpseudocode}

\newcommand{\comment}[1]{}

\newcommand\blfootnote[1]{%
  \begingroup
  \renewcommand\thefootnote{}\footnote{#1}%
  \addtocounter{footnote}{-1}%
  \endgroup
}

\title{Non-Gaussian Distribution Steering in Nonlinear Dynamics with Conjugate Unscented Transformation\blfootnote{An earlier version of this paper was presented as paper 25-611 at the 2025 AAS/AIAA Astrodynamics Specialist Conference \cite{Qi-Stat-moment-conference}.}
}

\author{Daniel C. Qi\footnote{Ph.D. Student, School of Aeronautics and Astronautics. (Corresponding Author: qi85@purdue.edu)} and Kenshiro Oguri\footnote{Assistant Professor, School of Aeronautics and Astronautics. Senior Member AIAA.}}
\affil{Purdue University, West Lafayette, Indiana, 47907}
\author{Puneet Singla\footnote{Harry and Arlene Schell Professor of Engineering, Department of Aerospace Engineering. Associate Fellow AIAA.}}
\affil{The Pennsylvania State University, University Park, Pennsylvania, 16802}
\author{Maruthi R. Akella\footnote{Cockrell Family Endowed Chair Professor, Department of Aerospace Engineering and Engineering Mechanics. Fellow AIAA.}}
\affil{The University of Texas at Austin, Austin, Texas, 78712}

\begin{document}

\maketitle

\begin{abstract}
In highly nonlinear systems such as the ones commonly found in astrodynamics, Gaussian distributions generally evolve into non-Gaussian distributions. This paper introduces a method for effectively controlling non-Gaussian distributions in nonlinear environments using optimized linear feedback control. This paper utilizes Conjugate Unscented Transformation to quantify the higher-order statistical moments of non-Gaussian distributions. The formulation focuses on controlling and constraining the sigma points associated with the uncertainty quantification, which would thereby reflect the control of the entire distribution and constraints on the moments themselves. This paper develops an algorithm to solve this problem with sequential convex programming, and it is demonstrated through a two-body and three-body example. The examples show that individual moments can be directly controlled, and the moments are accurately approximated for non-Gaussian distributions throughout the controller's time horizon in nonlinear dynamics.
\end{abstract}

\section{Introduction}
Depending on the mission, the latency between ground control and the corresponding spacecraft may make fast-reaction maneuvers infeasible if control decisions were dictated by ground controllers. As a result, autonomous spacecrafts may require the ability to make onboard maneuver decisions, such as during the entry, descent, and landing of the Mars 2020 Perseverance rover \cite{mars-lander}. The challenge of nonlinear dynamics, coupled with limited onboard computing power, often hinders a spacecraft’s ability to compute effective control actions quickly. One proposed method includes reallocating the optimization process to ground systems and periodically uploading controller gains to the spacecraft \cite{Oguri-Chance-Paper}. This way, the required onboard computation is reduced and the spacecraft can still operate autonomously after the gain upload. This type of gain-scheduled approach has been applied to scenarios such as trajectory optimization, stationkeeping, and proximity operations \cite{Ridderhof-SCP, Benedikter-CovSteering, Oguri-Chance-Paper, Naoya-Sequential-Cov-Steering, Jerry-CovSteering, Ra-RPO}.

One method of generating these gains is through covariance steering. Covariance control, or covariance steering, is a method aimed at controlling the mean and covariance of a distribution. This idea has been around since the 1980s by Hotz and Skelton \cite{Hotz-Cov-Steering-OG}, and can be formulated in a manner compatible with convex optimization, control feedback, and probabilistic constraints \cite{Bakolas-Cov-Steering, Okamoto-Cov-Steering}. A significant downside of this approach is that it assumes the linear transformation of a Gaussian distribution. Nonlinear systems, commonly found in astrodynamic problems, are known to devolve Gaussian distributions into non-Gaussian ones \cite{erin-cislunar-OD, John-GMM, Sharad-UQ-CR3BP}. Thus, nonlinear simulations of the controller have shown discrepancies between the Gaussian-approximated and the actual distribution \cite{Naoya-Sequential-Cov-Steering, Jerry-CovSteering, Qi-Stationkeeping}. 

The logical next step is to extend the current Gaussian control framework to be compatible with non-Gaussian distributions. Non-Gaussian distribution control has been explored in the past using both mixture and moment models to approximate the non-Gaussian distribution \cite{Boone-GMM-Control, Wang-NonGaussian-Control}. However, these approaches are limited to nominal maneuver actions and do not incorporate a feedback policy. Recently, Gaussian mixture-based, feedback control policy has been proposed \cite{fife-GMM-steering}, which uses the mixture model to better approximate the mean and covariance of the non-Gaussian distribution. A limitation of this approach is that, currently, it requires the distribution to be merged back into a single Gaussian at each timestep, so constraints on higher-ordered moments like skewness cannot be applied. A method of enforcing a terminal Gaussian distribution constraint exists for Gaussian mixture control, but it only applies to linear systems \cite{Naoya-GMM-Terminal}. Regardless, all mixture-based steering approaches carry inherent properties such as splitting and merging of kernels, which adds complexity to the control optimization framework.

Another approach for controlling distributions involves minimizing or constraining a statistical quantity between the controlled probability density function and a desired probability density function, or probability density function matching. For example, statistical quantities involving the characteristic function \cite{Sivaramakrishnan-CF-control}, Kullback–Leibler divergence \cite{Herzallah-KL-divergence-control}, and Wasserstein distance \cite{Balci-Wasserstein-divergence-control} have been applied to both Gaussian and non-Gaussian steering problems; however, the mathematical derivations of these controllers are based only on linear stochastic systems, and extending them to nonlinear environments can be challenging. Furthermore, these approaches are typically designed to drive the system toward a single target density function. This can unnecessarily restrict the solution space, since multiple distributions may also satisfy the same set of mission requirements.

Similarly, another strategy for managing systems with non-Gaussian uncertainties is to aim at reducing the uncertainty induced by disturbances, without requiring explicit prior knowledge of the disturbance distribution. These controllers are known as minimum-entropy control. They primarily rely on a system model, but require data to estimate unknown disturbance statistics \cite{Petersen-entropy, Liu-Min-Entrophy-Control, Zhang-Min-Entrophy-Control}. These methods, while effective in their steady-state, are not inherently suited for developing spacecraft guidance policies along a trajectory where sparse measurement updates and control actions prevent the system from approaching the theoretical steady-state behavior. Moreover, the complexity of these controllers increases significantly in higher-dimensional systems \cite{Zhang-Min-Entrophy-Control}. These limitations restrict their applicability to spacecraft trajectory optimization.

Controlling \emph{non-Gaussian} distribution in \emph{nonlinear} systems for spaceflight applications requires a method capable of quantifying a distribution after nonlinear transformation and integrating with control frameworks. Its applicability to spacecraft trajectory optimization arises from the challenge of managing spacecraft with limited opportunities for navigation and control. This paper looks at Conjugate Unscented Transformation (CUT), which is a technique that utilizes discrete sigma points to approximate a distribution up to any finite number of moments \cite{Adurthi-CUT}. This type of uncertainty quantification has already been used for many applications in estimation and tracking \cite{Sharad-UQ-CR3BP, Adurthi-CUT-Tracking}. In control applications, representing distributions with discrete points rather than density functions enables the use of established deterministic trajectory methods, as each point can be treated as its own trajectory. Previous works have used CUT or other unscented transformations in the context of stochastic trajectory optimization \cite{Ozaki-UT-TrajOp, Nandi-CUT, Ross-CUT-Traj-Op, fife-GMM-steering-conference}, but this has not been generalized to control feedback and, more importantly, control of higher statistical moments.

This paper proposes a formulation for controlling a non-Gaussian distribution in nonlinear systems using a linear feedback controller and CUT. This paper's \emph{statistical moment steering} improves current distribution controllers by its ability to introduce control feedback, and most importantly, directly enforce constraints on the higher-order statistical moments (i.e., not just mean and covariance). A subsequent benefit to using CUT is its ability to better estimate the lower moments throughout the time horizon in a nonlinear system compared to just linear propagation of Gaussian distributions in typical covariance steering. The method involves finding a common gain for all CUT points while satisfying any moment constraints, which thereby corresponds to a gain applicable to the rest of the distribution. Sequential convex programming is used to solve this nonlinear problem.

This paper is organized as follows. First, Section~\ref{sec: background} presents all the fundamental statistical background needed for this paper. Section~\ref{sec: problem statement} outlines the optimal statistical moment steering problem and this paper's solution. Section~\ref{sec: scvx implementation} discusses statistical moment steering's implementation with a specific sequential convex optimization algorithm. Sections~\ref{sec: numerical example two body} and \ref{sec: numerical example three body} present two applications of this formulation to problems relevant to astrodynamics and spacecraft operations. Section~\ref{sec: discussion and future works} includes some remarks regarding this formulation as well as future directions for this type of research. Finally, Section~\ref{sec: conclusion} concludes the paper. 

\section{Background}\label{sec: background}

\subsection{Expectation and Statistical Moments}
Given a univariate random variable ${X}\in\mathbb{R}$, the expected value of ${X}$ can be represented as
\begin{equation}
    \mathbb{E}[\mathcal{G}({X})] = \int_{\mathbb{R}} \mathcal{G}(x) f_{X}(x) dx
\end{equation}
where $f_{X}(x)$ is the probability density function of ${X}$ and $\mathcal{G}(\cdot)$ a measurable function of ${X}$. The expectation integral can be used to calculate specific moments of the random variable as detailed in Table~\ref{tab: Statistical Moments}.  

\begin{table}[htbp]
	\fontsize{10}{10}\selectfont
    \caption{Statistical Moments of Univariate Random Variables}
   \label{tab: Statistical Moments}
        \centering 
   \begin{tabular}{l c } 
      \hline 
      Statistical Moments & Expectation Calculation \\
      \hline 
      $m$-th Raw Moment & $\mathbb{E}[{X}^m]$  \\
      $m$-th Central Moment & $\mathbb{E}[({X} - \mu)^m]$   \\
      $m$-th Standardized Moment  & $\mathbb{E}\left[\left(\frac{{X} - \mu}{\sigma}\right)^m\right]$   \\
      \hline
   \end{tabular}
\end{table}

Moments can help provide a quantitative value for characterizing the probability distribution of the random variable. More intuitively, they help determine the ``shape'' of the distribution. Certain moments carry greater significance and are distinguished by specific names as outlined in Table~\ref{tab: Moments Names}.  
\begin{table}[htbp]
	\fontsize{10}{10}\selectfont
    \caption{Terminology for Moments of Univariate Random Variables}
   \label{tab: Moments Names}
        \centering 
   \begin{tabular}{l  c  c  c } 
      \hline 
      $m$& $m$-th Raw Moment & $m$-th Central Moment & $m$-th Standardized Moment \\
      \hline 
      1 & Mean ($\mu$) & - & -  \\
      2 & - & Variance ($\sigma^2$) & -  \\
      3 & - & -  & Skewness ($\gamma$)   \\
      4 & - & -  & Kurtosis  ($\kappa$)\\
      \hline
   \end{tabular}
\end{table}

For multivariate random variables, the integrand of the expectation integral becomes a multidimensional function. Given a random vector $\boldsymbol{X}\in\mathbb{R}^{n\times 1}$, the definition of its mean is $\boldsymbol{\mu}=\mathbb{E}[\boldsymbol{X}]$ and its covariance matrix is $P = \mathbb{E}[(\boldsymbol{X}-\boldsymbol{\mu})(\boldsymbol{X}-\boldsymbol{\mu})^\top]$. However, its equivalence to the univariate standardized moments, such as skewness, is not uniquely defined, and has many different metrics based on its intended application \cite{Averous-skewness, Jammalamadaka-skewness}. The paper's interpretation of these parameters is given in the later sections.

\subsection{Conjugate Unscented Transformation}
Conjugate Unscented Transformation (CUT) \cite{Adurthi-CUT} approximates the multidimensional expectation integral by a summation of functions of discrete sigma points $\boldsymbol{x}^{(i)} \in \mathbb{R}^{n \times 1}$ with associated weights $w_i  \in \mathbb{R}$. This is shown in Eq.~\eqref{eq: CUT sigma point approx}.

\begin{equation} \label{eq: CUT sigma point approx}
    \mathbb{E}[\mathcal{G}(\boldsymbol{X})] \approx \sum^{n_s}_{i=1} w_i \mathcal{G}(\boldsymbol{x}^{(i)})
\end{equation}
In this paper, summations will only indicate the index to maintain concision. For example, $\sum^{n_s}_{i=1}$ is shortened to $\sum_i$. The calculation of $w_i$ and the number of sigma points $n_s$ depend on the size of the vector $n$ along with the highest moment desired to be estimated. 

A major advantage of CUT is its approximation of the expectation integral after nonlinear transformations. If $\boldsymbol{Y} = f(\boldsymbol{X})$ is a nonlinear transformation of a random vector and $f$ is a nice real-valued function, the corresponding $i$-th sigma point $\boldsymbol{x}^{(i)}$ and $\boldsymbol{y}^{(i)}$ can be related with the same nonlinear transformation.
\begin{equation}
     \boldsymbol{Y} = f(\boldsymbol{X})
     \qquad
     \rightarrow
     \qquad
     \boldsymbol{y}^{(i)} = f(\boldsymbol{x}^{(i)})
\end{equation}
Then $\boldsymbol{y}^{(i)}$ can be used to estimate $\mathbb{E}[f(\boldsymbol{X})]$. It should be noted that CUT is only an approximation of the first finite number of moments for the true distribution. In addition, different distributions can share the same values for a given moment, so one should be aware of the non-uniqueness of the estimated distribution.  

The initial sigma points must be sampled from an initial known distribution. This paper focuses on the Gaussian distribution as the initial distribution. Before describing the sampling of these points, the following important axes are defined first. The principal axes, denoted by $\boldsymbol{\sigma}_i$, are defined as the positive and negative standard basis vectors given in $\mathbb{R}^{n}$. For example, in $\mathbb{R}^{3}$ the principal axis are
\begin{equation}
    \boldsymbol{\sigma}_i \in
\left\{
\begin{bmatrix} 1 \\ 0 \\ 0 \end{bmatrix},
\begin{bmatrix} 0 \\ 1 \\ 0 \end{bmatrix},
\begin{bmatrix} 0 \\ 0 \\ 1 \end{bmatrix},
\begin{bmatrix} -1 \\ 0 \\ 0 \end{bmatrix},
\begin{bmatrix} 0 \\ -1 \\ 0 \end{bmatrix},
\begin{bmatrix} 0 \\ 0 \\ -1 \end{bmatrix}
\right\}
\end{equation}

The next important set of axis is the $m$-th conjugate axes, denoted by $\boldsymbol{c}^{(m)}_i$, with $m \leq n$. These axes are constructed from all the combinations of principal axes, including the sign permutations, with $m$ axes taken at a time. For example, in $\mathbb{R}^{3}$ the $2$-nd and $3$-rd conjugate axes are
\begin{equation}
\begin{aligned}
\boldsymbol{c}^{(2)}_i &\in
\left\{
\begin{bmatrix} 1 \\ 1 \\ 0 \end{bmatrix},
\begin{bmatrix} 1 \\ -1 \\ 0 \end{bmatrix},
\begin{bmatrix} 1 \\ 0 \\ 1 \end{bmatrix},
\begin{bmatrix} 1 \\ 0 \\ -1 \end{bmatrix},
\begin{bmatrix} 0 \\ 1 \\ 1 \end{bmatrix},
\begin{bmatrix} 0 \\ 1 \\ -1 \end{bmatrix},
\begin{bmatrix} -1 \\ 1 \\ 0 \end{bmatrix},
\begin{bmatrix} -1 \\ -1 \\ 0 \end{bmatrix},
\begin{bmatrix} -1 \\ 0 \\ 1 \end{bmatrix},
\begin{bmatrix} -1 \\ 0 \\ -1 \end{bmatrix},
\begin{bmatrix} 0 \\ -1 \\ 1 \end{bmatrix},
\begin{bmatrix} 0 \\ -1 \\ -1 \end{bmatrix}
\right\} \\
\boldsymbol{c}^{(3)}_i &\in
\left\{
\begin{bmatrix} 1 \\ 1 \\ 1 \end{bmatrix},
\begin{bmatrix} -1 \\ 1 \\ 1 \end{bmatrix},
\begin{bmatrix} 1 \\ -1 \\ 1 \end{bmatrix},
\begin{bmatrix} 1 \\ 1 \\ -1 \end{bmatrix},
\begin{bmatrix} -1 \\ -1 \\ 1 \end{bmatrix},
\begin{bmatrix} 1 \\ -1 \\ -1 \end{bmatrix},
\begin{bmatrix} -1 \\ 1 \\ -1 \end{bmatrix},
\begin{bmatrix} -1 \\ -1 \\ -1 \end{bmatrix}
\right\}
\end{aligned}
\end{equation}

Depending on the CUT order, the sigma points will lie on any of these axes and be scaled accordingly. This process is described next.

\subsubsection{Fourth-Order Conjugate Unscented Transformation of Gaussians}
The 4th-order CUT of Gaussians, or CUT-4G, approximates the distribution's moments up to its 4th-order, and the initial sigma points are sampled from a Gaussian distribution. The 4th-order CUT is chosen as a good starting point because it is the minimum CUT order required to estimate the higher-order moments after covariance. 

Given a standard normal random vector $\boldsymbol{X}\sim\mathcal{N}(\vec{0}, I_{n})$, the unscaled sigma points are split into two groups: points that lie on the principal axes $\boldsymbol{\sigma}_j \in \mathbb{R}^{n \times 1}$ and points that lie on the $n$-th conjugate axes $\boldsymbol{c}^{(n)}_k \in \mathbb{R}^{n \times 1}$. They are then scaled by $r_i$ to be the sigma points for $\boldsymbol{X}$, denoted by $\boldsymbol{x}^{(i)}$:
\begin{equation} \label{eq: CUT points types}
    \boldsymbol{x}^{(i)} \in \{r_1 \boldsymbol{\sigma}_j,r_2 \boldsymbol{c}^{(n)}_k \}
    \qquad
    \begin{array}{l}
    j = 1,2,\ldots,2n \\
    k =1,2,\ldots,2^{n}
    \end{array}
\end{equation}
where $w_1$ and $w_2$ corresponds to the weights of $r_1 \boldsymbol{\sigma}_j$ and $r_2 \boldsymbol{c}^{(n)}_k$ respectively. Thus, the total number of sigma points for the 4th-order standard normal case is $n_s = 2n + 2^{n}$. The scaling variables $r_i$ and the associated weights $w_i$ are calculated by the following:
\begin{equation} \label{eq: CUT points r and w}
\begin{aligned}
   r_1 = \sqrt{\frac{n+2}{2}}, & \quad&
   r_2 = \sqrt{\frac{n+2}{n -2}}, & \quad&
   w_1 = \frac{4}{(n+2)^2}, & \quad&
   w_2 = \frac{(n-2)^2}{2^{n}(n+2)^2} 
\end{aligned}
\end{equation}

To calculate the sigma points for any Gaussian $\boldsymbol{Y}\sim\mathcal{N}(\boldsymbol{\mu},\Sigma)$ with corresponding sigma points $\boldsymbol{y}^{(i)}$, the $i$-th sigma point can be related by
\begin{equation}
     \boldsymbol{y}^{(i)} = \Sigma^{1/2} \boldsymbol{x}^{(i)} + \boldsymbol{\mu}
\end{equation}
where $\Sigma =  \Sigma^{1/2}( \Sigma^{1/2})^\top$. Note that Eq.~\eqref{eq: CUT points types} and \eqref{eq: CUT points r and w} apply only to the 4th-order CUT of Gaussians. 

\subsubsection{Sixth-Order Conjugate Unscented Transformation of Gaussians for \texorpdfstring{$n\leq 6$}{n<=6}}
The 6th-order CUT of Gaussians, or CUT-6G, approximates the distribution's moments up to its 6th-order, and is theoretically more accurate than the approximations from 4th-order CUT. This section also assumes that the initial sigma points are sampled from a Gaussian distribution. 

Given a standard normal random vector $\boldsymbol{X}\sim\mathcal{N}(\vec{0}, I_{n})$ with $n\leq 6$, the unscaled sigma points are split into four groups: the central weighted sigma point $\boldsymbol{x}_0 = \vec{0} \in \mathbb{R}^{n \times 1}$, points that lie on the principal axes $\boldsymbol{\sigma}_j \in \mathbb{R}^{n \times 1}$, points that lie on the $n$-th conjugate axes $\boldsymbol{c}^{(n)}_k \in \mathbb{R}^{n \times 1}$, and points that lie on the 2nd conjugate axes $\boldsymbol{c}^{(2)}_k \in \mathbb{R}^{n \times 1}$. Similarly, they are then scaled by $r_i$ to be the sigma points for $\boldsymbol{X}$:
\begin{equation} \label{eq: CUT6 points types}
    \boldsymbol{x}^{(i)} \in \{
    \boldsymbol{x}_0,
    r_1 \boldsymbol{\sigma}_j,
    r_2 \boldsymbol{c}^{(n)}_k,
    r_3 \boldsymbol{c}^{(2)}_l 
    \}
    \qquad
    \begin{array}{l}
    j = 1,2,\ldots,2n \\
    k = 1,2,\ldots,2^{n} \\
    l = 1,2,\ldots,2n(n-1) \\
    \end{array}
\end{equation}
where $w_0$, $w_1$, $w_2$, and $w_3$ corresponds to the weights of 
$\boldsymbol{x}_0$, $r_1 \boldsymbol{\sigma}_j$, $r_2 \boldsymbol{c}^{(n)}_k$, and $r_3 \boldsymbol{c}^{(2)}_l$ respectively. Thus, the total number of sigma points for the 6th-order standard normal case is $n_s = 2n^2 + 2^n +1$ for $n\leq 6$. Unlike the 4th-order case, the scaling variables $r_i$ are not calculated with analytical expressions but rather by numerically solving a system of equations:
\begin{equation} \label{eq: CUT6 points r}
\begin{aligned}
  2(8-n)a_1^2 + a_2^2 +2a_3^2(n-1)&= 1\\
  2(8-n)a_1 + a_2 +2a_3(n-1)&= 3\\
  a_2 + 2a_3 &= 1\\
\end{aligned}
\end{equation}
where $r_1 = 1/\sqrt{a_1}$, $r_2 = 1/\sqrt{a_2}$, and $r_3 = 1/\sqrt{a_3}$. Then the weights can be calculated as 
\begin{equation} \label{eq: CUT6 points w}
\begin{aligned}
  w_1 = \frac{8-n}{r_1^6}, & \quad&
  w_2 = \frac{1}{2^n r_2^6}, & \quad&
  w_3 = \frac{1}{2r_3^6}, & \quad&
1-2nw_1 -2^n w_2 -2n(n-1)w_3 = w_0
\end{aligned}
\end{equation}

Note that the calculation of sigma points for this section requires the dimensionality to be $n\leq 6$, although the 6th-order CUT for $n > 6$ can be found in Ref.~\citenum{Adurthi-CUT}. The 6th-order CUT was chosen as the maximum CUT order analyzed in this paper, but the rest of this paper's process can be generalized to an arbitrarily high order if the computational abilities of the optimizer allow for it. For more details on sigma points calculations, higher-order sigma points, and sampling from other types of distributions, refer to Ref.~\citenum{Adurthi-CUT}.

\section{Problem Statement: Convex Formulations for Optimal Statistical Moment Steering}\label{sec: problem statement}
Let times $t_0$ to $t_f$ be discretized into $N-1$ number of segments, or $N$ number of nodes. Then, $t_0 < t_1 <\ldots<t_{N-1}=t_f$. At each time, the spacecraft's state takes the form of the random vector $\boldsymbol{X}_k \in \mathbb{R}^{n_x \times 1}$ where $n_x$ is the length of the state vector. The goal is to apply control actions $\boldsymbol{U}_k$ such that $\boldsymbol{X}_k$ satisfies any inequality or equality constraints on its statistical moments under nonlinear dynamics for all $k$. It is assumed that $\boldsymbol{U}_k$ is some function of $\boldsymbol{X}_k$ to introduce feedback. This general problem is outlined in Eq.~\eqref{eq: problem formulation nonconvex}.

\begin{equation} \label{eq: problem formulation nonconvex}
\begin{aligned}
\min_{
\{\boldsymbol{X}_k,\boldsymbol{U}_k\}_{
k\in\mathbb{Z}_{0:N-1}}
}
& J\left(\{\boldsymbol{X}_k,\boldsymbol{U}_k\}_{
k\in\mathbb{Z}_{0:N-1}}\right) &\qquad& \text{(Objective Function)}\\
\text{s.t. } & \boldsymbol{X}_{0}\sim\mathcal{D}_0
&\qquad& \text{(Initial Distribution)} 
\\
& \boldsymbol{X}_{k+1} = \phi(\boldsymbol{X}_k,\boldsymbol{U}_k),
&\quad {\forall k\in\mathbb{Z}_{0:N-2}} \quad
& \text{(Dynamics and Control)} \\
~
& g_k\left(\mathbb{E}[f^{(m)}(\boldsymbol{X}_k)] \right)\leq 0,
&\quad {\forall k\in\mathbb{Z}_{0:N-1}} \quad
& \text{(Moment Inequality Constraint)} \\
~
& h_k\left(\mathbb{E}[f^{(m)}(\boldsymbol{X}_k)] \right)= 0,
&\quad {\forall k\in\mathbb{Z}_{0:N-1}}\quad
& \text{(Moment Equality Constraint)}
\end{aligned}
\end{equation}
where $\mathcal{D}_0$ is an arbitrary initial distribution, $\phi(\cdot)$ is the solution flow under control, $g_k(\cdot)$ is any inequality constraint at $t_k$, $h_k(\cdot)$ is any equality constraint at $t_k$, and $f^{(m)}(\cdot)$ is an arbitrary function that relates $\boldsymbol{X}_k$ to any statistical moment. The notation $\mathbb{Z}_{a:b}$ is used to represent the set of integers between and including $a$ and $b$. This paper investigates a linear feedback control in the following form:
\begin{equation}
\begin{aligned}
    \boldsymbol{U}_k = \bar{\boldsymbol{u}}_k + K_k(\boldsymbol{X}_k - \boldsymbol{\mu}_k) &\qquad& \in \mathbb{R}^{n_u \times 1}
\end{aligned}
\end{equation}
where $\bar{\boldsymbol{u}}_k$ is the nominal, feedforward control action and $K_k$ is a feedback gain for the deviation from the mean trajectory $\boldsymbol{\mu}_k$ at $t_k$. This leads to an objective function to accommodate the feedback policy:
\begin{equation}
\min_{
\{\boldsymbol{X}_k,\boldsymbol{U}_k\}_{
k\in\mathbb{Z}_{0:N-1}}
}
J\left(\{\boldsymbol{X}_k,\boldsymbol{U}_k\}_{
k\in\mathbb{Z}_{0:N-1}}\right)
 \qquad
 \rightarrow
 \qquad
\min_{
\{\boldsymbol{X}_k,\bar{\boldsymbol{u}}_k,K_k\}_{
k\in\mathbb{Z}_{0:N-1}}
}
J\left(\{\boldsymbol{X}_k,\bar{\boldsymbol{u}}_k, K_k\}_{
k\in\mathbb{Z}_{0:N-1}}\right)
\end{equation}

\begin{figure}[!htb]
	\centering\includegraphics[width=\textwidth]{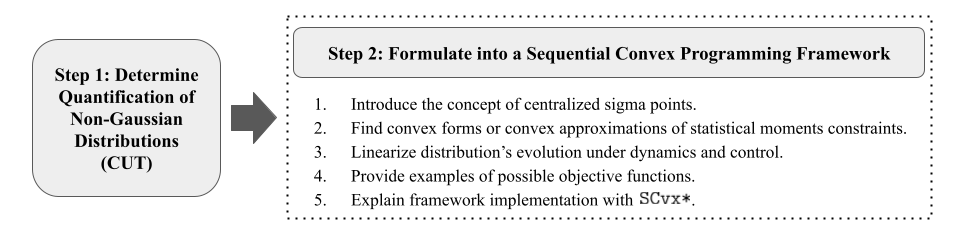}
	\caption{Flowchart of methodology for statistical moment steering.}
	\label{fig: Paper Flowchart}
\end{figure}

This paper takes a sequential convex programming approach to solving Eq.~\eqref{eq: problem formulation nonconvex} and uses a specific algorithm called \texttt{SCvx*} \cite{Oguri-SCVSstar}. There are numerous nonconvex elements to the problem, so the following sections outline techniques for converting the problem into one that is solvable with convex optimization. An outline of this section is shown in Figure~\ref{fig: Paper Flowchart}, which lists the step-by-step processes taken to develop statistical moment steering. These steps are described in more detail in the following sections.

\subsection{Centralized Sigma Points}
The main focus is to control the random variable for state $\boldsymbol{X}_k$ under some constraints on its statistical moments. However, linearizing statistical moments about this value may lead to cumbersome mathematical expressions. Instead, a new ``centralized'' random variable $\boldsymbol{Z}_k$ is defined: 
\begin{equation}
    \boldsymbol{Z}_k = \boldsymbol{X}_k - \mathbb{E}[\boldsymbol{X}_k]
\end{equation}
where it can be seen that $\mathbb{E}[\boldsymbol{Z}_k] = 0$. As a result, $m$-th central moment of $\boldsymbol{X}_k$ is equivalent to the $m$-th raw moment of $\boldsymbol{Z}_k$. This fact is leveraged to simplify the expressions during linearization of the higher-ordered moments. 

Firstly, the variational relationship between the state and the centralized-state random variable is computed. Let the state and centralized sigma point at the $k$-th instance be aggregated into a single vector denoted by $\boldsymbol{x}_k$ and $\boldsymbol{z}_k$ respectively.  
\begin{equation}
    \begin{aligned}
 \boldsymbol{x}_k =
 \begin{bmatrix}
     \boldsymbol{x}_k^{(1)} \\
     \vdots
     \\
     \boldsymbol{x}_k^{(n_s)} 
 \end{bmatrix}
&\qquad&
\boldsymbol{z}_k =
 \begin{bmatrix}
     \boldsymbol{z}_k^{(1)} \\
     \vdots
     \\
     \boldsymbol{z}_k^{(n_s)} 
 \end{bmatrix}
 &\qquad&
 \in\mathbb{R}^{n_x n_s \times 1}
    \end{aligned}
\end{equation}

Let $\boldsymbol{x}^{(i)}_k = E_i \boldsymbol{x}_k$, $\boldsymbol{z}^{(i)}_k =E_i \boldsymbol{z}_k$, where $E_i$ is a matrix that selects the $i$-th sigma point from the aggregated sigma point vector. It is seen that
\begin{equation}
    \boldsymbol{z}_{k}^{(i)} =\boldsymbol{x}_{k}^{(i)} - \sum_i w_i \boldsymbol{x}_{k}^{(i)}
\end{equation}

With the aggregated sigma point form,
\begin{equation} \label{eq: centralized sigma points}
\begin{aligned} 
    \boldsymbol{z}_{k} &= \boldsymbol{x}_{k} - \bar{I}\sum_i w_i E_i \boldsymbol{x}_{k} 
    = \left(I_{n_x n_s} - \bar{I}\sum_i w_i E_i \right)
    \boldsymbol{x}_{k} = A^{(z)} \boldsymbol{x}_{k}
\end{aligned}
\end{equation}
where $\bar{I} = [I_{n_x}\;I_{n_x}\ldots\;I_{n_x}]^\top$. It can be seen that the relationship between the aggregated forms of the state and centralized sigma points is both linear and time invariant.

\subsection{Convex Forms of Statistical Moments}
Let ($^*$) denote the reference value of a parameter. Then, $\boldsymbol{x}_{k} = \boldsymbol{x}_{k}^* + \delta \boldsymbol{x}_{k}$ and $\boldsymbol{z}_{k} = \boldsymbol{z}_{k}^* + \delta \boldsymbol{z}_{k}$. It can be seen that $\delta \boldsymbol{z}_{k} =  A^{(z)} \delta \boldsymbol{x}_{k}$ due to their linear relationship.

\subsubsection{Mean}
The state mean at $t_k$ is defined as $\boldsymbol{\mu}_k = \mathbb{E}[\boldsymbol{X}_k]$. Calculating with the state's sigma points,
\begin{equation}
    \begin{aligned}
        f^{(\mu)} (\boldsymbol{x}_k)&= \boldsymbol{\mu}_k= \sum_i w_i \boldsymbol{x}^{(i)}_k = \sum_i w_i E_i \boldsymbol{x}_k= \left(\sum_i w_i E_i \right) \boldsymbol{x}_k = A^{(\mu)} \boldsymbol{x}_k 
    \end{aligned}
\end{equation}
This shows that the relationship between the aggregated forms of the state and the state mean is also linear and time invariant. It follows that 
\begin{equation}  \label{eq: mean}
    \boldsymbol{\mu}_k = A^{(\mu)}(\boldsymbol{x}_{k}^* + \delta \boldsymbol{x}_{k}) = \boldsymbol{\mu}_k^* + A^{(\mu)} \delta\boldsymbol{x}_k 
    \qquad
    \in \mathbb{R}^{n_x \times 1}
\end{equation}

\subsubsection{Covariance Matrix}
The state covariance at $t_k$ is defined as $P_k = \mathbb{E}[(\boldsymbol{X}_k-\boldsymbol{\mu}_k)(\boldsymbol{X}_k-\boldsymbol{\mu}_k)^\top]$. This can be more conveniently represented as $P_k = \mathbb{E}[\boldsymbol{Z}_k\boldsymbol{Z}_k^\top]$.
\begin{equation}
    \begin{aligned}
        f^{(P)} (\boldsymbol{x}_k) &= P_k =\sum_i w_i (\boldsymbol{z}^{(i)}_k) (\boldsymbol{z}^{(i)}_k)^\top
        =\sum_i w_i E_i \boldsymbol{z}_k \boldsymbol{z}_k^\top E_i^\top
    \end{aligned}
\end{equation}

Assuming small variation such that $\delta \boldsymbol{z}_k \delta \boldsymbol{z}_k^\top \approx 0$,
\begin{equation}
\boldsymbol{z}_k \boldsymbol{z}_k^\top=
     (\boldsymbol{z}_k^*+\delta \boldsymbol{z}_k)(\boldsymbol{z}_k^*+\delta \boldsymbol{z}_k)^\top \approx
     \boldsymbol{z}_k^*{\boldsymbol{z}_k^*}^\top +\boldsymbol{z}_k^* \delta {\boldsymbol{z}_k}^\top
     +\delta \boldsymbol{z}_k{\boldsymbol{z}_k^*}^\top 
\end{equation}

It can be seen that 
\begin{equation}
    \begin{aligned}
P_k
 &\approx P_k^* +
\underbrace{\sum_i w_i E_i \boldsymbol{z}_k^*\delta \boldsymbol{z}_k^\top E_i^\top}_{\triangleq \delta P_k}
+
\underbrace{\sum_i w_i E_i \delta \boldsymbol{z}_k {\boldsymbol{z}_k^*}^\top E_i^\top}_{= \delta P_k^\top} 
    \end{aligned}
\end{equation}

The linearization still results in a symmetric matrix. However, covariance matrices must be positive-semidefinite, which linearization does not guarantee. An additional convex PSD constraint is placed to ensure that the linearized covariance matrix is still valid:
\begin{equation} \label{eq: Pk}
\begin{aligned}
P_k 
\approx P_k^* + \delta P_k\biggr\rvert_{\boldsymbol{z}_k^*}  + \delta P_k^\top\biggr\rvert_{\boldsymbol{z}_k^*} 
\qquad
\in \mathbb{R}^{n_x \times n_x} &\qquad,\qquad&
P_k \succeq 0
\end{aligned}
\end{equation}

\subsubsection{Convex Square-root Covariance Relationship}
The previous section presented a full covariance approach, but this introduces nonconvexity and potential numerical issues. Since the covariance $P_k$ depends on the square of the optimization variable, solvers may become unstable when $\boldsymbol{z}_k$ is either very small or very large. Previous works remedy the nonconvexity with linear matrix inequality convexifications when the covariance is propagated directly with linearized dynamics (rather than indirectly through sigma points) \cite{Bakolas-Cov-Steering, Naoya-Sequential-Cov-Steering, Benedikter-CovSteering}, and by introducing scaling variables to mitigate numerical issues \cite{Naoya-Sequential-Cov-Steering}. This paper takes an approach by considering only the square root of the covariance. The covariance matrix can be decomposed by its square root $P_k^{1/2}$ with the relationship.
\begin{equation}
    P_k = (P_k^{1/2})(P_k^{1/2})^\top
\end{equation}

From before, it is shown that
\begin{equation}
    \begin{aligned}
        P_k &=\sum_i w_i (\boldsymbol{z}^{(i)}_k) (\boldsymbol{z}^{(i)}_k)^\top = 
        w_1 (\boldsymbol{z}^{(1)}_k) (\boldsymbol{z}^{(1)}_k)^\top
        +
        w_2 (\boldsymbol{z}^{(2)}_k) (\boldsymbol{z}^{(2)}_k)^\top
        +
        \ldots
    \end{aligned}
\end{equation}

It can be seen that $P_k^{1/2}$ can be found by
\begin{equation}\label{eq: Pk_sqrt} 
    f^{(P^{1/2})} (\boldsymbol{x}_k) = P_k^{1/2} 
    =
    \begin{bmatrix}
    \sqrt{w_1}  \boldsymbol{z}^{(1)}_k
    &
    \sqrt{w_2}  \boldsymbol{z}^{(2)}_k
    &
    \ldots
    \end{bmatrix}
    \qquad
\in \mathbb{R}^{n_x \times n_s}
\end{equation}

This shows that $P_k^{1/2}$ is linear with respect to $\boldsymbol{z}_k$, and thus linear with respect to $\boldsymbol{x}_k$ as well. An important relationship is that 
\begin{equation}\label{eq: Pk_sqrt norm} 
    \sqrt{\lambda_{\max}(P_k)} = \norm{P_k^{1/2}}_2
\end{equation}
where $\lambda_{\max}\left(\cdot\right)$ returns the largest magnitude eigenvalue of the input. This relationship shows that the maximum variance along the principal directions of covariance can be determined from just its square-root form. Unlike the full covariance approach, $P_k^{1/2}$ is inherently a convex expression and requires no linearization, and is also the same order as the optimization variable $\boldsymbol{z}_k$. If the full covariance is not needed, instead of Eq.~\eqref{eq: Pk}, it is recommended to use this expression Eq.~\eqref{eq: Pk_sqrt} and \eqref{eq: Pk_sqrt norm} since this expression is inherently linear and convex.

\subsubsection{Skewness}
As mentioned before, measures of skewness and higher-order statistical moments are not uniquely defined for multivariate distributions. Some measures can involve tensor operations \cite{Jammalamadaka-skewness} or re-defining the notation of standardized moments altogether \cite{Averous-skewness}. For simplicity, this paper sticks with the univariate definition of skewness, and considers only the skewness along the vector basis directions. Consider the skewness along the $j$-th axis at $t_k$:

\begin{equation}
    \gamma_{j,k} = \mathbb{E}\left[\left(\frac{X_{j,k} - \mu_{j,k}}{\sigma_{j,k}}\right)^3\right] =
    \frac{\mathbb{E}\left[\left(X_{j,k} - \mu_{j,k}\right)^3\right]}{\mathbb{E}\left[\left(X_{j,k} - \mu_{j,k}\right)^2\right]^{3/2}} 
    =
    \mathbb{E}\left[Z_{j,k}^3\right]
    \mathbb{E}\left[Z_{j,k}^2\right]^{-3/2}
\end{equation}

Let $e_j$ be the matrix that selects the $j$-th's elements from an $n_x$-dimensional vector, or $z^{(i)}_{j,k}=e_j\boldsymbol{z}^{(i)}_k =e_jE_i \boldsymbol{z}_k$. The function that calculates the skewness along each element using sigma points $f^{(\gamma)} (\boldsymbol{x}_k) =\boldsymbol{\gamma}_k$ can be defined, where the $j$-th element of $\boldsymbol{\gamma}_k$ is calculated with sigma points:
\begin{equation}
\gamma_{j,k} 
=
\left( \sum_i w_i (z^{(i)}_{j,k})^3\right)
\left( 
\sum_i w_i (z^{(i)}_{j,k})^2
\right)^{-3/2}
=
\left( \sum_i w_i (e_j E_i\boldsymbol{z}_k)^3\right)
\left( 
\sum_i w_i (e_j E_i\boldsymbol{z}_k)^2
\right)^{-3/2}
\end{equation}

Linearizing skewness yields,
\begin{equation}
    \gamma_{j,k} \approx \gamma_{j,k}^* + \frac{\partial \gamma_{j,k}}{\partial \boldsymbol{z}_k} \biggr\rvert_{\boldsymbol{z}_k^*} \delta \boldsymbol{z}
\end{equation}

Taking the partial derivative,
\begin{equation}
\frac{\partial \gamma_{j,k}}{\partial \boldsymbol{z}_k} 
=
 3 \left( \sum_i w_i (e_j E_i\boldsymbol{z}_k)^2 e_j E_i\right)
\mathbb{E}\left[Z_{j,k}^2\right]^{-3/2}
-3
\mathbb{E}\left[Z_{j,k}^3\right]
\mathbb{E}\left[Z_{j,k}^2\right]^{-5/2}
 \left( \sum_i w_i (e_j E_i\boldsymbol{z}_k) e_j E_i\right) \\
\end{equation}

A matrix $A^{(\gamma)} = \left[ 
\left(\frac{\partial \gamma_{1,k}}{\partial \boldsymbol{z}_k}\right)^\top 
\;\ldots\;
\left(\frac{\partial \gamma_{n_x,k}}{\partial \boldsymbol{z}_k}\right)^\top
\right]^\top$ can be defined to compute the entire skewness vector. While the expression may seem cumbersome, Appendix~\ref{appendix: computation of gamma matrix} presents an efficient computation of this matrix. The linearized skewness is then
\begin{equation}
    \boldsymbol{\gamma}_{k} \approx \boldsymbol{\gamma}_{k}^* + A^{(\gamma)} \biggr\rvert_{\boldsymbol{z}_k^*}\delta \boldsymbol{z}_k     \qquad
    \in \mathbb{R}^{n_x \times 1}
\end{equation}

\subsubsection{m-th Standardized Moment}
The linearization process for skewness, when considering its value along the basis vector direction, can be generalized to any $m$-th standardized moment when $m \geq 3$.

\begin{equation}
    {^mC_{j,k}} = \mathbb{E}\left[\left(\frac{X_{j,k} - \mu_{j,k}}{\sigma_{j,k}}\right)^m\right] =
    \mathbb{E}\left[Z_{j,k}^m\right]\mathbb{E}\left[Z_{j,k}^2\right]^{-m/2}
\end{equation}

The corresponding function involving sigma points $f^{(^mC)} (\boldsymbol{x}_k) = {^m\boldsymbol{C}_{k}}$ can be defined, and its partial derivative with respect to $\boldsymbol{z}_k$ is as follows:
\begin{equation}
\begin{aligned}
\frac{\partial {^mC_{j,k}}}{\partial \boldsymbol{z}_k} 
&=
 m \left( \sum_i w_i (e_j E_i\boldsymbol{z}_k)^{m-1} e_j E_i\right)
\mathbb{E}\left[Z_{j,k}^2\right]^{-m/2}
-m
\mathbb{E}\left[Z_{j,k}^m\right]
\mathbb{E}\left[Z_{j,k}^2\right]^{-m/2-1}
 \left( \sum_i w_i (e_j E_i\boldsymbol{z}_k) e_j E_i\right) \\
 \end{aligned}
\end{equation}

A matrix $A^{(^m C)}$ can be constructed similarly to $A^{(\gamma)}$ using the same efficient method from Appendix~\ref{appendix: computation of gamma matrix}. The linearized equation for the $m$-th standardized moment is then
\begin{equation}
    {^m\boldsymbol{C}_{k}} \approx {^m\boldsymbol{C}_{k}}^* + A^{(^m C)} \biggr\rvert_{\boldsymbol{z}_k^*}\delta \boldsymbol{z}_k     \qquad
    \in \mathbb{R}^{n_x \times 1}
\end{equation}

\subsection{Linearization of Sigma Point Dynamics}
The random variables associated with state and control are written in terms of their sigma points. 
\begin{equation}
    \boldsymbol{X}_{k+1} = \phi(\boldsymbol{X}_{k}, \boldsymbol{U}_k) \qquad \rightarrow \qquad \boldsymbol{x}^{(i)}_{k+1} = \phi(\boldsymbol{x}^{(i)}_{k}, \boldsymbol{u}^{(i)}_k) 
\end{equation}

Feedback control can be expressed more conveniently in terms of $\boldsymbol{Z}$. Thus, at the $i$-th sigma point, its corresponding control is
\begin{equation} \label{eq: control sigma points}
\begin{aligned}
\boldsymbol{U}_k=\bar{\boldsymbol{u}}_k + K_k\boldsymbol{Z}_k 
\qquad \rightarrow \qquad
\boldsymbol{u}^{(i)}_k=\bar{\boldsymbol{u}}_k + K_k\boldsymbol{z}^{(i)}_k 
\end{aligned}
\end{equation}
This form of control is nonconvex due to the $K_k\boldsymbol{z}^{(i)}_k $ term: the optimization variable $K_k$ is multiplied by $\boldsymbol{z}^{(i)}_k$, which is an affine function of the other optimization variable $\boldsymbol{x}_k$. Assuming small variation $\delta K_k \delta\boldsymbol{z}^{(i)}_{k} \approx 0$,

\begin{equation} \label{eq: linearized feedback policy}
\begin{aligned}
\boldsymbol{u}^{(i)}_k&=
(\bar{\boldsymbol{u}}_k^* + \delta\bar{\boldsymbol{u}}_k) 
+ (K_k^*+\delta K_k)({\boldsymbol{z}^{(i)*}_k} + \delta \boldsymbol{z}^{(i)}_k) \\
& \approx (\bar{\boldsymbol{u}}_k^* + \delta\bar{\boldsymbol{u}}_k) + K_k^* {\boldsymbol{z}^{(i)*}_k} + K_k^*\delta \boldsymbol{z}^{(i)}_k + \delta K_k {\boldsymbol{z}^{(i)*}_k} = \boldsymbol{u}_k^{(i)*} + \delta\bar{\boldsymbol{u}}_k + K_k^*\delta \boldsymbol{z}^{(i)}_k + \delta K_k {\boldsymbol{z}^{(i)*}_k}
\end{aligned}
\end{equation}

The sigma point dynamics can be linearized and discretized about a reference sigma point trajectory. This process is the same as typical convex trajectory optimization \cite{Oguri-Chance-Paper, Naoya-Sequential-Cov-Steering}.
\begin{equation}
    \begin{aligned}
\boldsymbol{x}^{(i)}_{k+1} &\approx A^{(i)}_k\boldsymbol{x}^{(i)}_{k} + B^{(i)}_k\boldsymbol{u}^{(i)}_k + \boldsymbol{c}^{(i)}_k\\
    \end{aligned}
\end{equation}

The solution flow at the $i$-th sigma point can be linearized about ${\boldsymbol{x}^{(i)*}}$ and written in terms of the feedback from Eq.~\eqref{eq: linearized feedback policy}, 
\begin{equation} \label{eq: linearized sigma point dynamics}
\begin{aligned}
    {\boldsymbol{x}^{(i)*}_{k+1}} + \delta \boldsymbol{x}^{(i)}_{k+1} 
    &\approx A^{(i)}_k 
    ({\boldsymbol{x}^{(i)*}_{k}} + \delta \boldsymbol{x}^{(i)}_{k}) + B_k 
    (
     \boldsymbol{u}_k^{(i)*} + \delta\bar{\boldsymbol{u}}_k + K_k^*\delta \boldsymbol{z}^{(i)}_k + \delta K_k {\boldsymbol{z}^{(i)*}_k}
    )+ \boldsymbol{c}^{(i)}_k
    \\
\end{aligned}
\end{equation}

It is important to note that the dynamics of each sigma point are linearized independently from each other (i.e., not linearized about a common mean trajectory). This is to ensure that nonlinear dynamics are captured by the distribution even with linearization of dynamics for individual sigma point, so in most cases $A^{(i)}_k \neq A^{(j)}_k$ when $i\neq j$. This also applies to $B^{(i)}_k$ and $\boldsymbol{c}^{(i)}_k$.

\subsection{Objective Functions for Control Cost Distributions}
A common objective function is the minimization of fuel costs or control action. In the problem of distribution steering, the total control cost is itself a random variable, so its minimization is not well defined and can hold many interpretations. Some stochastic control algorithms minimize the feedforward term \cite {fife-GMM-steering} or consider some information from the control covariance \cite{Ridderhof-SCP, fife-GMM-steering-conference, Benedikter-CovSteering, Ozaki-UT-TrajOp}. This paper takes a percentile approach for fuel minimization. This approach focuses on minimizing the $99$-th percentile of the total fuel cost, which is known as ``$\Delta V_{99}$'' \cite{Oguri-Chance-Paper, Naoya-Sequential-Cov-Steering, Jerry-CovSteering}. 
\begin{equation}
\begin{aligned}
    \Delta V_{99,k} &\triangleq Q_{\norm{\boldsymbol{U}_k}_2}(0.99)
    \\
   \Delta V_{99} &\triangleq Q_{\sum_k \norm{\boldsymbol{U}_k}_2}(0.99) 
\end{aligned}
\end{equation}
where $Q_{\norm{\boldsymbol{U}_k}_2}(p)$ and $Q_{\sum_k \norm{\boldsymbol{U}_k}_2}(p)$ is the quantile function for fuel cost at $t_k$ and total fuel costs respectively. Previous works assume that $\boldsymbol{U}_k$ is a Gaussian vector, so then a chi-squared quantile function can be used to upper-bound the $99$-th percentile of $\norm{\boldsymbol{U}_k}_2$ \cite{Oguri-Chance-Paper, Naoya-Sequential-Cov-Steering}. Under the Gaussian approximation, $\boldsymbol{U}_k$ is fully characterized by its mean and covariance: recall that $\boldsymbol{U}_k$ is a function of $\boldsymbol{Z}_k$ from Eq.~\eqref{eq: control sigma points}. Since $\mathbb{E}[\boldsymbol{Z}_k] = 0$ and $\mathbb{E}[\boldsymbol{Z}_k\boldsymbol{Z}_k^\top] = P_k$,
\begin{equation}
\begin{aligned}
    \mathbb{E}[\boldsymbol{U}_k] = \bar{\boldsymbol{u}}_k, 
    \qquad & \qquad
    \text{Cov}(\boldsymbol{U}_k) \triangleq P_{u_k} =K_k P_k K_k^\top,
    \qquad & \qquad
     P_{u_k}^{1/2} =K_k P_k^{1/2}
\end{aligned}
\end{equation}

Then, an upper-bound can be provided on the fuel cost quantile functions.
\begin{equation} \label{eq: deltaV 99}
\begin{aligned}
    \Delta V_{99,k} &\leq \norm{\bar{\boldsymbol{u}}_k}_2 + \sqrt{Q_{\chi^2_{n_u}}(0.99)} \norm{P_{u_k}^{1/2}}_2 
    \\
   \Delta V_{99} &\leq 
   \Delta V_{99,\text{ub}} = \sum_k \norm{\bar{\boldsymbol{u}}_k}_2 + \sqrt{Q_{\chi^2_{n_u}}(0.99)} \norm{P_{u_k}^{1/2}}_2 
\end{aligned}
\end{equation}
where $\Delta V_{99,\text{ub}}$ is the $\Delta V_{99}$ upper-bound, $Q_{\chi^2_{n_u}}(p)$ is the quantile function for a chi-squared distribution with degree of freedom $n_u$. In \texttt{MATLAB}, this function is $\texttt{chi2inv}(p,n_u)$. However, Eq.~\eqref{eq: deltaV 99} is not convex since both $K_k$ and $P_k$ are functions of optimization variables. Inexact linearization from perturbing the optimization variable yields
\begin{equation} \label{eq: deltaV 99 convex}
    \Delta V_{99,\text{ub}} \approx \sum_k \norm{\bar{\boldsymbol{u}}_k^* + \delta\bar{\boldsymbol{u}}_k}_2 + \sqrt{Q_{\chi^2_{n_u}}(0.99)} \norm{(K_k^* + \delta K_k)(P_k^*)^{1/2}}_2
\end{equation}
Note that the exact linearization of the $KP_k^{1/2}$ term in this expression requires tensor operations or the flattening of the matrices down to vectors, which can result in cumbersome mathematical expressions. This was not pursued since \texttt{SCvx*} has the ability to handle inexact linearization \cite{Naoya-Sequential-Cov-Steering}, and improved mathematical readability was prioritized over exact gradient accuracy.

Another useful metric is the expected value for fuel consumption, which can be calculated with the corresponding control sigma points. Since the $l^2$-norm is a nonlinear transformation and $\mathbb{E}\left[\norm{\boldsymbol{U}_k}_2\right] \neq \norm{\bar{\boldsymbol{u}}_k}_2$, sigma points are used to approximate this expectation.
\begin{equation} \label{eq: expected fuel cost}
\begin{aligned}
    \mathbb{E}\left[\sum_k \norm{\boldsymbol{U}_k}_2\right] &= \sum_k \sum_i w_i \norm{\boldsymbol{u}^{(i)}_k}_2
    =\sum_k\sum_i w_i \norm{\bar{\boldsymbol{u}}_k + K_k\boldsymbol{z}^{(i)}_k}_2
\end{aligned}
\end{equation}
Note that Eq.~\eqref{eq: expected fuel cost} still requires linearization for it to be a convex objective function. Both Eq.~\eqref{eq: deltaV 99} and Eq.~\eqref{eq: expected fuel cost} are different interpretations of a cost function involving the minimization of fuel costs. Eq.~\eqref{eq: deltaV 99} can be viewed as minimizing the maximum fuel cost up to a certain confidence level. While $\Delta V_{99,\text{ub}}$ is an upper-bounding function, it still requires a Gaussian assumption for the control distribution, so the accuracy of this bound may vary. Previous works \cite{Oguri-Chance-Paper, Naoya-Sequential-Cov-Steering}, along with upcoming results in Sections~\ref{sec: numerical example two body} and \ref{sec: numerical example three body}, show that this bound is reasonable. On the other hand, Eq.~\eqref{eq: expected fuel cost} and its use of sigma points can directly approximate its parameter (the actual expected value for fuel cost) without the Gaussian assumption. However, this formulation lacks the ability to minimize fuel costs from the worst-case scenarios. The selection of a formulation for the minimum-fuel objective function is left to the user's discretion.

\subsection{Summary of Convex Formulation}
Eq.~\eqref{eq: problem formulation convex} presents the convexified problem of Eq.~\eqref{eq: problem formulation nonconvex}. The convex objective function for the upper-bound to $\Delta V_{99}$ is found in Eq.~\eqref{eq: deltaV 99 convex}. As previously noted, there are many interpretations of ``minimum fuel'' since fuel cost is now represented as a distribution, and a linearized version of Eq.~\eqref{eq: expected fuel cost} can be used to minimize the expected fuel cost.

\begin{equation} \label{eq: problem formulation convex}
\begin{aligned}
\min_{\{\delta \boldsymbol{x}_k,\delta\bar{\boldsymbol{u}}_k,\delta K_k\}_{k\in\mathbb{Z}_{0:N-1}}} 
&  J_{\text{cvx}}\left(\{\delta \boldsymbol{x}_k,\delta\bar{\boldsymbol{u}}_k,\delta K_k\}_{k\in\mathbb{Z}_{0:N-1}}\right) &\qquad& \text{(Convex Objective Function)}\\
\text{s.t. }
&\boldsymbol{x}^{*}_{0} + \delta \boldsymbol{x}_{0} \leftarrow \text{CUT} 
&
& \text{(Initial Distribution Sampled with CUT)} \\
& \delta \boldsymbol{x}^{(i)}_{k+1} \leftarrow \text{Eq.~\eqref{eq: linearized sigma point dynamics}}, 
&{\forall k\in\mathbb{Z}_{0:N-2},\forall i\in\mathbb{Z}_{1:n_s}} \quad
& \text{(Linearized Sigma Point Dynamics)} \\
~
& g_{\text{cvx},k}\left(f^{(m)}_{\text{affine}}(\delta\boldsymbol{x}_k) \right)\leq 0, 
&{\forall k\in\mathbb{Z}_{0:N-1}} \quad
& \text{(Convex Moment Inequality Constraint)} \\
~
& h_{\text{affine},k}\left(f^{(m)}_{\text{affine}}(\delta\boldsymbol{x}_k) \right)= 0, 
&{\forall k\in\mathbb{Z}_{0:N-1}} \quad
& \text{(Affine Moment Equality Constraint)} \\
~
& f^{(m)}_{\text{affine}}(\delta \boldsymbol{x}_k) \leftarrow \text{Table~\ref{tab: moments linearized}} & & \text{(Linearized Statistical Moments)}
\end{aligned}
\end{equation}

\begin{table}[htbp]
	\fontsize{10}{10}\selectfont
    \caption{Summary of Linearized Forms of Statistical Moments}
   \label{tab: moments linearized}
    \centering 
   \begin{tabular}{l c  c  c} 
      \hline 
      Statistical Moment & Expectation Form & Aggregated Sigma Point Form & Originally Linear? \\
      \hline
      Mean ($\boldsymbol{\mu}_k$)  & $\mathbb{E}[\boldsymbol{X}_k]$ & $\boldsymbol{\mu}_k^* + A^{(\mu)} \delta\boldsymbol{x}_k$ & Yes\\
      Covariance ($P_k$) & $\mathbb{E}[\boldsymbol{Z}_k\boldsymbol{Z}_k^\top]$  & $P_k^* + \delta P_k\biggr\rvert_{\boldsymbol{z}_k^*}  + \delta P_k^\top\biggr\rvert_{\boldsymbol{z}_k^*} \succeq 0$ & No \\
    Square-root Covariance ($P_k^{1/2}$) & - &    $\begin{bmatrix}
    \sqrt{w_1}  \boldsymbol{z}^{(1)}_k
    &
    \sqrt{w_2}  \boldsymbol{z}^{(2)}_k
    &
    \ldots
    \end{bmatrix}$ & Yes\\
      Skewness ($\boldsymbol{\gamma}_{k}$) & $\mathbb{E}\left[Z_{j,k}^3\right]
    \mathbb{E}\left[Z_{j,k}^2\right]^{-3/2}$  & $\boldsymbol{\gamma}_{k}^* + A^{(\gamma)} \biggr\rvert_{\boldsymbol{z}_k^*}\delta \boldsymbol{z}_k$ & No\\
      $m$-th Standardized Moment (${^m\boldsymbol{C}_{k}}$)& $\mathbb{E}\left[Z_{j,k}^m\right]\mathbb{E}\left[Z_{j,k}^2\right]^{-m/2}$   &  ${^m\boldsymbol{C}_{k}}^* + A^{(^m C)} \biggr\rvert_{\boldsymbol{z}_k^*}\delta \boldsymbol{z}_k$ & No \\
      \hline
   \end{tabular}
\end{table}

Table~\ref{tab: moments linearized} lists the convex forms of the statistical moments. Since the expressions for the statistical moments are all affine with respect to the optimization variables, $g_{\text{cvx},k}(\cdot)$ can be any convex function and $h_{\text{affine},k}(\cdot)$ must be affine. Other nonconvex expressions for constraints not considered in this paper can also be included, but a similar linearization process needs to be performed. As a reminder, most of the moments in Table~\ref{tab: moments linearized} focus on optimizing the centralized sigma points rather than the state sigma points due to their simpler expressions. These two types of sigma points are related by the linear relationship $\delta \boldsymbol{z}_{k} = A^{(z)} \delta \boldsymbol{x}_{k}$. 

\section{Implementation with \texttt{SCvx*}}\label{sec: scvx implementation}
Since the problem is inherently nonconvex, sequential convex programming (SCP) is used to obtain a solution. This paper utilizes an SCP algorithm known as \texttt{SCvx*}, which has theoretical guarantees for convergence to a feasible local solution \cite{Oguri-SCVSstar}. While this paper's framework with the CUT points holds for any SCP algorithm or nonlinear optimization method, \texttt{SCvx*} was chosen for its aforementioned convergence properties.

The crux of this section is to determine the slack variable assignment to the nonconvex constraints needed for \texttt{SCvx*}. The optimization variables in this problem are the deviation of state sigma points $\delta\boldsymbol{x}_k$, nominal control $\delta\bar{\boldsymbol{u}}_k$, and gain $\delta K_k$. As stated in Ref.~\citenum{Oguri-SCVSstar}, only nonconvex constraints require slack variables. If $g_k (\delta\boldsymbol{x}_k) \leq 0$ and $h_k (\delta\boldsymbol{x}_k) = 0$ are originally nonconvex constraints, slack variables can be assigned to remove the strict zero equality/inequality of the linearized $\Tilde{g}_k(\delta\boldsymbol{x}_k)$ and $\Tilde{h}_k(\delta\boldsymbol{x}_k)$ constraint functions.
\begin{equation}\label{eq: scvx slack var}
\Tilde{g}_k(\delta\boldsymbol{x}_k) \leq \zeta_j \quad,\quad \zeta_j \geq 0
\qquad\qquad
\Tilde{h}_k(\delta\boldsymbol{x}_k) = \xi_i 
\end{equation}

The slack variables are used in the calculation of a penalty function within each convex optimization iteration.
\begin{equation}\label{eq: scvx star penalty}
    P_{\texttt{SCvx*}}(w_p, \boldsymbol{\xi}, \boldsymbol{\lambda}, \boldsymbol{\zeta}, \boldsymbol{\mu})
    =
    \boldsymbol{\lambda} \cdot \boldsymbol{\xi} + \frac{w_p}{2} \boldsymbol{\xi}\cdot \boldsymbol{\xi}
    +
    \boldsymbol{\mu} \cdot [\boldsymbol{\zeta}]_+ + \frac{w_p}{2} [\boldsymbol{\zeta}]_+ \cdot [\boldsymbol{\zeta}]_+
\end{equation}
where $w_p$ is the penalty weight, $\boldsymbol{\xi}$ and $\boldsymbol{\zeta}$ are vectors containing all the slack equality and inequality constraints, $\boldsymbol{\lambda}$ and $ \boldsymbol{\mu}$ are the Lagrange multiplier vectors corresponding to the slack equality and inequality constraints, and the function $[\cdot]_+ = \max\{0,\cdot\}$ is performed element-wise. The penalty weight and Lagrange multipliers are updated throughout the SCP algorithm to ensure feasibility guarantees to the original nonconvex problem. Other solver parameters are needed for the rest of the \texttt{SCvx*} algorithm: convergence criterion $\{ \epsilon_{\text{opt}},\epsilon_{\text{feas}}\}$, solution acceptance thresholds $\{\eta_0,\eta_1,\eta_2\}$, parameters for trust region update $\{\alpha_1,\alpha_2\}$, parameters for lagrange multiplier update $\{\beta_{\texttt{SCvx*}},\gamma_{\texttt{SCvx*}}\}$, minimum and maximum trust regions $\{\Delta_{\text{TR},\min},\Delta_{\text{TR},\max}\}$, maximum penalty $w_{p,\max}$, and initial trust region and penalty values ($\Delta_{\text{TR}}^{(1)}$ and $w^{(1)}_p$ respectively). A pseudocode of the \texttt{SCvx*} algorithm applied to statistical moment steering is shown in Algorithm~\ref{code: SCVX code}. For more details on \texttt{SCvx*}'s implementation in trajectory optimization problems, refer to Ref.~\citenum{Naoya-Sequential-Cov-Steering}, and for the original algorithm, refer to Ref.~\citenum{Oguri-SCVSstar}.

\begin{algorithm}[!htb]
\caption{Optimal Statistical Moment Steering via \texttt{SCvx*}\label{code: SCVX code}}
\begin{algorithmic}[1]
\Require Initial reference sigma points $\{\boldsymbol{x}^*_k\}$ and control $\{\bar{\boldsymbol{u}}_k^*, K_k^*\}$ for all $k$.
\State Compute $A^{(z)}$ and $A^{(\mu)}$ 
\While{iterations don’t exceed the maximum}
    \If{first iteration or the reference was updated in the previous iteration}
        \State Compute linearized sigma point dynamics $\{A^{(i)}_k, B^{(i)}_k, \boldsymbol{c}^{(i)}_k\}$, and the reference moments $\{\boldsymbol{\mu}_k^*, \boldsymbol{\gamma}_{k}^*, {^m\boldsymbol{C}_{k}}^*\}$ along with their Jacobians $\{A^{(\gamma)}_k,A^{(^m C)}_k\}$ from $\{\boldsymbol{x}^*_k, \bar{\boldsymbol{u}}_k^*, K_k^*\}$
    \EndIf
    \State $\{\delta \boldsymbol{x}_k,\delta\bar{\boldsymbol{u}}_k,\delta K_k\} \gets$ solve convex subproblem from Eq.~\eqref{eq: problem formulation convex}, with slack variables from Eq.~\ref{eq: scvx slack var} and additional penalties from Eq.~\eqref{eq: scvx star penalty}
    \If{acceptance conditions met}
        \State $\{\boldsymbol{x}^*_k, \bar{\boldsymbol{u}}_k^*, K_k^*\}
        \gets \{\boldsymbol{x}^*_k + \delta \boldsymbol{x}_k, 
        \bar{\boldsymbol{u}}_k^*+ \delta\bar{\boldsymbol{u}}_k, 
        K_k^* + \delta K_k\}$ 

    \If{convergence criteria met}
        \State \Return $\{\boldsymbol{x}^*_k, \bar{\boldsymbol{u}}_k^*, K_k^*\}$
    \EndIf
        
        \State Update penalty weight $\{w\}$, Lagrange multipliers $\{\boldsymbol{\lambda}, \boldsymbol{\mu}\}$ \Comment{multiplier update, Algorithm 1 from Ref.~\citenum{Oguri-SCVSstar}}
    \EndIf
    \State Update trust region $\Delta_{TR}$ \Comment{trust region update, Eq.~(57) from Ref.~\citenum{Naoya-Sequential-Cov-Steering}}
\EndWhile
\end{algorithmic}
\end{algorithm}

\section{Numerical Example: Unskewing Distributions in Two-body Dynamics}\label{sec: numerical example two body}
Gaussian assumptions are known to break down into highly-skewed distributions under natural two-body dynamics. This section presents an example in which a distribution skewed by two-body motion is subsequently unskewed by statistical moment steering during an orbital transfer with impulsive control and 4th-order CUT. 

\subsection{Problem Setup}
This case features a two-body system of a spacecraft around Earth ($\mu \approx 398600$ km$^3$/sec$^2$) transferring from one orbit to another. The problem is partitioned into two phases: $0=T_{0^-} < T_0 < T_f = 2T_0$, where $t\in [T_{0^-},T_0)$ is the uncontrolled portion, $t\in [T_0,T_f]$ is the controlled portion, and $T_0$ being equal to the period of the initial orbit. The controlled portion is discretized into $N=9$. The sigma points corresponding to a ``pre-initial'' Gaussian distribution at $T_{0^-}$ are propagated uncontrolled to $T_0$. Due to the dynamical nature of two-body motion, the distribution at $T_0$ will be non-Gaussian. In other words, the pre-initial Gaussian is propagated uncontrolled for one period of the initial orbit, and then statistical moment steering has an additional one period of the original orbit to correct the skewness during the orbital transfer. The initial conditions for this pre-initial distribution are given in Table~\ref{tab: 2BP ICs}. Note that the mean parameters in Table~\ref{tab: 2BP ICs} correspond to a circular orbit with a radius of 8000 km and 30 degrees of inclination, and the propagation time is one period of this initial circular orbit. Table~\ref{tab: 2BP target} lists the target state that the spacecraft must reach at the end of the control, which corresponds to a circular orbit with a radius of 9000 km and 60 degrees of inclination. The matrices $H_r$ and $H_v$ select the corresponding position and velocity components from a state vector, respectively. 
\begin{table}[htbp]
	\fontsize{10}{10}\selectfont
    \caption{Parameters for Gaussian Distribution at Pre-Initial Time}
   \label{tab: 2BP ICs}
        \centering 
   \begin{tabular}{l  c  c} 
      \hline 
      Parameter & Value & Units\\
      \hline 
      Pre-initial Position Mean ($H_r \boldsymbol{\mu}_{0^-}$)& [8000 \quad 0 \quad 0]$^\top$ &[km]\\
      Pre-initial Velocity Mean ($H_v \boldsymbol{\mu}_{0^-}$) &[0 \quad 6.1130 \quad 3.5293]$^\top$ & [km/s]\\
      Pre-initial Position $3\sigma$& 50 &[km]\\
      Pre-initial Velocity $3\sigma$& 0.01 & [km/s]\\
      Uncontrolled Propagation Time ($T_0 - T_{0^-}$)& 1.9781 &  [hours] \\
      \hline
   \end{tabular}
\end{table}

\begin{table}[htbp]
	\fontsize{10}{10}\selectfont
    \caption{Final Target Parameters}
   \label{tab: 2BP target}
        \centering 
   \begin{tabular}{l  c  c} 
      \hline 
      Parameter & Value & Units\\
      \hline 
      Target Position Mean ($H_r \boldsymbol{\mu}_{f}$)& [9000 \quad 0 \quad 0]$^\top$ &[km]\\
      Target Velocity Mean ($H_v \boldsymbol{\mu}_{f}$) &[0 \quad 3.327 \quad 5.763]$^\top$ & [km/s]\\
      Control Time ($T_f - T_0$)& 1.9781 &  [hours] \\
      Final Skewness Constraint ($\epsilon_{\gamma}$)&  0.01 &  - \\
      \hline
   \end{tabular}
\end{table}

\begin{figure}[htb]
	\centering\includegraphics[width=0.9\textwidth]{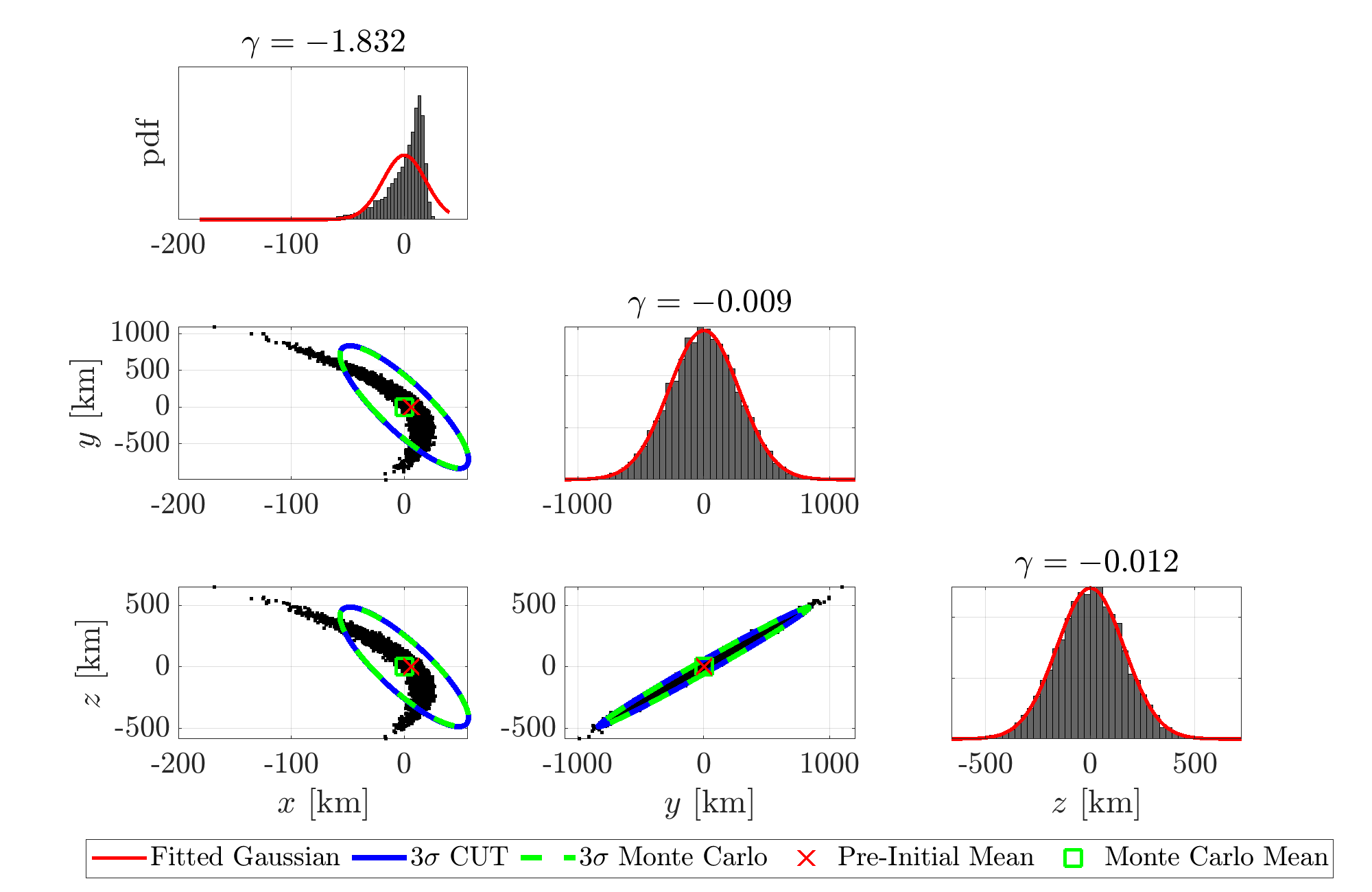}
	\caption{Monte Carlo ($n_{\text{samples}}=10,000$) for two-body example: Initial non-Gaussian, highly skewed, distribution. \emph{Axes are not equalized to better show skewness. Origin normalize to mean predicted by CUT.}}
	\label{fig: Two-Body, initial distribution}
\end{figure}

Figure~\ref{fig: Two-Body, initial distribution} shows the initial distribution that is to be unskewed by statistical moment steering. The initial distribution can be seen to be highly skewed. Eq.~\eqref{eq: Two-Body Original Formulation} shows the original problem and Eq.~\eqref{eq: Two-Body CVX Formulation} shows the convex subproblem in \texttt{SCvx*} form for this two-body example. 
\begin{equation} \label{eq: Two-Body Original Formulation}
\begin{aligned}
\min_{
\substack{
\{\boldsymbol{x}_k\}_{
k\in\mathbb{Z}_{0:N-1}} \\
\{\bar{\boldsymbol{u}}_k, K_k\}_{
k\in\mathbb{Z}_{0:N-2}}
}}
& \Delta V_{99,\text{ub}} \\
\text{s.t. } 
& \boldsymbol{x}_{0} \leftarrow \phi_{0}\left(\text{CUT}_{4G}\left(\mathcal{N}(\boldsymbol{\mu}_{0^-},P_{0^-})\right)\right)\\
& \boldsymbol{x}^{(i)}_{k+1} = \phi\left(\boldsymbol{x}^{(i)}_{k}, \bar{\boldsymbol{u}}_k, K_k\right), 
& \quad {\forall k\in\mathbb{Z}_{0:N-2}, \forall i\in\mathbb{Z}_{1:n_s}}\\
& f^{(\mu)} (\boldsymbol{x}_{N-1}) = \boldsymbol{\mu}_{f} \\
& \norm{H_r f^{(\gamma)} (\boldsymbol{x}_{N-1})}_{\infty}
\leq \epsilon_{\gamma} 
&\text{(Final Skewness Constraint)}
\end{aligned}
\end{equation} 

\begin{subequations}\label{eq: Two-Body CVX Formulation}
\begin{align}
\min_{
\substack{
\{\delta\boldsymbol{x}_k\}_{
k\in\mathbb{Z}_{0:N-1}} \\
\{\delta\bar{\boldsymbol{u}}_k,\delta K_k\}_{
k\in\mathbb{Z}_{0:N-2}} \\
\{\boldsymbol{\xi},\zeta\}
}
} & 
\text{Convexified }\Delta V_{99,\text{ub}} + P_{\texttt{SCvx*}}(w_p, \boldsymbol{\xi},\boldsymbol{\lambda}, \zeta, \mu) \\
\text{s.t. } 
& \boldsymbol{x}_k = \boldsymbol{x}_k^* + \delta \boldsymbol{x}_k,
& {\forall k\in\mathbb{Z}_{0:N-1}}\\
& \boldsymbol{u}^{(i)}_k = \boldsymbol{u}_k^{(i)*} + \delta\bar{\boldsymbol{u}}_k + K_k^*\delta \boldsymbol{z}^{(i)}_k + \delta K_k {\boldsymbol{z}^{(i)*}_k},
& {\forall k\in\mathbb{Z}_{0:N-2}, \forall i\in\mathbb{Z}_{1:n_s}}\\
~
& \boldsymbol{x}_{0} \leftarrow \phi_{0}\left(\text{CUT}_{4G}\left(\mathcal{N}(\boldsymbol{\mu}_{0^-},P_{0^-})\right)\right)\\
& \boldsymbol{x}^{(i)}_{k+1} = A^{(i)}_k\boldsymbol{x}^{(i)}_{k} + B^{(i)}_k\boldsymbol{u}^{(i)}_k + \boldsymbol{c}^{(i)}_k + \boldsymbol{\xi}^{(i)}_k, &{\forall k\in\mathbb{Z}_{0:N-2}, \forall i\in\mathbb{Z}_{1:n_s}} \\
& \boldsymbol{\xi}^{(i)}_k = 0, &{\forall k\in\mathbb{Z}_{0:N-3}, \forall i\in\mathbb{Z}_{1:n_s}}
\label{eq: Two-Body CVX Formulation, slack}
\\
& \boldsymbol{\mu}_{N-1}^* + A^{(\mu)} \delta\boldsymbol{x}_{N-1} = \boldsymbol{\mu}_{f} \\
& \norm{H_r \left( 
\boldsymbol{\gamma}_{N-1}^* + A^{(\gamma)} \biggr\rvert_{\boldsymbol{z}_{N-1}^*}\delta \boldsymbol{z}_{N-1}
\right)}_{\infty} \leq \epsilon_{\gamma} + \zeta, 
&\zeta \geq 0 \\
& \norm{\boldsymbol{\delta x}_k}_{\infty} \leq \Delta_{TR}, & {\forall k\in\mathbb{Z}_{0:N-1}}\\
& \norm{\delta K_k}_{\infty} \leq \Delta_{TR}, & {\forall k\in\mathbb{Z}_{0:N-2}} \\
& \boldsymbol{\xi} = 
\left[
\boldsymbol{\xi}_0^\top
,\ldots,
\boldsymbol{\xi}_{N-2}^\top 
\right]^\top
\end{align}
\end{subequations}
where the process of generating initial sigma points by uncontrolled propagation from $T_{0^-}$ to $T_0$ is denoted by $\phi_{0}(\cdot)$, and the initial sigma points at $T_0$ are the initial sigma points for convex optimization. The 4th-order CUT is used to sample the Gaussian at the pre-initial time. Only two moment constraints are placed: final mean and final skewness. The final mean constraint ensures that the distribution's final mean is on the target orbit. The skewness constraint ensures that the maximum absolute skewness along the positional axes is smaller than some small $\epsilon_{\gamma}$ value. Although an affine equality constraint to ensure zero skewness is still allowable in a convex form (i.e., $\boldsymbol{\gamma}_{N-1} = \vec{0}$), some issues arise. Firstly, due to the imposed linear mapping between state and control, it is likely difficult to achieve zero skewness in nonlinear problems. Secondly, while CUT provides more accurate moment estimates than traditional methods in nonlinear systems (e.g., linear covariance), it is still an approximation, and any imposed equality constraints are unlikely to hold exactly. Lastly, achieving high statistical confidence in confirming zero skewness is likely unattainable, as even a Monte Carlo simulation is also just an approximation of the density function. Thus, $\epsilon_{\gamma} = 0.01$ is introduced in this problem to show that skewness can still be reduced to a significant degree.

Note that this problem purposely has no constraint on covariance. This showcases the formulation's ability to directly control different orders of moments while neglecting others if they are not needed. In contrast, previous works on distribution steering still require covariance to be explicitly calculated, such as indirectly controlling skewness with linear covariance steering with minimum nonlinear error \cite{Qi-Stationkeeping} or neglecting skewness altogether and focusing more on higher-order calculations of covariance \cite{Naoya-GMM-Terminal, Ozaki-UT-TrajOp, Nandi-CUT, Ross-CUT-Traj-Op, fife-GMM-steering-conference}. 

A comment can be made on the assignment of slack variables needed for \texttt{SCvx*} in Eq.~\eqref{eq: Two-Body CVX Formulation, slack}. Having an ``overly-slacked'' convex subproblem can affect the quality of the convex solution and the convergence rate of the SCP. A careful balance must be struck to ensure that sufficient slack variables are introduced to prevent artificial infeasibility without over-relaxing the problem to the point where convergence is impeded. For the sigma point dynamics constraint, slack is permitted only between the final two nodes, a heuristic choice that was found to improve convergence. However, all violations of the other nonconvex constraint must still be penalized according to Eq.~\eqref{eq: scvx star penalty} when evaluating the nonlinear problem during the step acceptance conditions of \texttt{SCvx*}.

\subsection{Numerical Considerations}
\begin{table}[htbp]
	\fontsize{10}{10}\selectfont
    \caption{\texttt{SCvx*} Parameters for Two-Body Example}
   \label{tab: 2BP SCvx Params}
    \centering 
    \small
   \begin{tabular}{l c c c c c c c c} 
      \hline 
      Parameter & 
      $\{ \epsilon_{\text{opt}},\epsilon_{\text{feas}}\}$ & $\{\eta_0,\eta_1,\eta_2\}$&
      $\{\alpha_1,\alpha_2\}$&
      $\{\beta_{\texttt{SCvx*}},\gamma_{\texttt{SCvx*}}\}$&
      $\Delta_{\text{TR}}^{(1)}$&
      $\{\Delta_{\text{TR},\min},\Delta_{\text{TR},\max}\}$&  
      $w^{(1)}_p$ &
      $w_{p,\max}$\\
      \hline 
      Value & 
    $\{ 10^{-4},10^{-6}\}$ & $\{1,0.85,0.1\}$&
      $\{2,3\}$&
      $\{1.5,0.99\}$&
      $0.5$& 
      $\{10^{-10},20\}$& 
      $100$ &
      $10^{10}$\\
      \hline
   \end{tabular}
\end{table}

Table~\ref{tab: 2BP SCvx Params} lists the numerical parameters used for \texttt{SCvx*} in the two-body example. The initial reference is computed with the following procedure. The initial reference mean $\boldsymbol{\mu}_{\text{guess},k}$ for each node corresponds to a linearly interpolated point between the initial and final states in Keplerian orbital elements, which is then converted back to Cartesian coordinates. The initial reference covariances $P_{\text{guess},k}$ are computed via a ``scaled linear covariance'' approach outlined in Appendix~\ref{appendix: Scaled Linear Covariance}. In short, the previous timestep's covariance $P_{\text{guess},k-1}$ is propagated under the linearized dynamics around $\boldsymbol{\mu}_{\text{guess},k-1}$ under zero control input, and then scaled accordingly to become $P_{\text{guess},k}$. The rationale for this approach is to prevent the initial reference distribution from becoming overly dispersed, while still accounting for deformation effects from natural dynamics in order to help reduce control cost. The initial reference sigma points are sampled from these Gaussians of $\mathcal{N}(\boldsymbol{\mu}_{\text{guess},k}, P_{\text{guess},k})$, except for the sigma point at $t_0$ since it is already predefined. 

Another consideration is the numerical scaling of the problem. Typical numerical solvers require a reasonably scaled problem such that the optimization variables are of similar magnitude. All units are scaled by their corresponding characteristic quantities: the characteristic length is $5000$ km, and the characteristic time is the reciprocal of mean motion with a semi-major axis equivalent to the characteristic length. The \texttt{SCvx*} parameters and convergence properties can vary based on different scaling methods. 

\subsection{Results and Discussion}
This SCP problem converged after 41 iterations and took about 24 minutes to solve using \texttt{MOSEK} with \texttt{CVX} on \texttt{MATLAB} R2024a.\footnote{Running on Snapdragon(R) X Elite - X1E78100 - Qualcomm(R) Oryon(TM) CPU 3.42 GHz}. The convergence profile of the \texttt{SCvx*} algorithm for this two-body example is shown in Figure~\ref{fig: Two-Body, scvx convergence}. It can be seen that from the \texttt{SCvx*} algorithm, not every iteration from solving the convex subproblem was accepted. Figure~\ref{fig: Two-Body, transfers} shows the optimized mean trajectory relative to the initial and final orbits, while Figure~\ref{fig: Two-Body, CUT points} complements this by visualizing how the CUT points evolve under control along the trajectory.

\begin{figure}[!htb]
	\centering\includegraphics[width=0.6\textwidth]{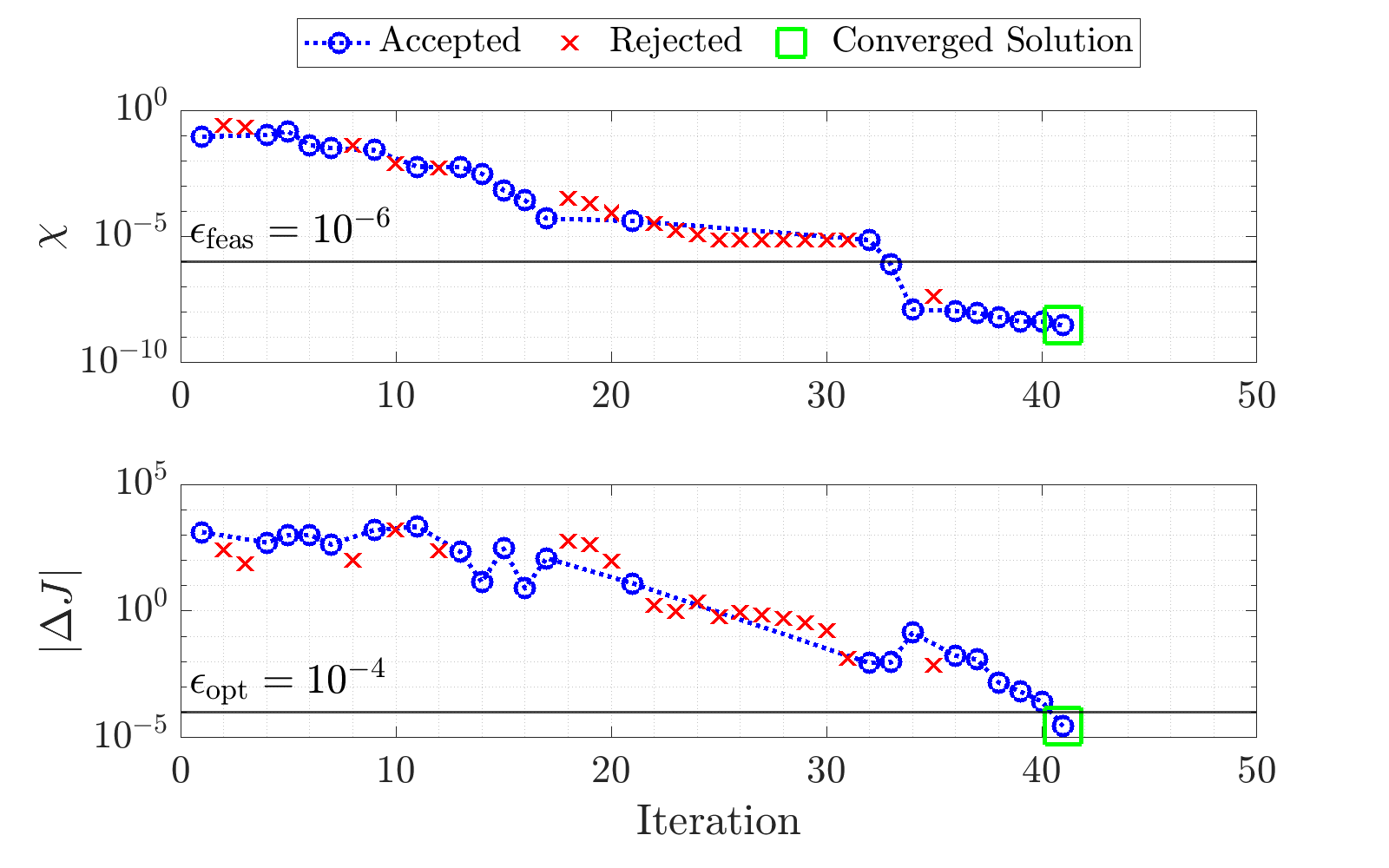}
	\caption{Convergence profile of \texttt{SCvx*} for two-body example. \emph{Y-axis in log scale.}}
	\label{fig: Two-Body, scvx convergence}
\end{figure}

\begin{figure}[!htb]
  \centering 
    \subcaptionbox
    {Initial and Final Orbits.}
    {\includegraphics[width=0.49\textwidth]{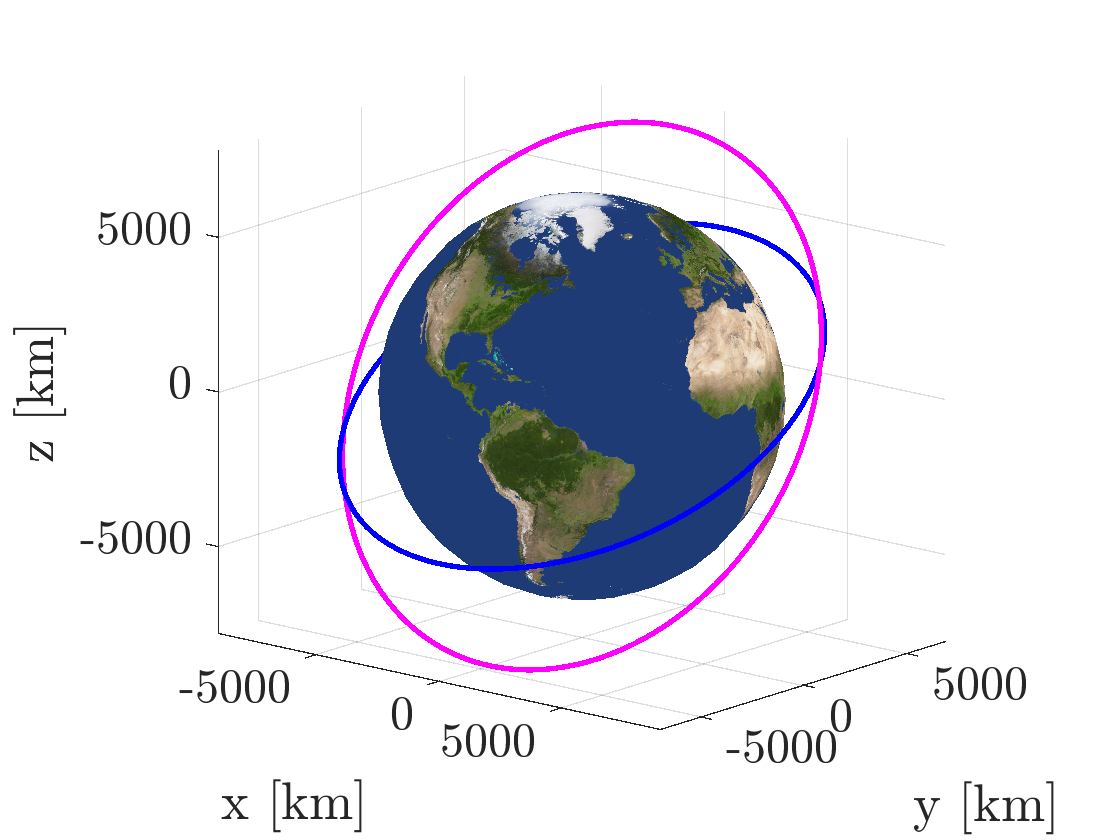}}
    \hskip 0.1truein
    \subcaptionbox
    {Initial and Final Orbits with Transfer.}
    {\includegraphics[width=0.49\textwidth]{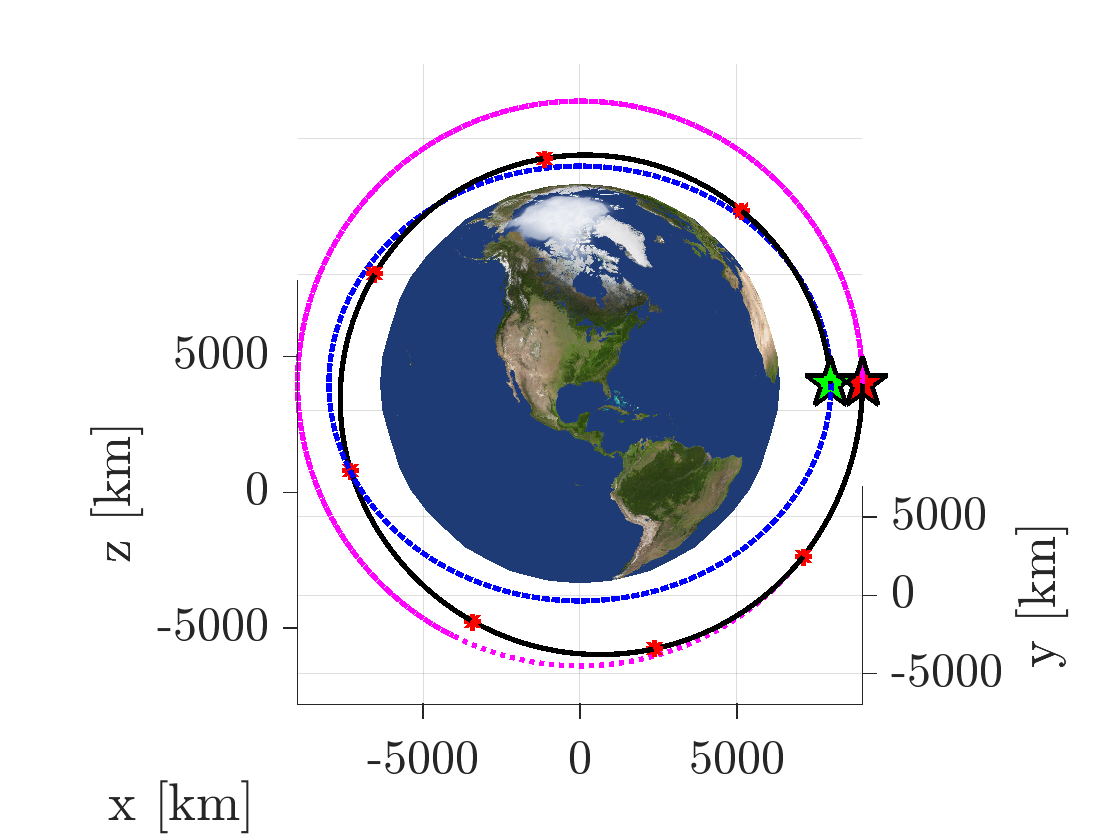}}
    \caption{Left: initial orbit in blue, final orbit in magenta. Right: transfer trajectory with control nodes marked in red and initial/final points marked with a green/red $\star$.}
    \label{fig: Two-Body, transfers}
\end{figure}

\begin{figure}[!htb]
  \centering 
    \subcaptionbox
    {Transfer Trajectory.}
    {\includegraphics[width=0.49\textwidth]{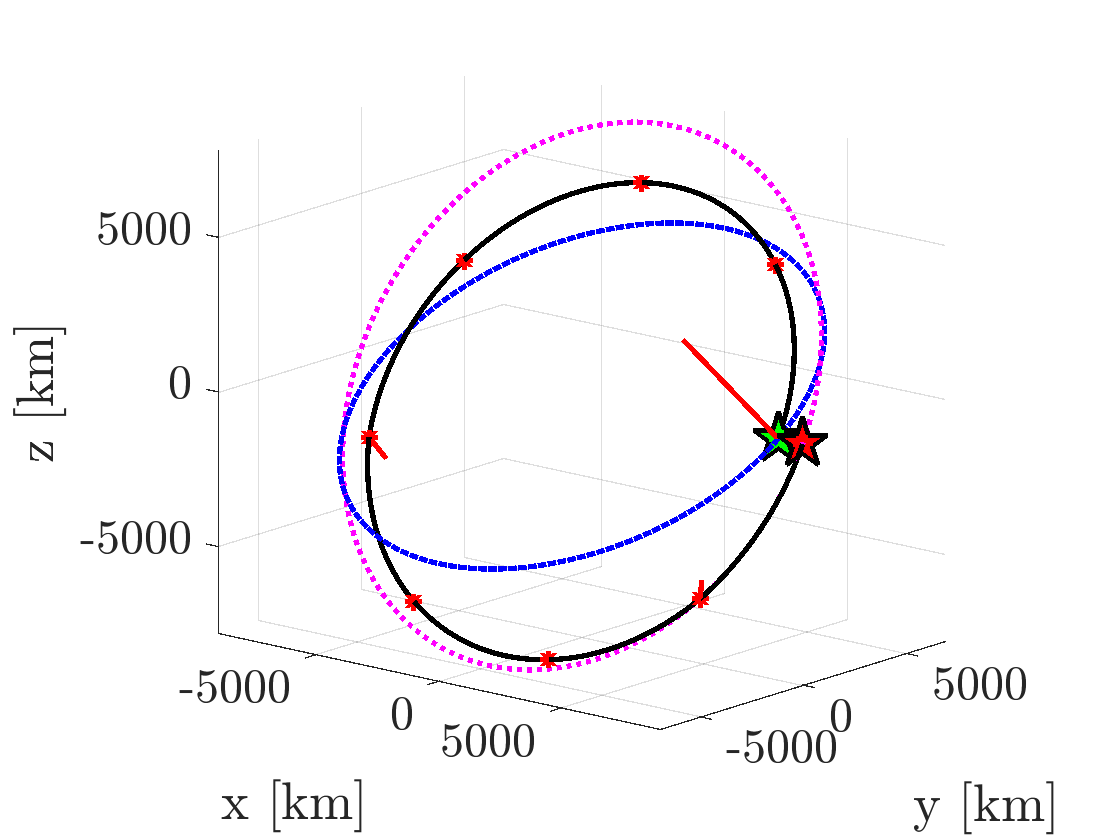}}
    \hskip 0.1truein
    \subcaptionbox
    {CUT Points Under Control.}
    {\includegraphics[width=0.49\textwidth]{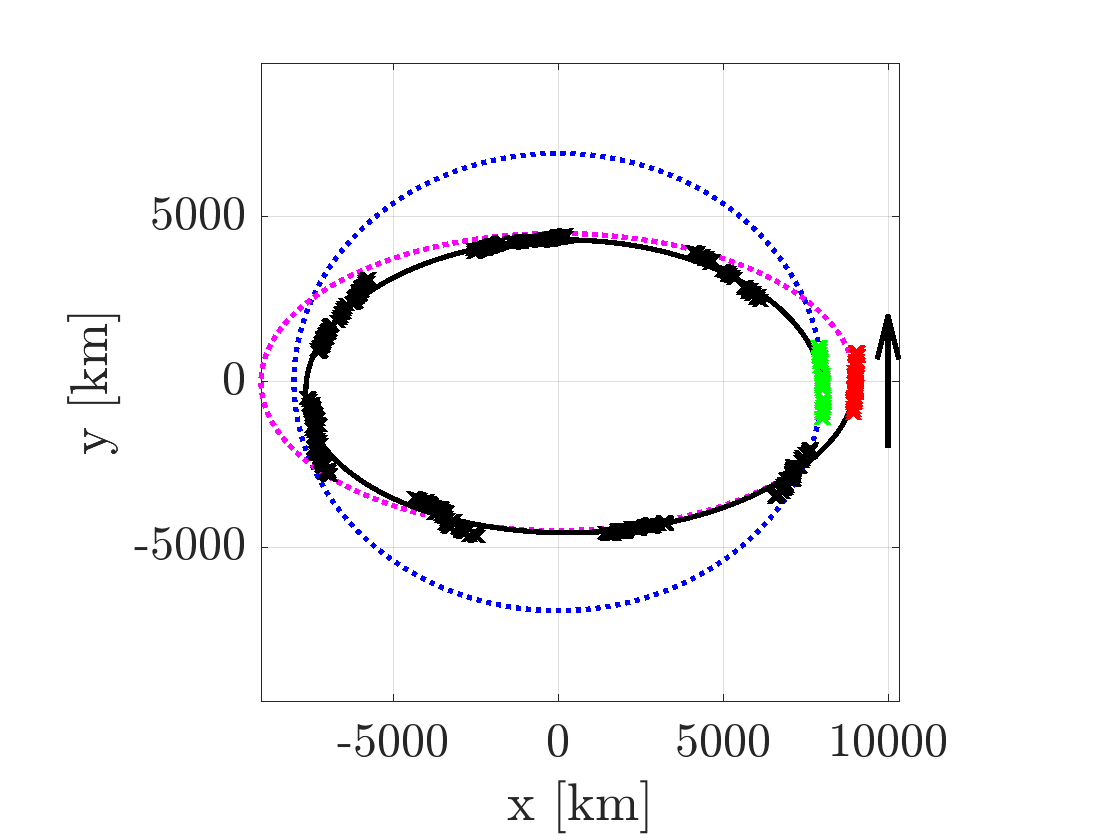}}
    \caption{Left: transfer trajectory with nominal control vectors marked in red and initial/final points marked with a green/red $\star$. Right: CUT points ($\times$) along transfer trajectory, with initial/final CUT points marked in green/red. \emph{Deviations enlarged $2{\times}$ to better show the individual points.}}
    \label{fig: Two-Body, CUT points}
\end{figure}

\begin{figure}[!htbp]
  \centering 
    \subcaptionbox
    {With $\bar{\boldsymbol{u}}_k$ Only.\label{fig: Two-Body, Results ubar only}}
    {\includegraphics[width=0.9\textwidth]{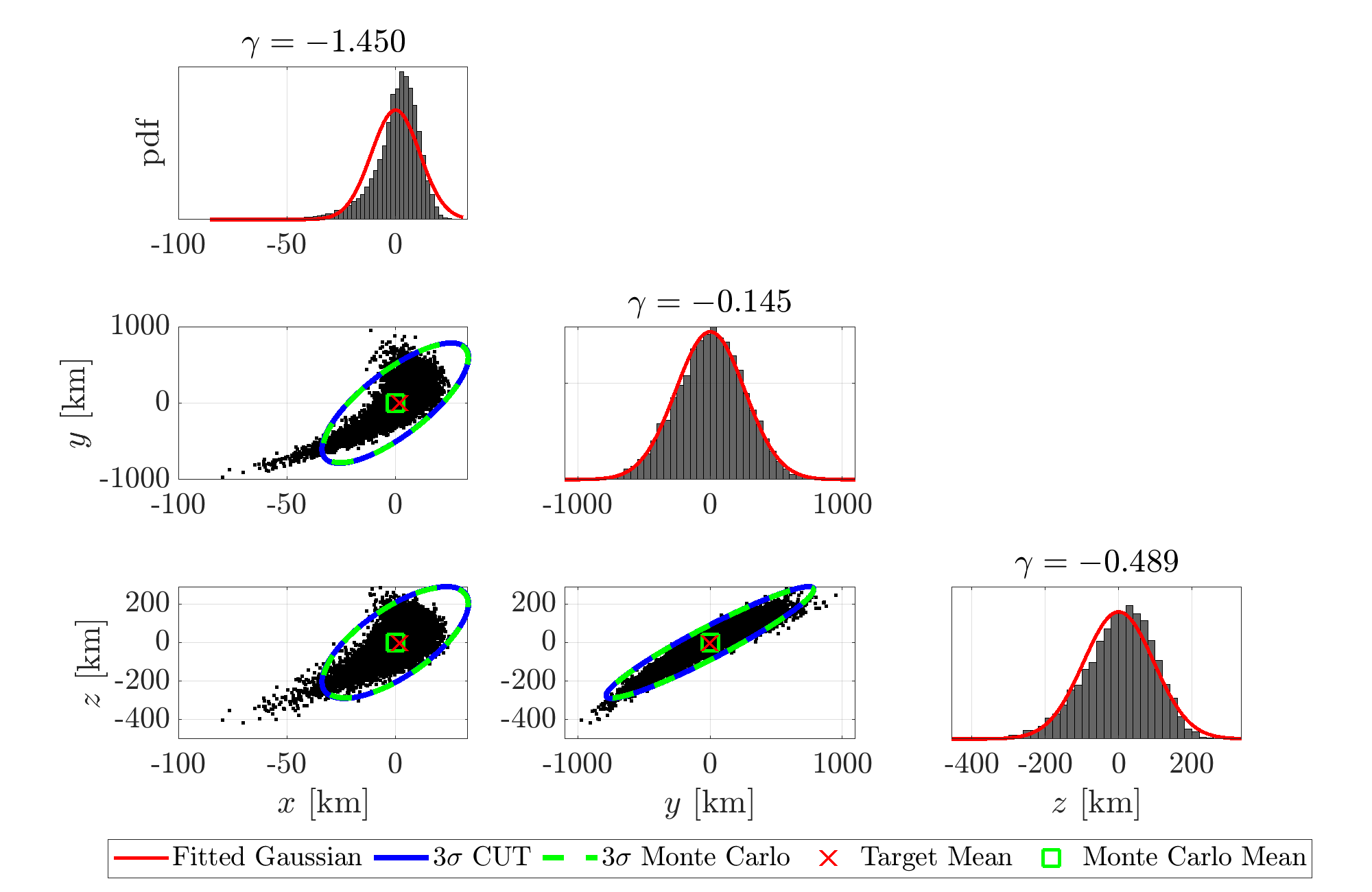}}
    \hskip 0.1truein
    ~
    \subcaptionbox
    {With Statistical Moment Steering.\label{fig: Two-Body, Results CUT Steering}}
    {\includegraphics[width=0.9\textwidth]{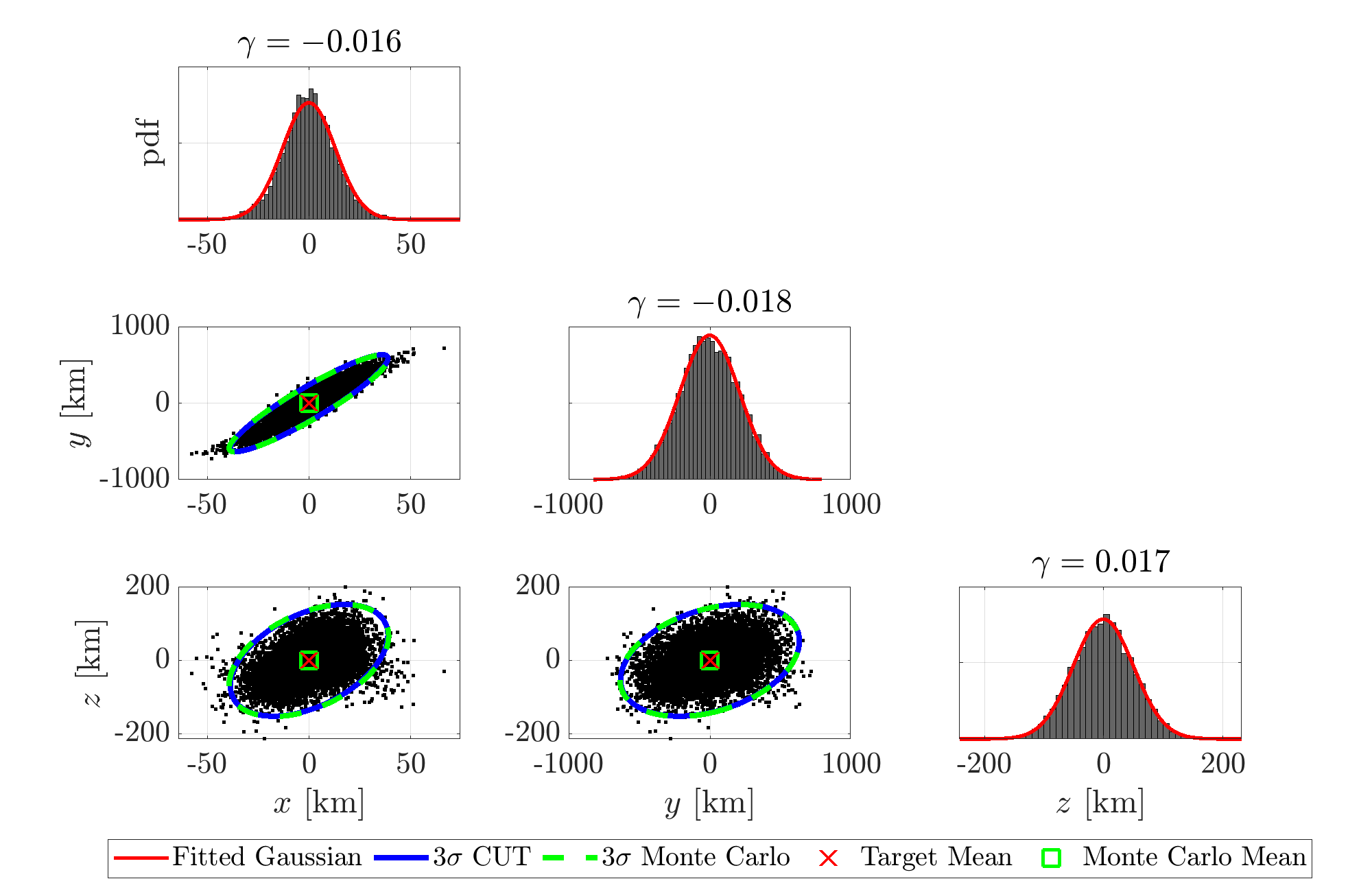}}
    \caption{Monte Carlo ($n_{\text{samples}}=10,000$) for two-body example: Distribution at terminal time with nominal control actions compared with one with statistical moment steering. \emph{Axes are not equalized to better show skewness. Origin normalize to mean predicted by CUT.}}
    \label{fig: Two-Body, Results}
\end{figure}

Figure~\ref{fig: Two-Body, Results} shows the effects of statistical moment steering. Figure~\ref{fig: Two-Body, Results ubar only} shows the distribution if only the feedforward control action is applied, which is equivalent to typical deterministic trajectory optimization. It can be seen that the distribution remains skewed if no feedback is applied, and any unskewing actions are a result of the feedback action from statistical moment steering. In this case, the skewness in the x-axis direction is most affected by the dynamics. Furthermore, the mean of the distribution is not aligned with the target mean as a result of the distribution's transformation through nonlinear dynamics. Figure~\ref{fig: Two-Body, Results CUT Steering} presents the unskewed distribution from statistical moment steering. It can be seen that all axes, emphasizing the x-axis, are near-symmetric. The optimizer satisfies the $\epsilon_{\gamma}=0.01$ inequality constraint, though the nonlinear Monte Carlo exhibits minor violations. This outcome reflects either the approximation limits of CUT or inaccuracies from the Monte Carlo, as discussed before: since the density function of the true distribution is not explicitly calculated, the skewness from both CUT and the large Monte Carlo is still only an approximation of the true skewness value. The main takeaway is that skewness is significantly reduced. Furthermore, the other statistical moments, such as mean and covariance, are accurately predicted by the CUT points: Figure~\ref{fig: Two-Body, Results CUT Steering} shows that the final mean constraint is satisfied, and the $3\sigma$ ellipses from CUT align with that of Monte Carlo.

\begin{figure}[!htb]
	\centering\includegraphics[width=0.9\textwidth]{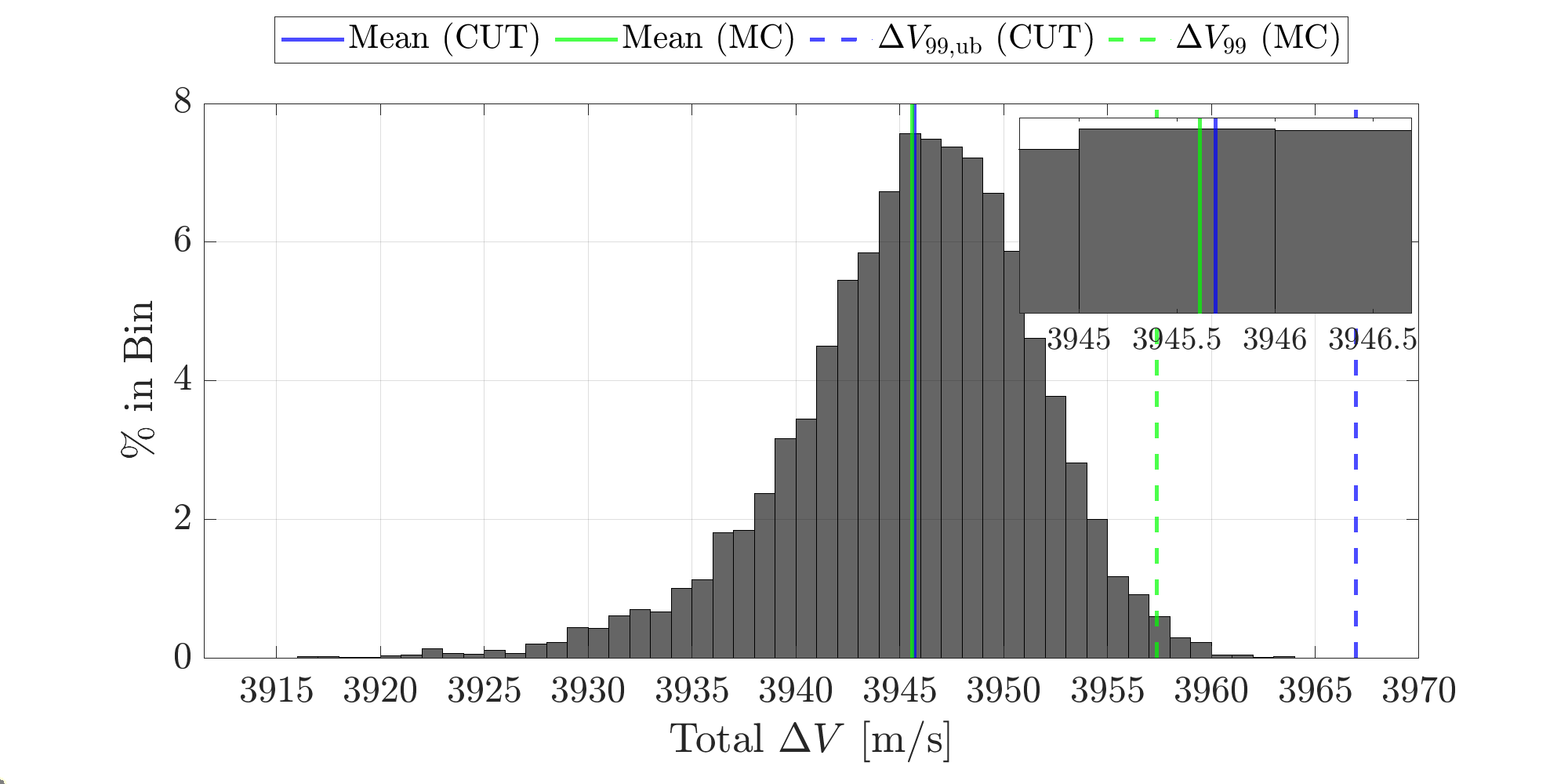}
	\caption{Monte Carlo ($n_{\text{samples}}=10,000$) for two-body example: Histogram of Total $\Delta V$ Costs.}
	\label{fig: Two-Body, DeltaV Histogram}
\end{figure}

\begin{figure}[!htb]
  \centering 
    \subcaptionbox
    {Time History of $\mathbf{U}_k$.\label{fig: Two-Body, DeltaV Hist, total}}
    {\includegraphics[width=0.49\textwidth]{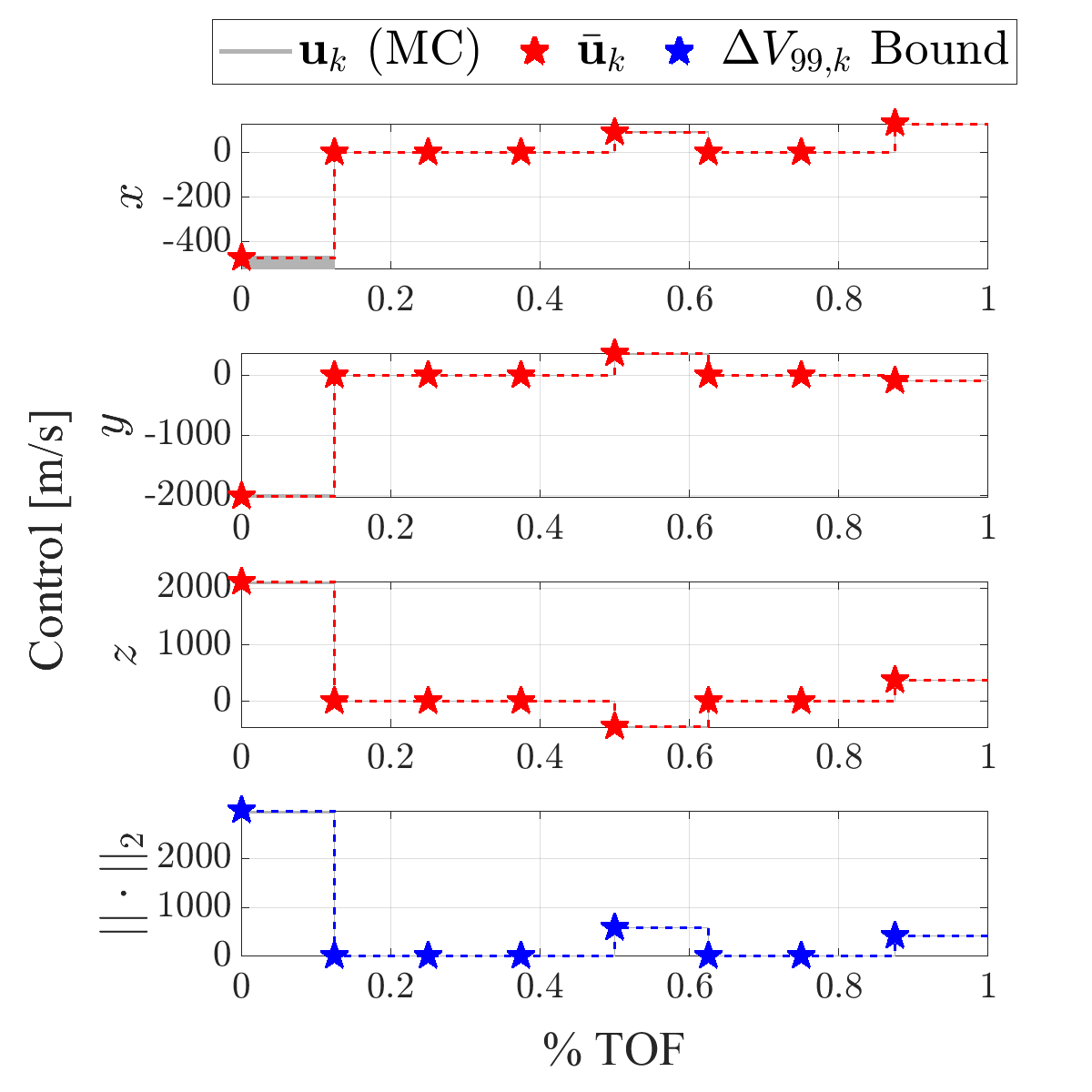}}
    \hskip 0.1truein
    \subcaptionbox
    {Time History of $\mathbf{U}_{\text{fb},k} = K_k(\boldsymbol{X}_k - \boldsymbol{\mu}_k)$.
    \label{fig: Two-Body, DeltaV Hist, fb only}
    }
    {\includegraphics[width=0.49\textwidth]{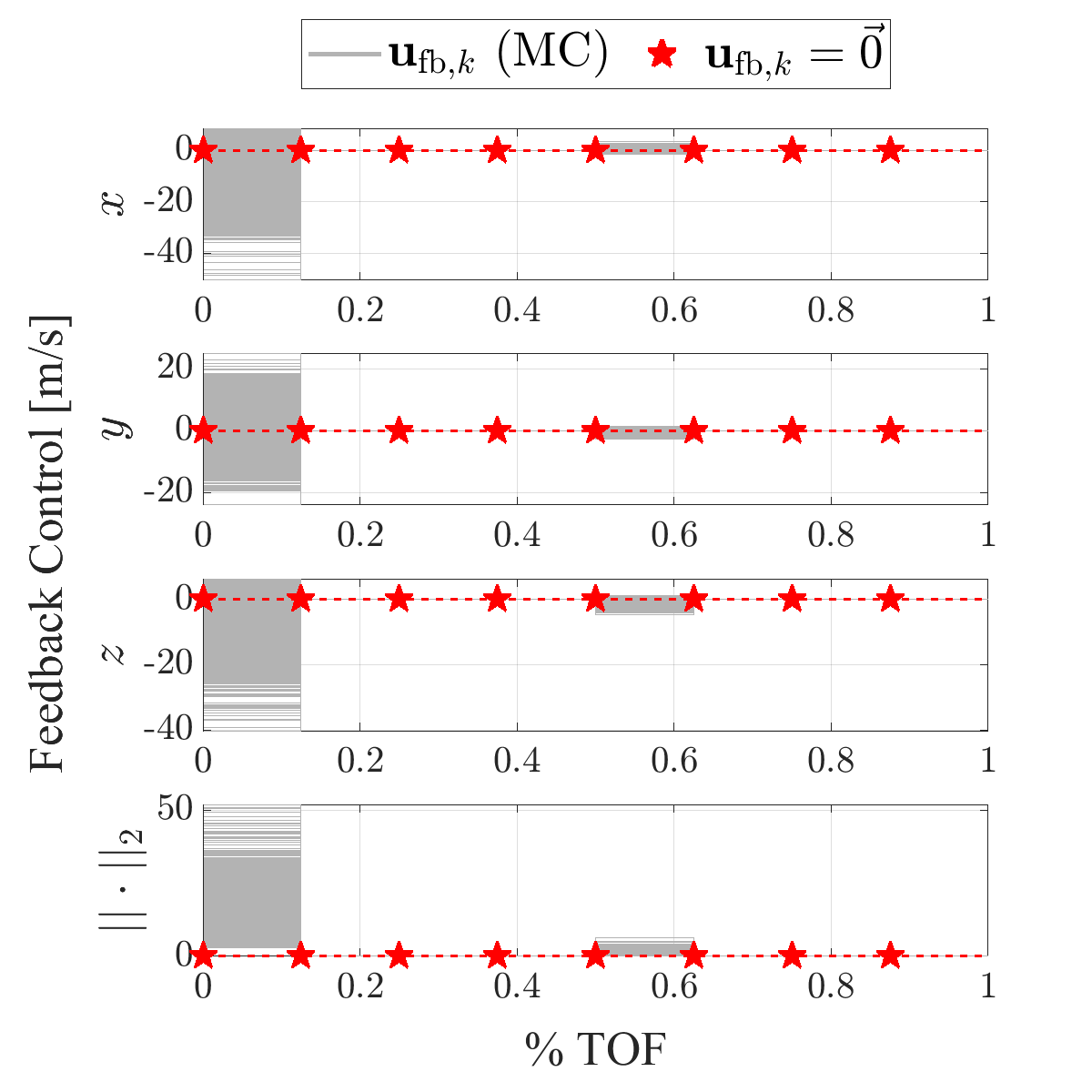}}
    \caption{Monte Carlo ($n_{\text{samples}}=10,000$) for two-body example: Time history of total control and feedback contributions from statistical moment steering.}
    \label{fig: Two-Body, DeltaV Hist}
\end{figure}

Figure~\ref{fig: Two-Body, DeltaV Histogram} shows the total cost, and Figure~\ref{fig: Two-Body, DeltaV Hist} shows the time history of the maneuvers. Firstly, it can be seen from both figures that $\Delta V_{99,\text{ub}}$ is a reasonable upper-bound to the actual $99$-th percentile of $\Delta V$ cost for a single $t_k$ and total $\Delta V$ cost. As noted earlier, $\Delta V_{99,\text{ub}}$ is obtained using CUT under the Gaussian assumption for the control distribution, where CUT provides the corresponding mean and covariance. This value serves only as an upper bound under that assumption, and in this case, the Gaussian approximation appears to be reasonable. Secondly, the expected value for the fuel costs calculated with Eq.~\eqref{eq: expected fuel cost} using CUT accurately captures the average fuel costs from the Monte Carlo as shown in Figure~\ref{fig: Two-Body, DeltaV Histogram}. This shows that CUT is still reasonably accurate even after a nonlinear transformation with the $l^2$-norm. Lastly, in this case of a two-body transfer, the nominal maneuvers are the primary contributor to the $\Delta V$ cost as seen by the difference in magnitudes in Figure~\ref{fig: Two-Body, DeltaV Hist, total} and ~\ref{fig: Two-Body, DeltaV Hist, fb only}. The contribution of feedback control is much smaller than that of the nominal control efforts for a trajectory transfer problem, but the inclusion of the feedback term is critical in shaping the distribution to satisfy the moment constraint. This illustrates the sensitivity of control actions in managing non-Gaussian systems within a nonlinear environment. Figures~\ref{fig: Two-Body, DeltaV Histogram} and \ref{fig: Two-Body, DeltaV Hist} demonstrate the validity of the $\Delta V_{99,\text{ub}}$ and expected fuel cost objective functions used for statistical moment steering.

\section{Numerical Example: Non-Gaussian Stationkeeping in Halo Orbit}\label{sec: numerical example three body}
It has been shown that unstable halo orbits can lead to the breakdown of linear covariance controllers \cite{Qi-Stationkeeping}. This is due to the high degrees of nonlinearity found in these orbits, meaning that the Gaussian assumption is less applicable in these environments. This section demonstrates that statistical moment steering can still function in these regions and is better equipped to handle distribution steering in nonlinear environments with impulsive control. 

\subsection{Problem Setup}
\begin{figure}[!htb]
  \centering 
    \subcaptionbox
    {Halo Family in Synodic Frame.}
    {\includegraphics[width=0.49\textwidth]{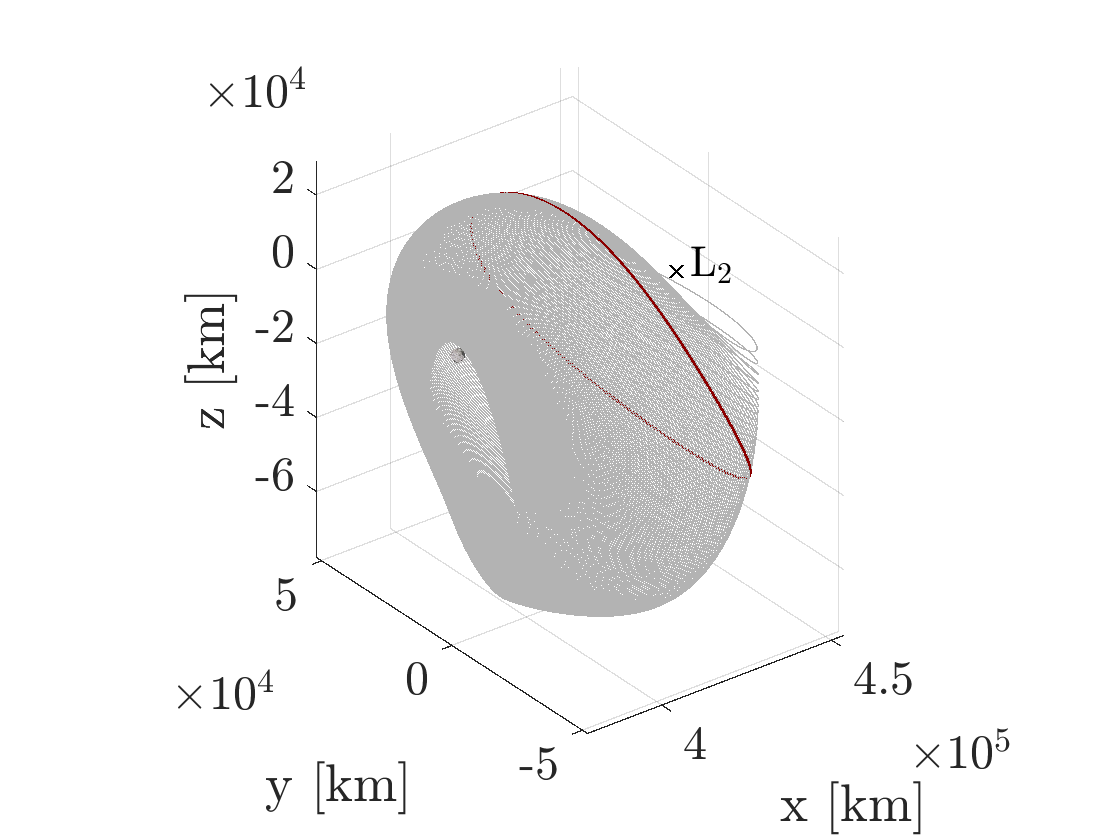}}
    \hskip 0.1truein
    \subcaptionbox
    {Time Constant of Halo Family.}
    {\includegraphics[width=0.49\textwidth]{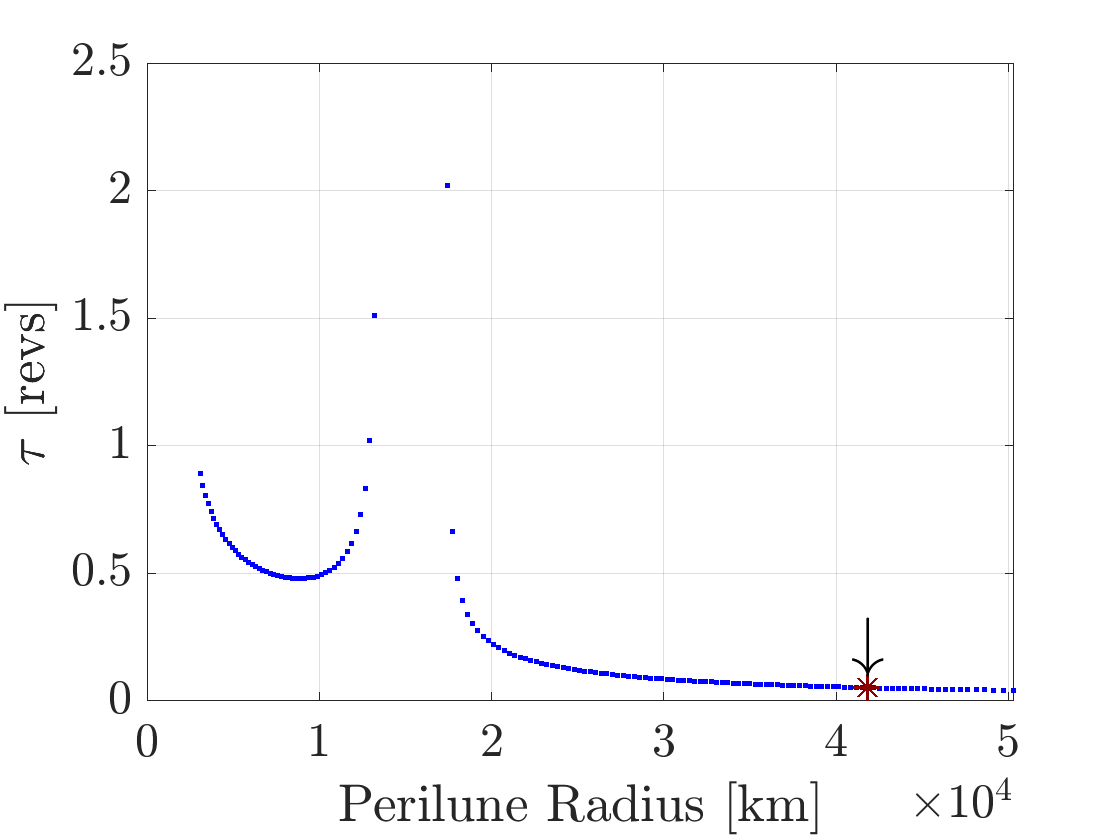}}
    \caption{Southern L$_2$ halo family in barycenter-centered synodic frame. Examined orbit in red.}
    \label{fig: CR3BP orbits}
\end{figure}

The dynamics are under the circular restricted three-body problem (CR3BP) assumption \cite{Zimovan-Spreen-Halo-Stability} for the Earth-Moon system.
\begin{equation}\label{eq: CR3BP EOM}
\begin{aligned}
\ddot{x} - 2\dot{y} = \frac{\partial U}{\partial x}, 
\qquad 
\ddot{y} + 2\dot{x} = \frac{\partial U}{\partial y}, 
\qquad 
\ddot{z} = \frac{\partial U}{\partial z}
\end{aligned}
\end{equation}
where $x,y,z$ are the nondimensional positions in the synodic frame, $U = \frac{1}{2}\left( x^2 + y^2\right)+\frac{1-\mu}{d}+\frac{\mu}{r}$ is the pseudo-potential function, $\mu$ is the nondimensional mass parameter (Earth-Moon system $\mu \approx 0.0122$), and $d = \sqrt{(x+\mu)^2 + y^2 + z^2}$ and $r = \sqrt{(x-1+\mu)^2 + y^2 + z^2}$. A metric used to quantify the instability of an orbit is with time constant $\tau$ [revs] \cite{Zimovan-Spreen-Halo-Stability}.
\begin{equation}
    \tau\text{ [revs] } = \frac{1}{\text{Re}\left[ \text{Ln}\left( \lambda_{\max}\left[\Phi(t + T,t)  \right] \right) \right]} \frac{1}{T}
\end{equation}
where $T$ is the period of the orbit and $\Phi(t + T,t)$ is the monodromy matrix of the orbit. For stable orbits this value is infinity, and for nearly stable orbits this value is greater than one. The more unstable an orbit, the lower its associated time constant. This case uses the same ``unstable'' halo orbit from Ref.~\citenum{Qi-Stationkeeping} as seen highlighted by red in Figure~\ref{fig: CR3BP orbits}. In Ref.~\citenum{Qi-Stationkeeping}, it has been shown that this orbit is difficult to control with linear covariance controllers for a given number of maneuvers due to the strong nonlinearities degrading the Gaussian assumption. The low time constant of this orbit supports this argument. 

\begin{table}[htbp]
	\fontsize{10}{10}\selectfont
    \caption{Parameters for Gaussian Distribution at Initial Time along with Constraints}
   \label{tab: CR3BP ICs}
        \centering 
   \begin{tabular}{l  c  c} 
      \hline 
      Parameter & Value & Units\\
      \hline 
      Initial State Mean ($\boldsymbol{\mu}_{0}$) & [1.1600 \quad
         0 \quad
   -0.1247 \quad
         0 \quad
   -0.2087 \quad
         0]$^\top$ & [n.d.] \\
      Initial Position $3\sigma$& 30 &[km]\\
      Initial Velocity $3\sigma$& 3 & [m/s]\\
      Simulation Time & 6.5379 & [n.d.] \\
    $3\sigma$ Constraint ($3\sqrt{\lambda_{r,\max}}$)& 2000 & [km] \\
    Final Skewness Constraint ($\epsilon_{\gamma}$)&  0.01 &  - \\
    Final Kurtosis Constraint ($\epsilon_{\kappa}$)&  0.5 &  - \\
    \hline
   \end{tabular}
\end{table}

The orbit is discretized into $N=19$ nodes starting at the halo orbit's apolune with a simulation time of twice the period of the halo orbit. A constraint on the maximum eigenvalue of the positional covariance $\lambda_{\max}(H_r P_k H_r^\top) \leq \lambda_{r,\max}$ is introduced. This value corresponds to a $3\sigma$ of $2000$ km, or $3\sqrt{\lambda_{r,\max}} = 2000$ km. The more detailed simulation parameters are provided in Table~\ref{tab: CR3BP ICs}. 

This halo stationkeeping example shows two sub-examples to demonstrate the versatility of statistical moment steering. Example~\ref{sec: halo CUT4} uses the 4th-order CUT to show that distributions can still be effectively steered in an orbit in which previous covariance controllers failed. Example~\ref{sec: halo CUT6} extends to the 6th-order CUT to accurately perform statistical moment steering on the higher moments of skewness and kurtosis.

\subsubsection{Non-Gaussian Covariance Constraints with 4th-order CUT}\label{sec: halo CUT4}
This example showcases the power of statistical moment steering in a highly nonlinear environment where typical linear covariance controllers break down. Eq.~\eqref{eq: CR3BP Original Formulation} shows the original problem and Eq.~\eqref{eq: CR3BP CVX Formulation} shows the convex subproblem in \texttt{SCvx*} form for this CR3BP example. For the convex subproblem, a slack variable is assigned for each segment of the sigma point dynamics. Recall that this is different than the two-body example, showing that the assignment of slack variables in the nonconvex elements is up to the user, as long as their violations are penalized later on in the step acceptance process of the \texttt{SCvx*} algorithm. Note that there are no intermediate mean constraints, as the optimal distributions do not necessarily imply that their mean will lie on the orbit itself. However, the positional covariance constraint is placed throughout the time horizon to help ensure that the distribution does not become too dispersed. 

\begin{equation} \label{eq: CR3BP Original Formulation}
\begin{aligned}
\min_{
\substack{
\{\boldsymbol{x}_k\}_{
k\in\mathbb{Z}_{0:N-1}} \\
\{\bar{\boldsymbol{u}}_k, K_k\}_{
k\in\mathbb{Z}_{0:N-2}}
}}
& \Delta V_{99,\text{ub}} \\
\text{s.t. }  
& \boldsymbol{x}_{0} \leftarrow 
\text{CUT}_{4G}\left(\mathcal{N}(\boldsymbol{\mu}_{0},P_{0})\right)\\
& \boldsymbol{x}^{(i)}_{k+1} = \phi\left(\boldsymbol{x}^{(i)}_{k}, \bar{\boldsymbol{u}}_k, K_k\right), 
& \quad {\forall k\in\mathbb{Z}_{0:N-2}, \forall i\in\mathbb{Z}_{1:n_s}}\\
& f^{(\mu)} (\boldsymbol{x}_{N-1}) = \boldsymbol{\mu}_{0} \\
& \norm{H_r f^{(P^{1/2})} (\boldsymbol{x}_k)}_{2} \leq \sqrt{\lambda_{r,\max}}, & \quad {\forall k\in\mathbb{Z}_{0:N-1}}\\
\end{aligned}
\end{equation} 

\begin{equation}\label{eq: CR3BP CVX Formulation}
\begin{aligned}
\min_{
\substack{
\{\delta\boldsymbol{x}_k\}_{
k\in\mathbb{Z}_{0:N-1}} \\
\{\delta\bar{\boldsymbol{u}}_k,\delta K_k\}_{
k\in\mathbb{Z}_{0:N-2}} \\
\{\boldsymbol{\xi}\}
}
} & 
\text{Convexified }\Delta V_{99,\text{ub}} + P_{\texttt{SCvx*}}(w_p, \boldsymbol{\xi},\boldsymbol{\lambda}) \\
\text{s.t. } 
& \boldsymbol{x}_k = \boldsymbol{x}_k^* + \delta \boldsymbol{x}_k,
& {\forall k\in\mathbb{Z}_{0:N-1}},\\
& \boldsymbol{u}^{(i)}_k = \boldsymbol{u}_k^{(i)*} + \delta\bar{\boldsymbol{u}}_k + K_k^*\delta \boldsymbol{z}^{(i)}_k + \delta K_k {\boldsymbol{z}^{(i)*}_k},
& {\forall k\in\mathbb{Z}_{0:N-2}, \forall i\in\mathbb{Z}_{1:n_s}}\\
& \boldsymbol{x}_{0} \leftarrow \text{CUT}_{4G}\left(\mathcal{N}(\boldsymbol{\mu}_{0},P_{0})\right)\\
& \boldsymbol{x}^{(i)}_{k+1} = A^{(i)}_k\boldsymbol{x}^{(i)}_{k} + B^{(i)}_k\boldsymbol{u}^{(i)}_k + \boldsymbol{c}^{(i)}_k + \boldsymbol{\xi}^{(i)}_k, &{\forall k\in\mathbb{Z}_{0:N-2}, \forall i\in\mathbb{Z}_{1:n_s}} \\
& \boldsymbol{\mu}_{N-1}^* + A^{(\mu)} \delta\boldsymbol{x}_{N-1} = \boldsymbol{\mu}_{0} \\
& \norm{H_r 
\left[\sqrt{w_1}  \boldsymbol{z}^{(1)}_k 
\; 
\sqrt{w_2}  \boldsymbol{z}^{(2)}_k
\; \ldots \right]
}_{2} 
\leq 
\sqrt{\lambda_{r,\max}},
& \quad {\forall k\in\mathbb{Z}_{0:N-1}} \\
& \norm{\boldsymbol{\delta x}_k}_{\infty} \leq \Delta_{TR}, & {\forall k\in\mathbb{Z}_{0:N-1}}\\
& \norm{\delta K_k}_{\infty} \leq \Delta_{TR}, & {\forall k\in\mathbb{Z}_{0:N-2}} \\
& \boldsymbol{\xi} = 
\left[
\boldsymbol{\xi}_0^\top
,\ldots,
\boldsymbol{\xi}_{N-2}^\top 
\right]^\top
\end{aligned}
\end{equation}

\subsubsection{Improving Gaussianity of Final Distribution with 6th-order CUT}\label{sec: halo CUT6}
It is well known that the Gaussian distributions have zero skewness and a kurtosis of three. Since statistical moment steering can directly control these parameters, this example shows that even in nonlinear dynamics, the final distribution can remain Gaussian-like if these moment constraints are applied. Note that the final distribution is not truly Gaussian, as exact Gaussianity would require alignment across an infinite number of moments. Nevertheless, by matching skewness and kurtosis, the distribution’s Gaussian-like characteristics are significantly improved. Eq.~\eqref{eq: CR3BP Original Formulation CUT6} shows the original problem and Eq.~\eqref{eq: CR3BP CVX Formulation CUT6} shows the convex subproblem in \texttt{SCvx*} form for this CR3BP example. For comparative purposes, this example presents converged solutions evaluated both \emph{with} and \emph{without} the kurtosis constraint to assess its impact on the resulting distribution.

\begin{equation} \label{eq: CR3BP Original Formulation CUT6}
\begin{aligned}
\min_{
\substack{
\{\boldsymbol{x}_k\}_{
k\in\mathbb{Z}_{0:N-1}} \\
\{\bar{\boldsymbol{u}}_k, K_k\}_{
k\in\mathbb{Z}_{0:N-2}}
}}
& \Delta V_{99,\text{ub}} \\
\text{s.t. }  
& \boldsymbol{x}_{0} \leftarrow 
\text{CUT}_{6G}\left(\mathcal{N}(\boldsymbol{\mu}_{0},P_{0})\right)\\
& \boldsymbol{x}^{(i)}_{k+1} = \phi\left(\boldsymbol{x}^{(i)}_{k}, \bar{\boldsymbol{u}}_k, K_k\right), 
& \quad {\forall k\in\mathbb{Z}_{0:N-2}, \forall i\in\mathbb{Z}_{1:n_s}}\\
& f^{(\mu)} (\boldsymbol{x}_{N-1}) = \boldsymbol{\mu}_{0} \\
& \norm{H_r f^{(P^{1/2})} (\boldsymbol{x}_k)}_{2} \leq \sqrt{\lambda_{r,\max}}, & \quad {\forall k\in\mathbb{Z}_{0:N-1}}\\
& \norm{H_r f^{(\gamma)} (\boldsymbol{x}_{N-1})}_{\infty} 
\leq \epsilon_{\gamma}
&\text{(Final Skewness Constraint)}\\ 
& \norm{H_r f^{(\kappa)} (\boldsymbol{x}_{N-1}) - \vec{3}}_{\infty} \leq \epsilon_{\kappa} 
&\text{(Final Kurtosis Constraint)}\\
\end{aligned}
\end{equation} 

\begin{subequations}\label{eq: CR3BP CVX Formulation CUT6}
\begin{align}
\min_{
\substack{
\{\delta\boldsymbol{x}_k\}_{
k\in\mathbb{Z}_{0:N-1}} \\
\{\delta\bar{\boldsymbol{u}}_k,\delta K_k\}_{
k\in\mathbb{Z}_{0:N-2}} \\
\{\boldsymbol{\xi},\boldsymbol{\zeta}\}
}} & 
\text{Convexified }\Delta V_{99,\text{ub}} + P_{\texttt{SCvx*}}(w_p, \boldsymbol{\xi}, \boldsymbol{\lambda}, \boldsymbol{\zeta}, \boldsymbol{\mu}) \\
\text{s.t. } 
& \boldsymbol{x}_k = \boldsymbol{x}_k^* + \delta \boldsymbol{x}_k,
& {\forall k\in\mathbb{Z}_{0:N-1}}\\
& \boldsymbol{u}^{(i)}_k = \boldsymbol{u}_k^{(i)*} + \delta\bar{\boldsymbol{u}}_k + K_k^*\delta \boldsymbol{z}^{(i)}_k + \delta K_k {\boldsymbol{z}^{(i)*}_k},
& {\forall k\in\mathbb{Z}_{0:N-2}, \forall i\in\mathbb{Z}_{1:n_s}}\\
& \boldsymbol{x}_{0} \leftarrow \text{CUT}_{6G}\left(\mathcal{N}(\boldsymbol{\mu}_{0},P_{0})\right)\\
& \boldsymbol{x}^{(i)}_{k+1} = A^{(i)}_k\boldsymbol{x}^{(i)}_{k} + B^{(i)}_k\boldsymbol{u}^{(i)}_k + \boldsymbol{c}^{(i)}_k + \boldsymbol{\xi}^{(i)}_k, &{\forall k\in\mathbb{Z}_{0:N-2}, \forall i\in\mathbb{Z}_{1:n_s}} \\
& \boldsymbol{\xi}^{(i)}_k = 0, &{\forall k\in\mathbb{Z}_{0:N-2}, \forall i\in\mathbb{Z}_{1:n_s}}
\label{eq: CR3BP CVX Formulation CUT6, slack}
\\
& \boldsymbol{\mu}_{N-1}^* + A^{(\mu)} \delta\boldsymbol{x}_{N-1} = \boldsymbol{\mu}_{0} \\
& \norm{H_r 
\left[\sqrt{w_1}  \boldsymbol{z}^{(1)}_k 
\; 
\sqrt{w_2}  \boldsymbol{z}^{(2)}_k
\; \ldots \right]
}_{2} 
\leq 
\sqrt{\lambda_{r,\max}}, 
& \quad {\forall k\in\mathbb{Z}_{0:N-1}} \\
& \norm{H_r \left( 
\boldsymbol{\gamma}_{N-1}^* + A^{(\gamma)} \biggr\rvert_{\boldsymbol{z}_{N-1}^*}\delta \boldsymbol{z}_{N-1}
\right)}_{\infty} \leq \epsilon_{\gamma} + \zeta^{(\gamma)}, 
&\zeta^{(\gamma)} \geq 0 \\
& \norm{H_r \left( 
\boldsymbol{\kappa}_{N-1}^* + A^{(\kappa)} \biggr\rvert_{\boldsymbol{z}_{N-1}^*}\delta \boldsymbol{z}_{N-1}
\right) - \vec{3}}_{\infty} \leq \epsilon_{\kappa} + \zeta^{(\kappa)}, 
&\zeta^{(\kappa)} \geq 0 \\
& \norm{\boldsymbol{\delta x}_k}_{\infty} \leq \Delta_{TR}, & {\forall k\in\mathbb{Z}_{0:N-1}}\\
& \norm{\delta K_k}_{\infty} \leq \Delta_{TR}, & {\forall k\in\mathbb{Z}_{0:N-2}} \\
& \boldsymbol{\xi} = 
\left[
\boldsymbol{\xi}_0^\top
,\ldots,
\boldsymbol{\xi}_{N-2}^\top 
\right]^\top,
\qquad
\boldsymbol{\zeta} = 
\left[
\zeta^{(\gamma)},\zeta^{(\kappa)}
\right]^\top
\end{align}
\end{subequations}
where $\vec{3}$ denotes a column vector with entries of only threes and $A^{(\kappa)} \triangleq A^{(^4 C)}$. The values of $\epsilon_{\gamma} = 0.01$ and $\epsilon_{\kappa} = 0.5$ are chosen to ensure that the moment deviations from that of a true-Gaussian remain small. The skewness and kurtosis moment constraints are formulated as an inequality rather than an equality due to the same reasons presented in the two-body example. The value of $\epsilon_{\kappa}$ is larger than $\epsilon_{\gamma}$, reflecting the increased difficulty in controlling higher-order moments. This limitation is examined in detail in later sections.

Note that Eq.~\eqref{eq: CR3BP CVX Formulation CUT6, slack} does not allow for slacking of the sigma point dynamics constraint. The solution from Example~\ref{sec: halo CUT4} serves as the initial reference, placing the reference already close to a feasible solution. Consequently, introducing slack variables in the dynamics is unnecessary and was found to degrade convergence due to excessive relaxation.

\subsection{Numerical Considerations}
\begin{table}[htbp]
	\fontsize{10}{10}\selectfont
    \caption{\texttt{SCvx*} Parameters for CR3BP Example}
   \label{tab: CR3BP SCvx Params}
    \centering 
    \small
   \begin{tabular}{l c c c c c c c c} 
      \hline 
      Parameter & 
      $\{ \epsilon_{\text{opt}},\epsilon_{\text{feas}}\}$ & $\{\eta_0,\eta_1,\eta_2\}$&
      $\{\alpha_1,\alpha_2\}$&
      $\{\beta_{\texttt{SCvx*}},\gamma_{\texttt{SCvx*}}\}$&
      $\Delta_{\text{TR}}^{(1)}$&
      $\{\Delta_{\text{TR},\min},\Delta_{\text{TR},\max}\}$&  
      $w^{(1)}_p$ &
      $w_{p,\max}$\\
      \hline 
      Value & 
    $\{ 10^{-4},10^{-7}\}$ & $\{1,0.2,0.1\}$&
      $\{3,2\}$&
      $\{1.5,0.99\}$&
      $0.1$& 
      $\{10^{-10},0.1\}$& 
      $100$ &
      $10^{10}$\\
      \hline
   \end{tabular}
\end{table}

Table~\ref{tab: CR3BP SCvx Params} lists the numerical parameters used for \texttt{SCvx*} for solving both of the examples from Example~\ref{sec: halo CUT4} and \ref{sec: halo CUT6}. In the case of the halo stationkeeping example, the initial reference mean $\boldsymbol{\mu}_{\text{guess},k}$ corresponds to the state along the halo orbit. For covariance, the same ``scaled linear covariance'' approach from the two-body example is used (see Appendix~\ref{appendix: Scaled Linear Covariance}). Likewise, the initial reference sigma points are sampled from these Gaussians of $\mathcal{N}(\boldsymbol{\mu}_{\text{guess},k}, P_{\text{guess},k})$.

Note that there is no additional numerical scaling of the problem since the CR3BP equation of motion from Eq.~\eqref{eq: CR3BP EOM} is already nondimensional.


\subsection{Results and Discussion for Example~\ref{sec: halo CUT4}}
This SCP problem converged after 45 iterations and took about 56 minutes to solve using \texttt{MOSEK} with \texttt{CVX} on \texttt{MATLAB} R2024a. It should be noted that this problem is a much larger optimization problem compared to the two-body example, with $N_{\text{CR3BP}}=19$ optimization nodes compared to $N_{\text{2BP}}=9$, hence the longer computation time. The convergence profile of the \texttt{SCvx*} algorithm for this CR3BP example is shown in Figure~\ref{fig: CR3BP, CUT4, scvx convergence}.

\begin{figure}[!htb]
	\centering\includegraphics[width=0.6\textwidth]{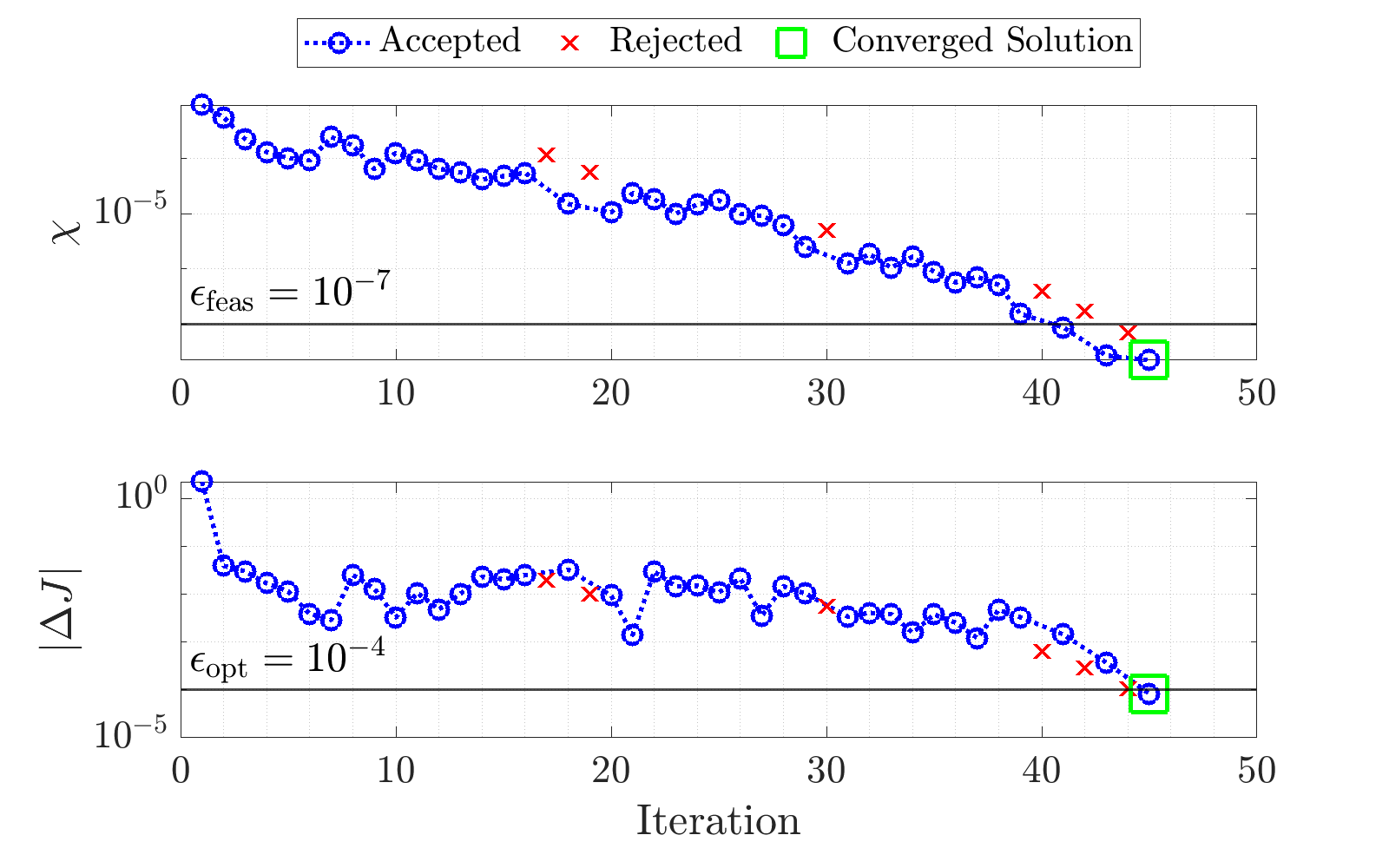}
	\caption{Convergence profile of \texttt{SCvx*} for CR3BP example with CUT4. \emph{Y-axis in log scale.}}
	\label{fig: CR3BP, CUT4, scvx convergence}
\end{figure}

\begin{figure}[!htb]
  \centering 
    \subcaptionbox
    {First Revolution.}
    {\includegraphics[width=0.49\textwidth]{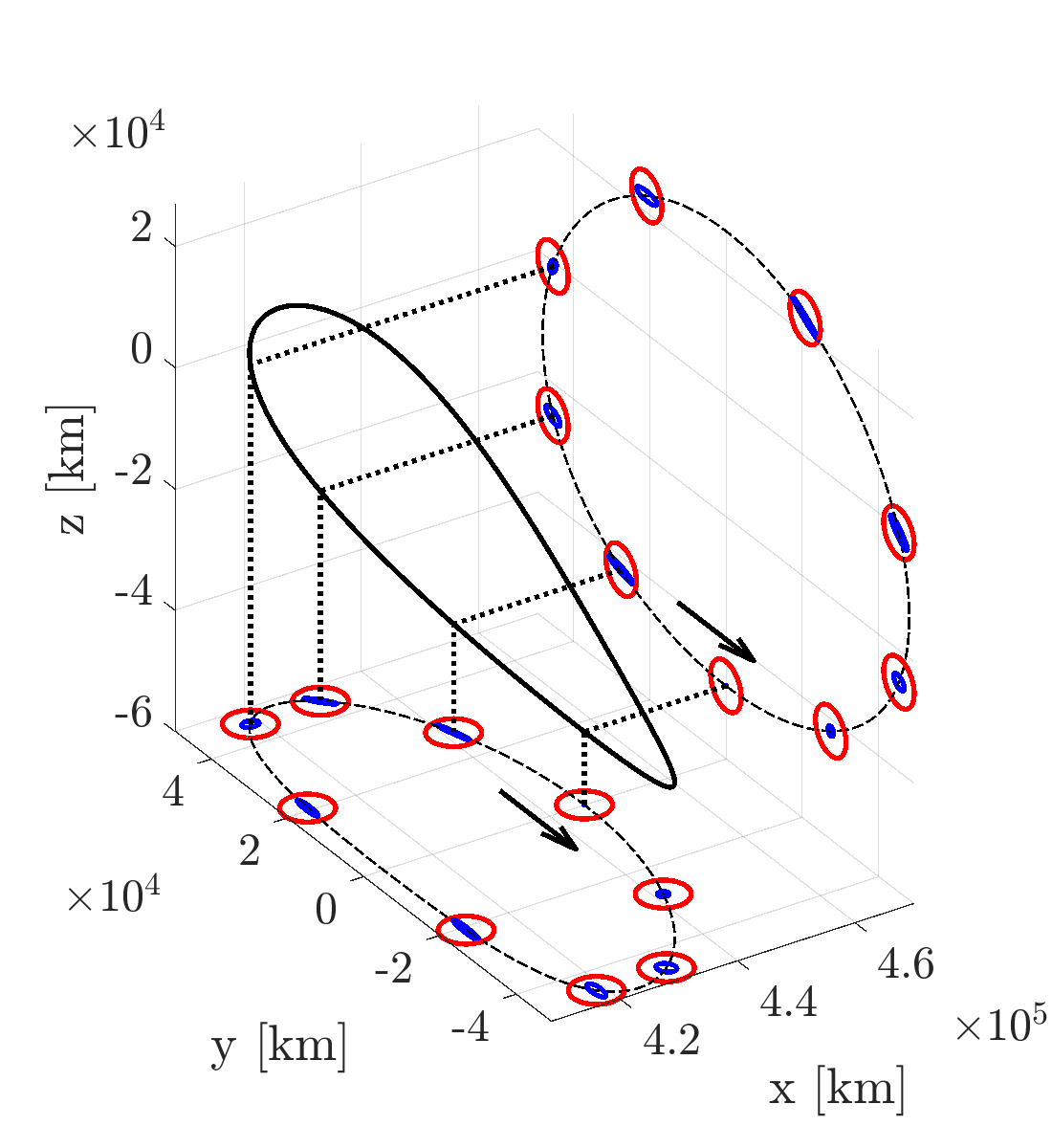}}
    \hskip 0.1truein
    \subcaptionbox
    {Second Revolution.}
    {\includegraphics[width=0.49\textwidth]{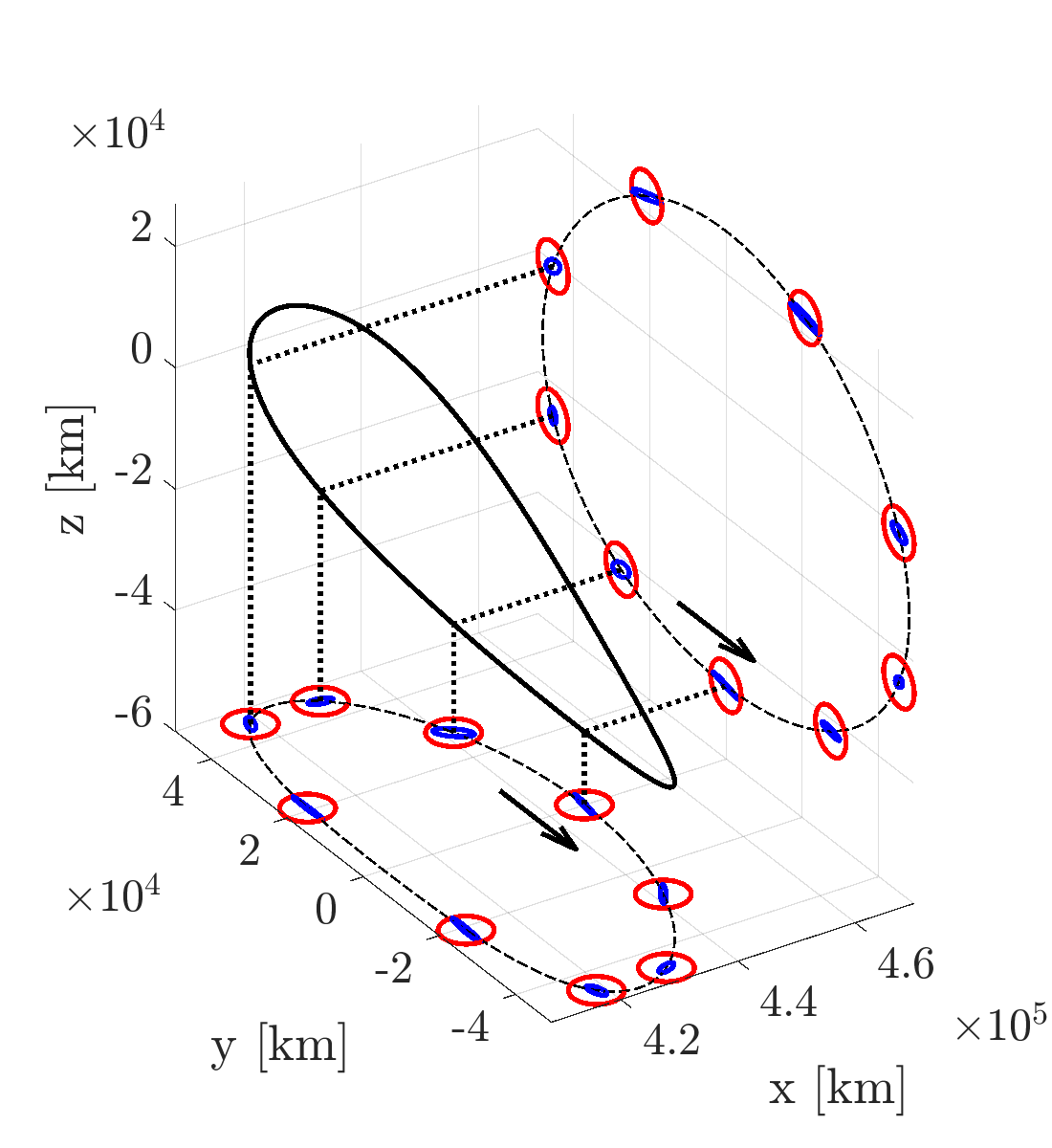}}
    \caption{Halo orbit with projected $3\sigma$ constraint in red and $3\sigma$ from optimized CUT points in blue. Trajectories start at the apolune node. \emph{All $3\sigma$ ellipses enlarged $2{\times}$ for visibility.}}
    \label{fig: CR3BP CUT4 Optimized}
\end{figure}

Figure~\ref{fig: CR3BP CUT4 Optimized} shows the evolution of the calculated $3\sigma$ ellipsoids from CUT with the intermediate $3\sigma$ constraints. The ellipsoids along the orbit are projected onto a plane and enlarged for better visualization. It can be seen that the constraint is satisfied throughout all nodes as expected. The final distribution from the Monte Carlo is shown in Figure~\ref{fig: CR3BP CUT4, final Distribution}. It can be seen that the final distribution is heavily skewed and definitively non-Gaussian. However, the covariance of the distribution aligns almost perfectly with the one predicted by CUT, and thus the Monte Carlo's covariance satisfies the imposed covariance constraints. This is in contrast to previous linear covariance controllers \cite{Qi-Stationkeeping}, which fail to both predict and enforce the imposed constraints for this halo orbit as it exhibits strong nonlinear behavior.

\begin{figure}[!htb]
	\centering\includegraphics[width=0.95\textwidth]{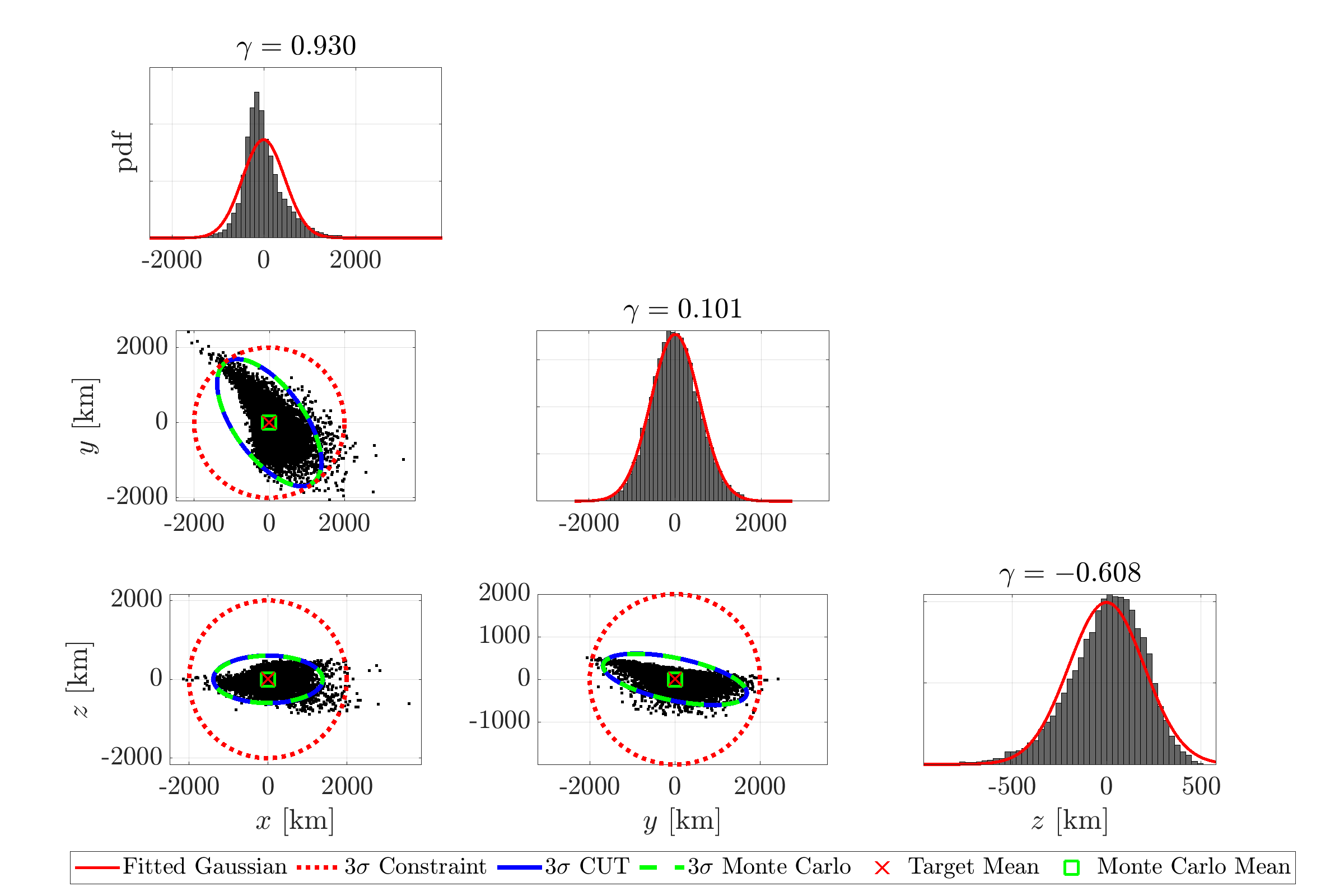}
	\caption{Monte Carlo ($n_{\text{samples}}=10,000$) for CR3BP example with CUT4: Final non-Gaussian distribution controlled by statistical moment steering. \emph{Origin normalize to mean predicted by CUT.}}
	\label{fig: CR3BP CUT4, final Distribution}
\end{figure}

\begin{figure}[!htb]
  \centering 
    \subcaptionbox
    {Time History of $\Delta V$.}
    {\includegraphics[width=0.4\textwidth]{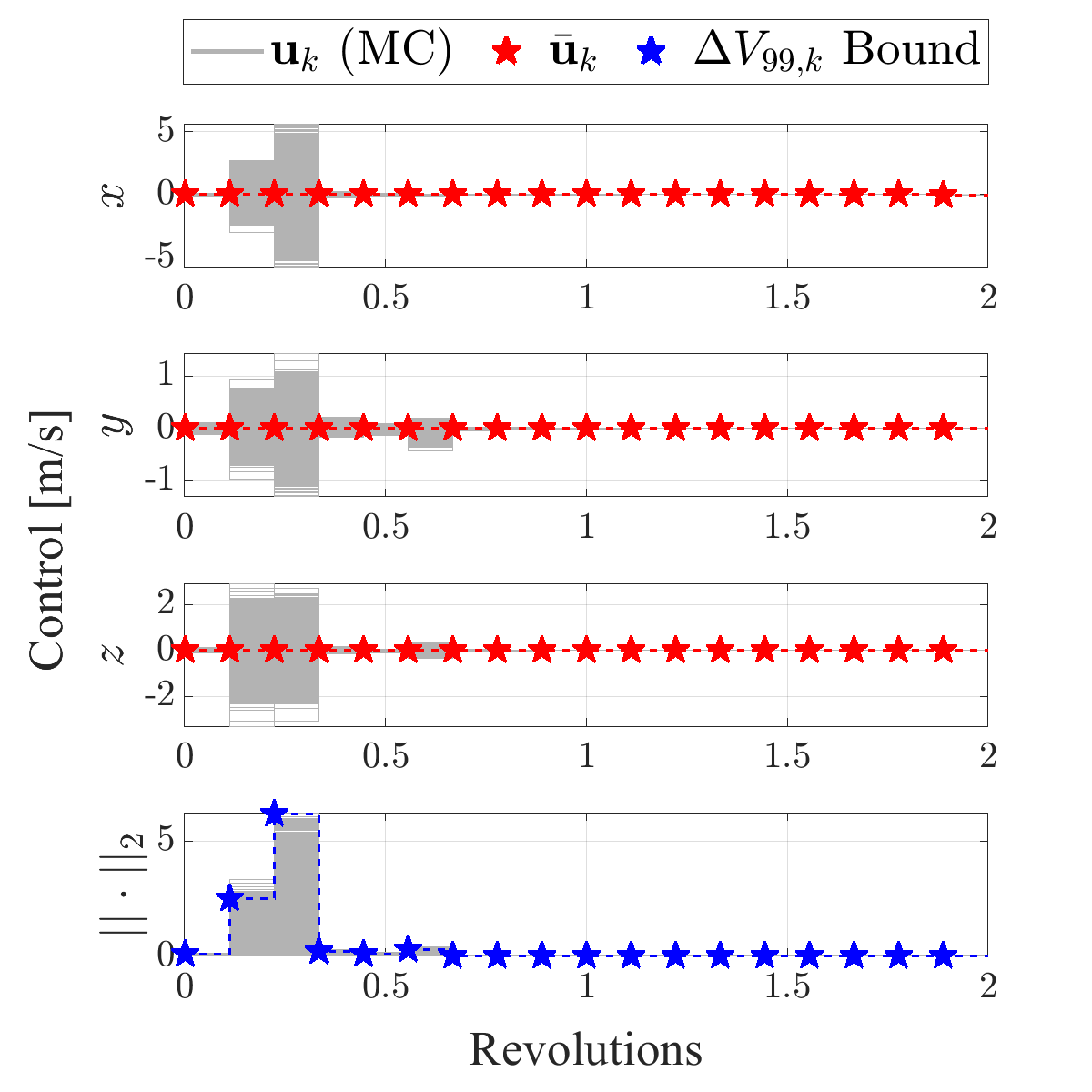}}
    \hskip 0.1truein
    \subcaptionbox
    {Histogram of Total $\Delta V$ Costs.}
    {\includegraphics[width=0.55\textwidth]{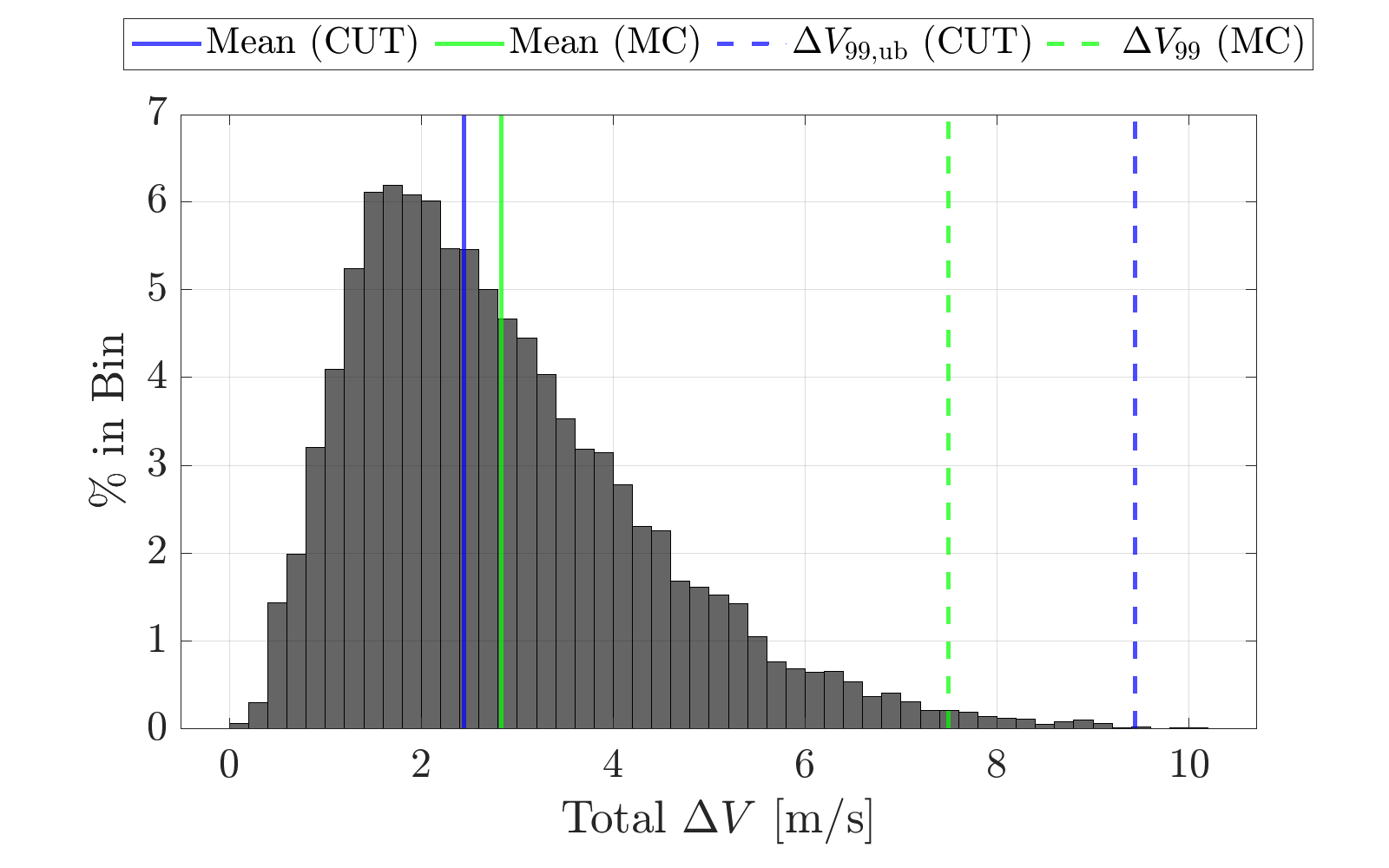}}
    \caption{Monte Carlo ($n_{\text{samples}}=10,000$) for CR3BP example with CUT4: $\Delta V$ costs for statistical moment steering of CR3BP example along with statistical parameters calculated using both CUT and Monte Carlo.}
    \label{fig: CR3BP CUT4, DeltaV}
\end{figure}

Figure~\ref{fig: CR3BP CUT4, DeltaV} shows both the time history of control as well as the total $\Delta V$ cost. Despite the non-Gaussian nature of the state distribution, $\Delta V_{99,\text{ub}}$ still provides a reasonable upper-bound to the actual $99$-th percentile of $\Delta V$ costs. The expected value for the fuel costs calculated by CUT also accurately captures the average fuel costs from the Monte Carlo, albeit a slightly worse prediction compared to the two-body example. 

\begin{figure}[!htb]
  \centering 
    \subcaptionbox
    {Mean.\label{fig: CR3BP CUT4, mean}}
    {\includegraphics[width=0.8\textwidth]{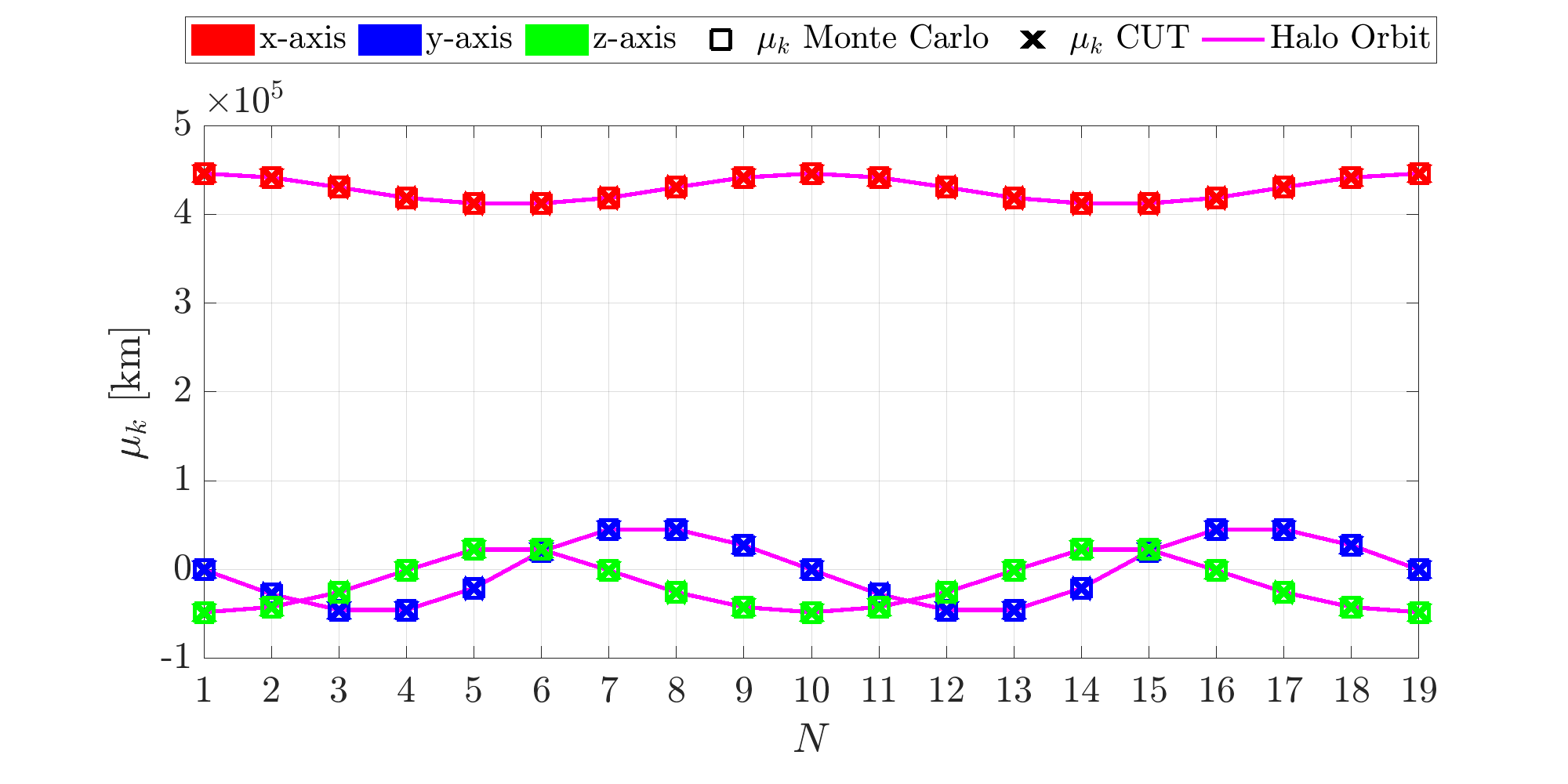}}
    \hskip 0.1truein
    \subcaptionbox
    {$3 \sigma$ (along basis axes).\label{fig: CR3BP CUT4, 3sigma}}
    {\includegraphics[width=0.8\textwidth]{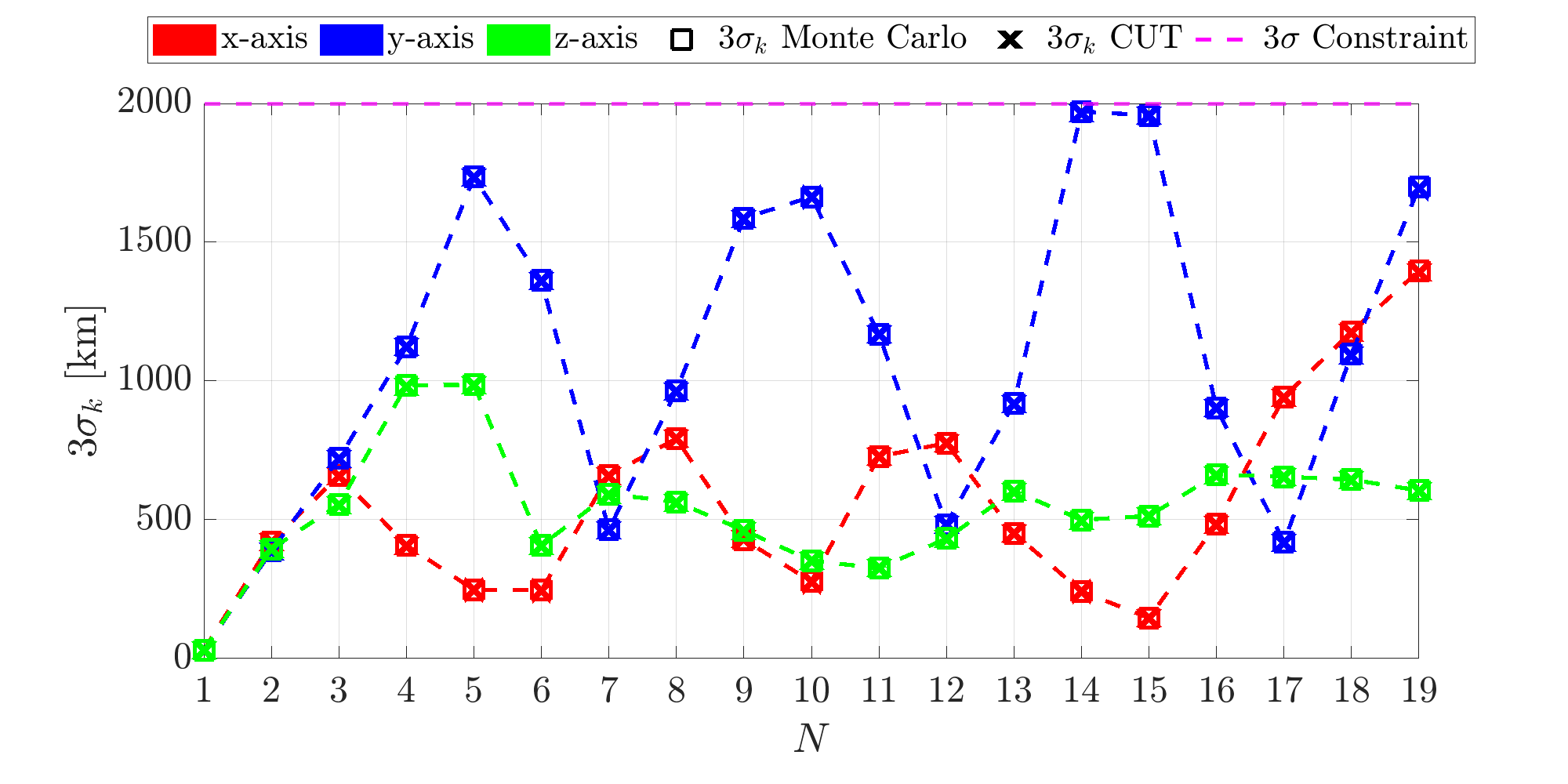}}
    \caption{Monte Carlo ($n_{\text{samples}}=10,000$) for CR3BP example with CUT4: Time history of lower moments predicted by CUT vs Monte Carlo.}
    \label{fig: CR3BP CUT4, lower moments}
\end{figure}

Figure~\ref{fig: CR3BP CUT4, lower moments} illustrates the accuracy of CUT in predicting the lower-order moments (mean and covariance) at each node, benchmarked against Monte Carlo results. Covariance is visualized using the $3\sigma$ bounds of its diagonal entries. It can be seen that both the mean and $3\sigma$ estimates are exactly aligned with those of the Monte Carlo. This represents a substantial improvement over the results in Ref.~\citenum{Qi-Stationkeeping}, where the halo orbit’s nonlinearity led to a breakdown in the predictive accuracy of the covariance controller’s linear approximation. This demonstrates that, even without considering its ability to control higher-order moments, statistical moment steering's capability to accurately predict the mean and covariance of non-Gaussian distributions in nonlinear environments marks an improvement over previous linear covariance controllers. 

\begin{figure}[!htb]
  \centering 
    \subcaptionbox
    {Skewness.\label{fig: CR3BP CUT4, skewness}}
    {\includegraphics[width=0.8\textwidth]{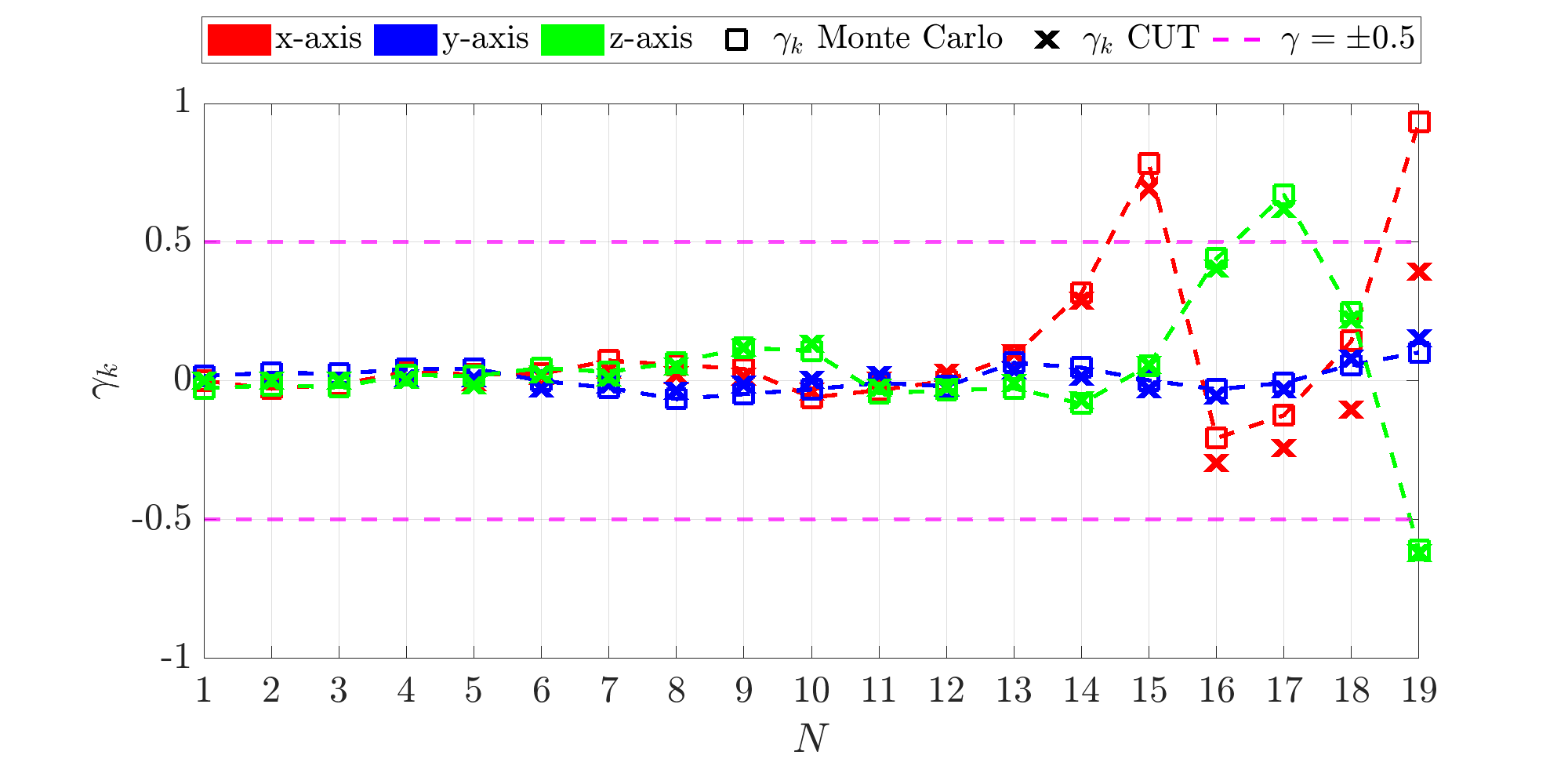}}
    \hskip 0.1truein
    \subcaptionbox
    {Kurtosis.\label{fig: CR3BP CUT4, kurt}}
    {\includegraphics[width=0.8\textwidth]{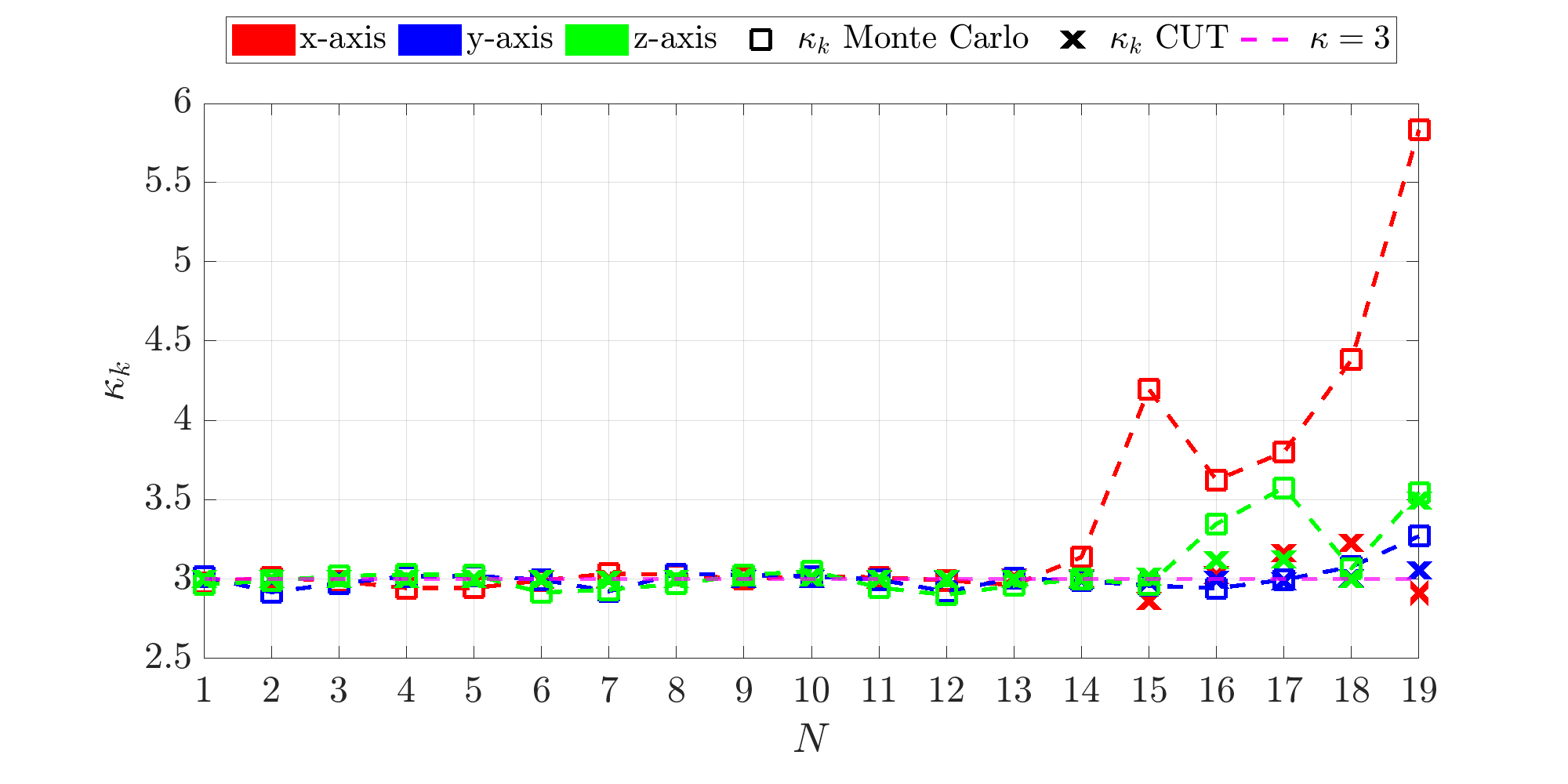}}
    \caption{Monte Carlo ($n_{\text{samples}}=10,000$) for CR3BP example with CUT4: Time history of higher moments predicted by CUT vs Monte Carlo.}
    \label{fig: CR3BP CUT4, higher moments}
\end{figure}

Figure~\ref{fig: CR3BP CUT4, higher moments} compares the predicted higher-order moments from CUT with those obtained via Monte Carlo simulations at each node. In Figure~\ref{fig: CR3BP CUT4, skewness}, skewness exceeding the line of $\gamma = \pm0.5$ can be viewed as non-Gaussian from a statistical sampling standpoint.\footnote{This value for skewness is not definitive. Some papers use $\pm0.5$ as a measure \cite{skewness-0.5}, while others use values much greater such as $\pm 1$ \cite{skewness-one}.} In Figure~\ref{fig: CR3BP CUT4, kurt}, $\kappa =3$ denotes the value of kurtosis for a true Gaussian possesses. It can be seen that there are multiple locations in which the distribution becomes non-Gaussian. This reinforces the key advantage of statistical moment steering: it remains effective without assuming Gaussianity, unlike conventional linear covariance controllers that rely on such assumptions for accurate performance.

The skewness prediction in the x-axis starts to become noticeably poor towards the last few nodes. Although some deviation between CUT and Monte Carlo can be attributed to statistical sampling errors, the y-axis and z-axis CUT predictions are still relatively good. These results indicate a potential breakdown in CUT’s estimation accuracy. But even with diminished predictive fidelity along the x-axis, the method continues to reflect the trends in skewness. Conversely, the kurtosis estimates produced by CUT deviate significantly from those obtained via Monte Carlo simulation. This highlights the fact that CUT is still only an approximation technique, and as the order of statistical moments increases, their estimation and control become increasingly challenging with lower-order CUT. This motivates the use of 6th-order CUT in the next section to control kurtosis.

\subsection{Results and Discussion for Example~\ref{sec: halo CUT6}}
As mentioned earlier, this example presents converged solutions of Eq.~\eqref{eq: CR3BP CVX Formulation CUT6} both with and without the kurtosis constraint. The SCP problem with the kurtosis constraint converged after 16 iterations and took about 63 minutes, and the problem without the kurtosis constraint converged after 16 iterations and took about 61 minutes. Both were solved using \texttt{MOSEK} with \texttt{CVX} on \texttt{MATLAB} R2024a. Both problems had similar computation time, and more similarities can also be seen in their convergence profiles shown in Figure~\ref{fig: CR3BP, CUT6, scvx convergence}. Due to the better initial reference, the problems converge in fewer iterations than in the previous 4th-order CUT example. Each iteration took longer to solve as a result of the greater number of sigma points (76 points for CUT-4G compared to 137 for CUT-6G in $\mathbb{R}^6$) needed for the 6th-order CUT, and thus a greater number of optimization variables for the convex subproblem.

\begin{figure}[!htb]
  \centering 
    \subcaptionbox
    {Without Kurtosis Constraint.}
    {\includegraphics[width=0.49\textwidth]{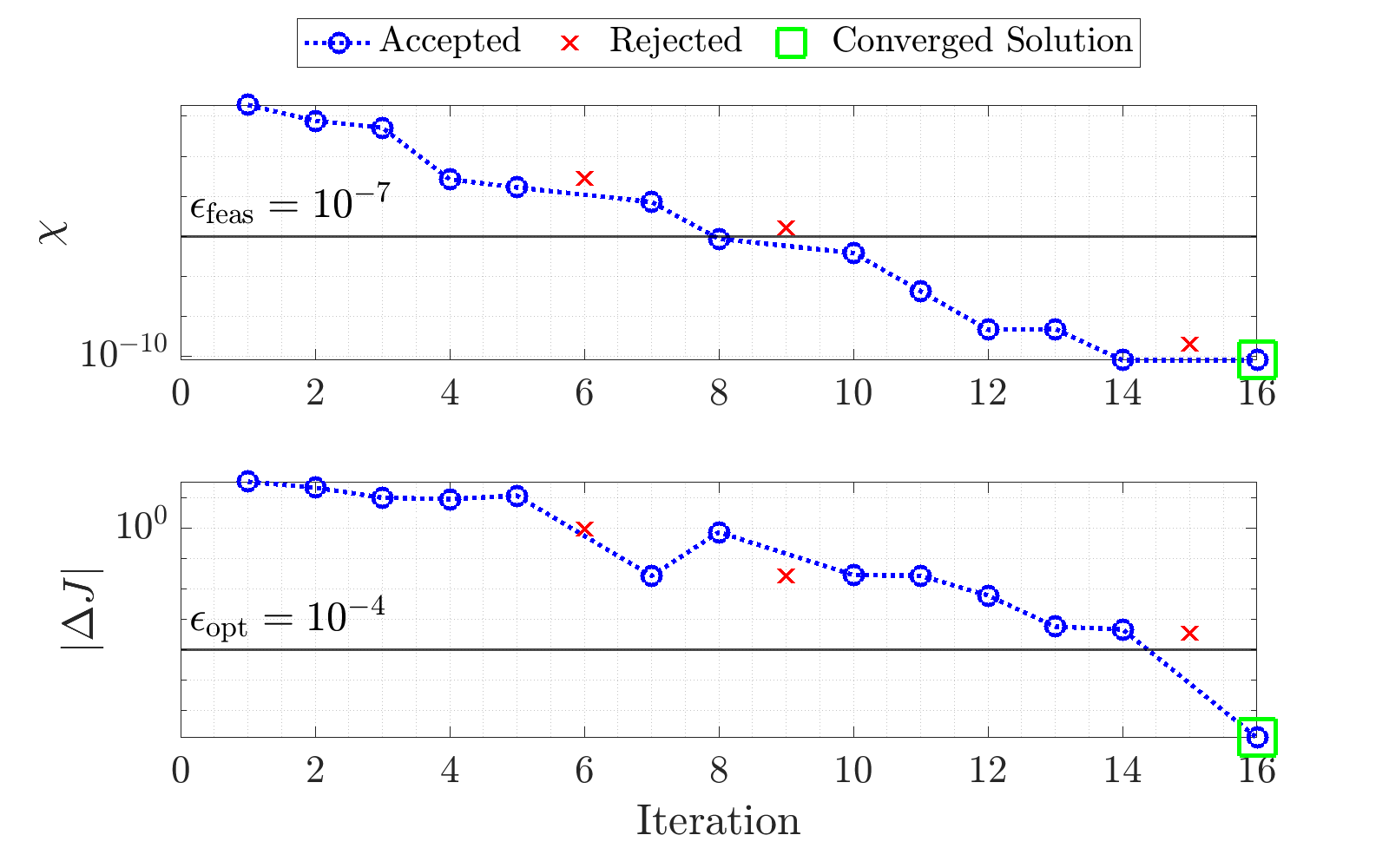}}
    \hskip 0.1truein
    \subcaptionbox
    {With Kurtosis Constraint.}
    {\includegraphics[width=0.49\textwidth]{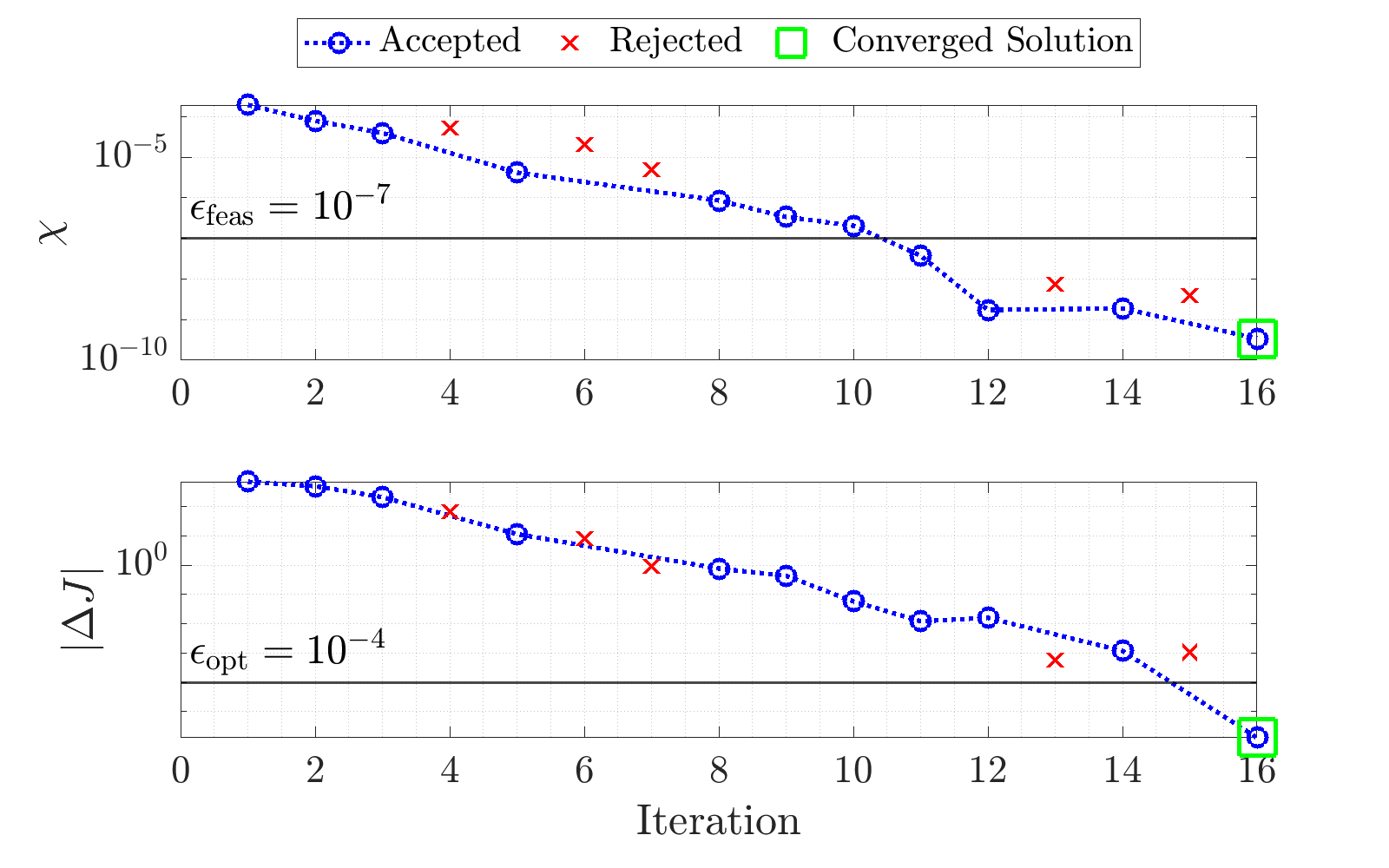}}
    \caption{Convergence profile of \texttt{SCvx*} for CR3BP example with CUT6. \emph{Y-axis in log scale.}}
    \label{fig: CR3BP, CUT6, scvx convergence}
\end{figure}

\begin{figure}[!htbp]
  \centering 
    \subcaptionbox
    {Without Kurtosis Constraint. \label{fig: CR3BP CUT6, CUT6 Results, kurt off}}
    {\includegraphics[width=0.9\textwidth]{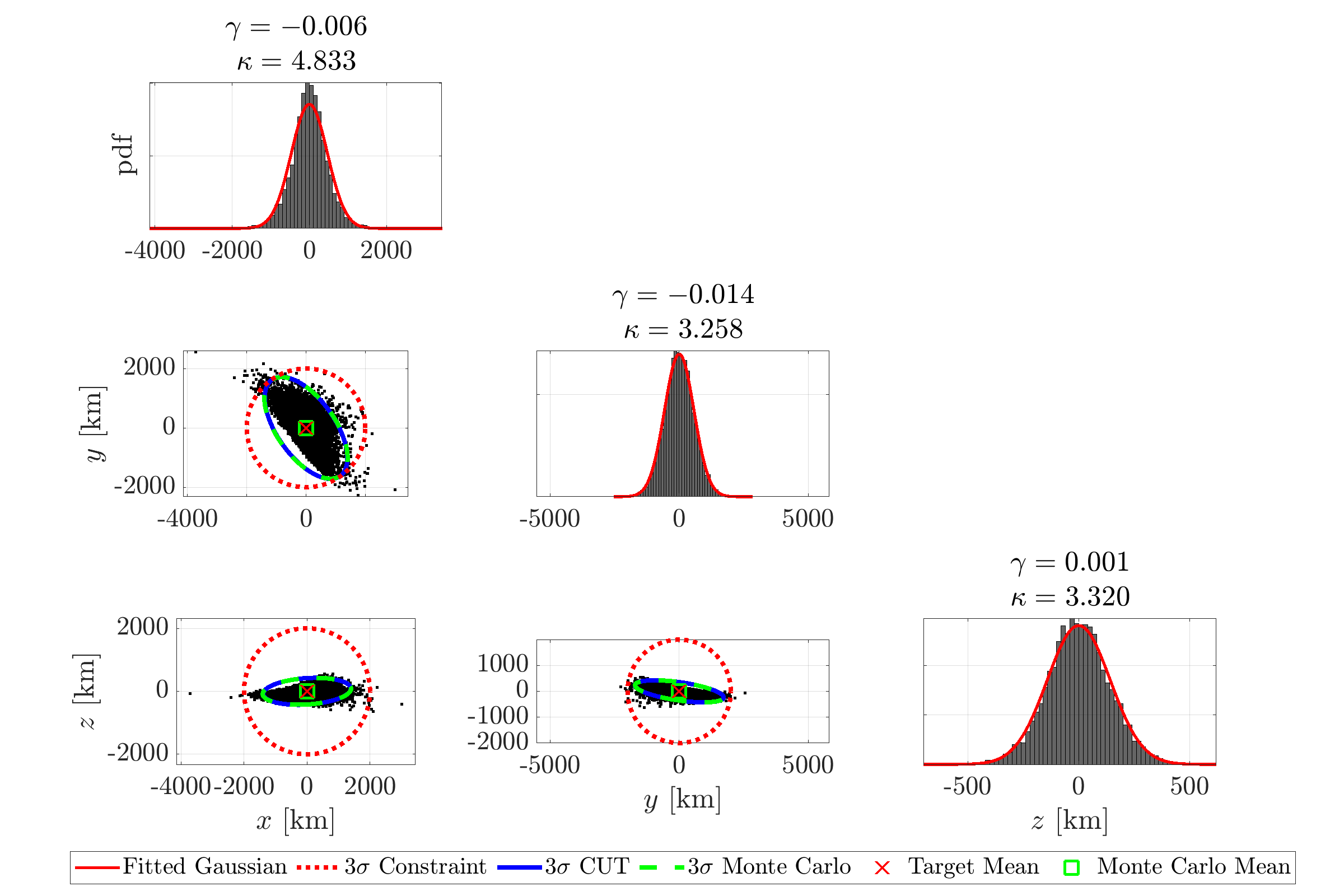}}
    \hskip 0.1truein
    ~
    \subcaptionbox
    {With Kurtosis Constraint. \label{fig: CR3BP CUT6, CUT6 Results, kurt on}}
    {\includegraphics[width=0.9\textwidth]{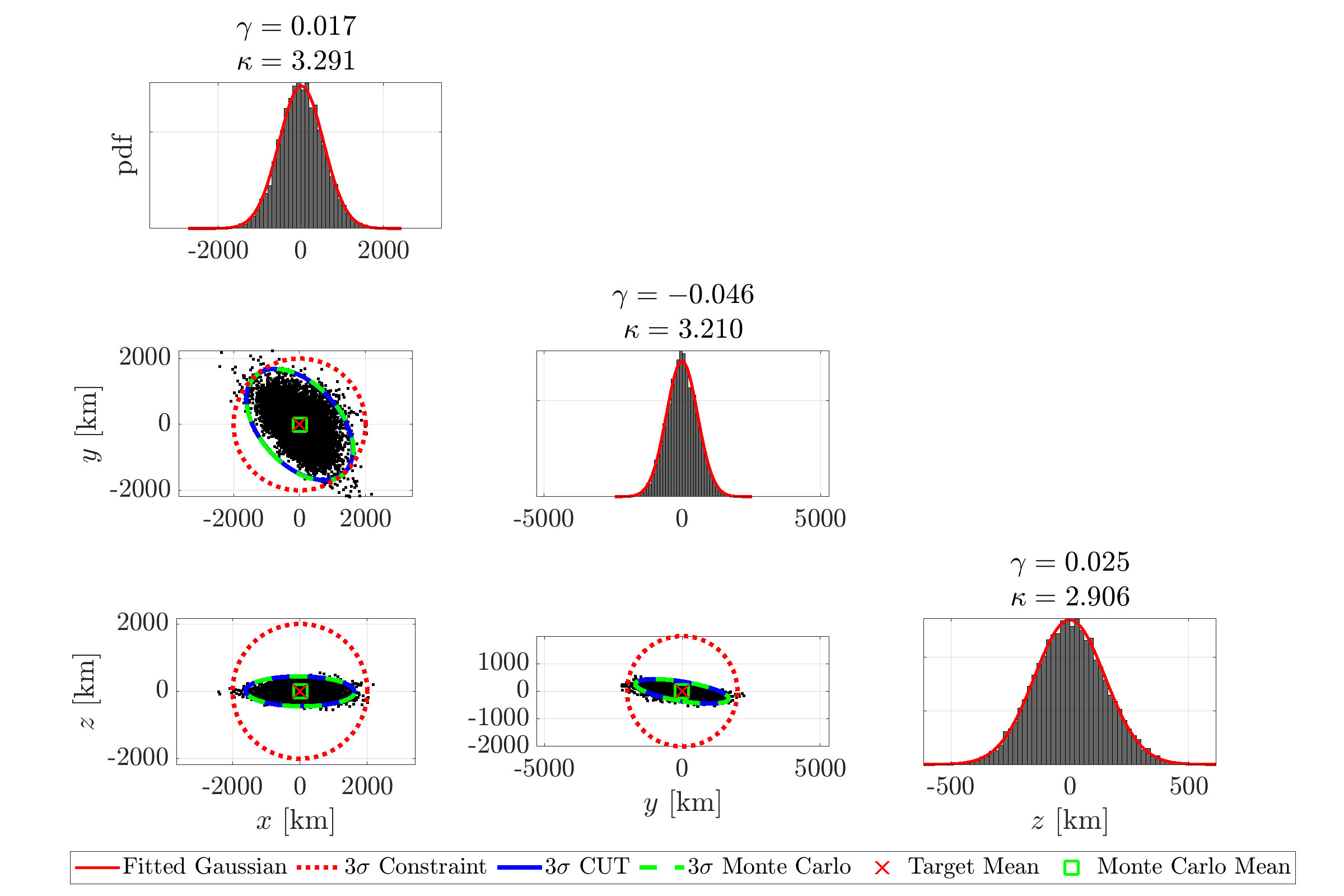}}
    \caption{Monte Carlo ($n_{\text{samples}}=10,000$) for CR3BP example with CUT6: Final unskewed distribution controlled by statistical moment steering with and without an additional kurtosis constraint. \emph{Origin normalize to mean predicted by CUT.}}
    \label{fig: CR3BP CUT6, CUT6 Results}
\end{figure}

The final distribution from the Monte Carlo is shown in Figure~\ref{fig: CR3BP CUT6, CUT6 Results}. Figure~\ref{fig: CR3BP CUT6, CUT6 Results, kurt off} shows the distribution after solving the problem without the kurtosis constraint (i.e., only skewness constraint), and Figure~\ref{fig: CR3BP CUT6, CUT6 Results, kurt on} shows the distribution after solving the problem with both skewness and kurtosis constraints. Comparing the plots, the kurtosis constraint does affect the final distribution to a noticeable degree in the x-axis. Despite all distributions being near-symmetric, the kurtosis in the x-axis is not aligned with that of a Gaussian if the kurtosis constraint is dropped. This is likely due to the stronger nonlinearities and longer time horizon, allowing the distribution to become more non-Gaussian in its higher moments. This indicates that, in some cases like the two-body example, a skewness constraint alone is sufficient for obtaining a good Gaussian fit; but in other cases, it may require higher-order constraints for an accurate Gaussian fit. It should be remembered that these distributions are theoretically still not Gaussian, but Gaussian-like in that the first couple of moments match those of a true Gaussian distribution.  

\begin{figure}[!htb]
  \centering 
    \subcaptionbox
    {Time History of $\Delta V$.}
    {\includegraphics[width=0.4\textwidth]{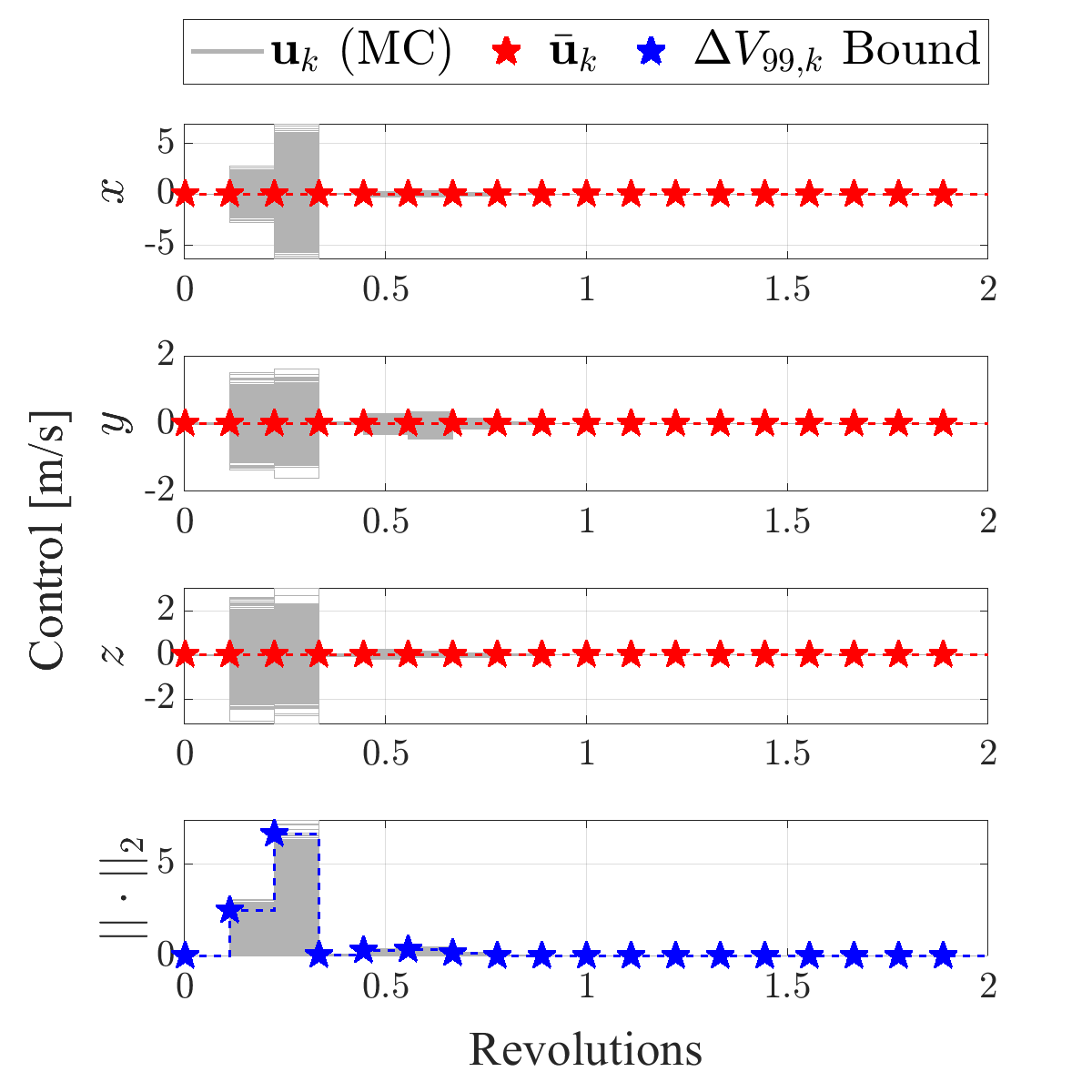}}
    \hskip 0.1truein
    \subcaptionbox
    {Histogram of Total $\Delta V$ Costs.}
    {\includegraphics[width=0.55\textwidth]{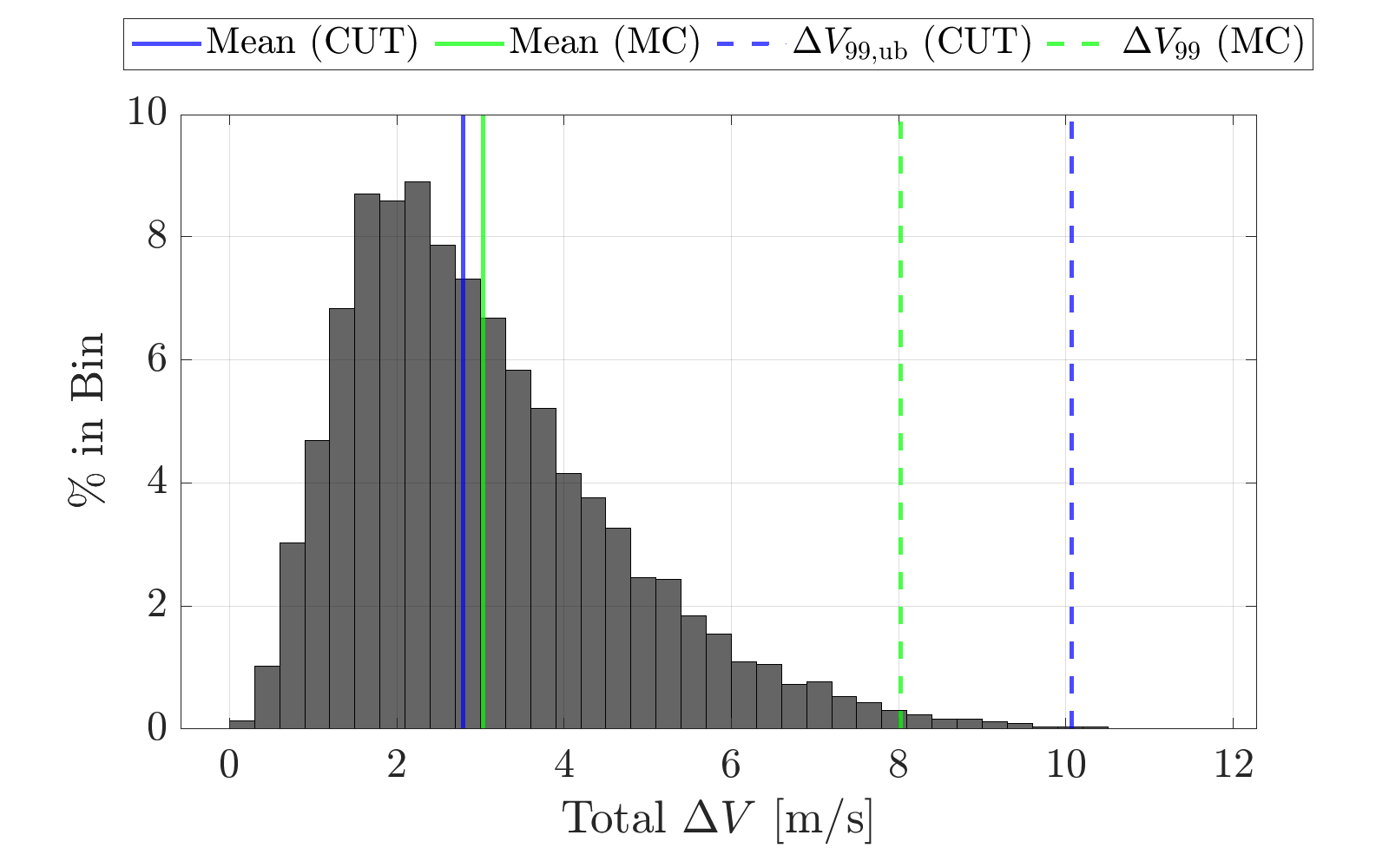}}
    \caption{Monte Carlo ($n_{\text{samples}}=10,000$) for CR3BP example with CUT6 and kurtosis constraint: $\Delta V$ costs for statistical moment steering along with statistical parameters calculated using both CUT and Monte Carlo.}
    \label{fig: CR3BP CUT6, DeltaV}
\end{figure}

Figure~\ref{fig: CR3BP CUT6, DeltaV} shows the $\Delta V$ histories and total fuel costs. The plots are nearly identical to the ones found in Figure~\ref{fig: CR3BP CUT4, DeltaV}, which can be attributed to the fact that the previous 4th-order CUT solution was the initial reference for this example. Still, the drastic difference in the final distribution again underscores the sensitivity of the system to small variations in nominal control or control gains. This behavior is consistent with the inherently chaotic nature of the three-body problem.

\begin{figure}[!htb]
  \centering 
    \subcaptionbox
    {Skewness.\label{fig: CR3BP CUT6, skewness}}
    {\includegraphics[width=0.8\textwidth]{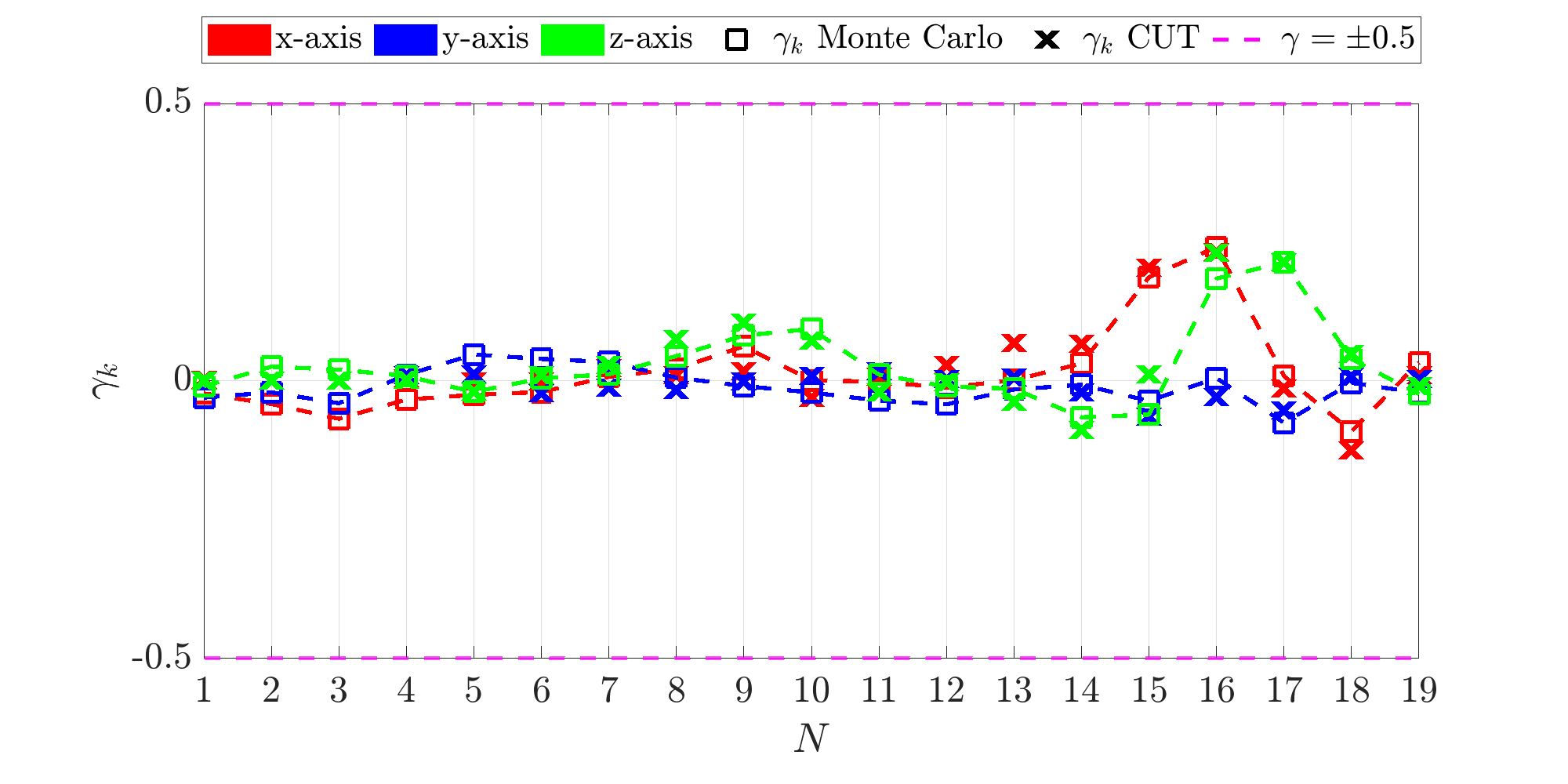}}
    \hskip 0.1truein
    \subcaptionbox
    {Kurtosis.\label{fig: CR3BP CUT6, kurt}}
    {\includegraphics[width=0.8\textwidth]{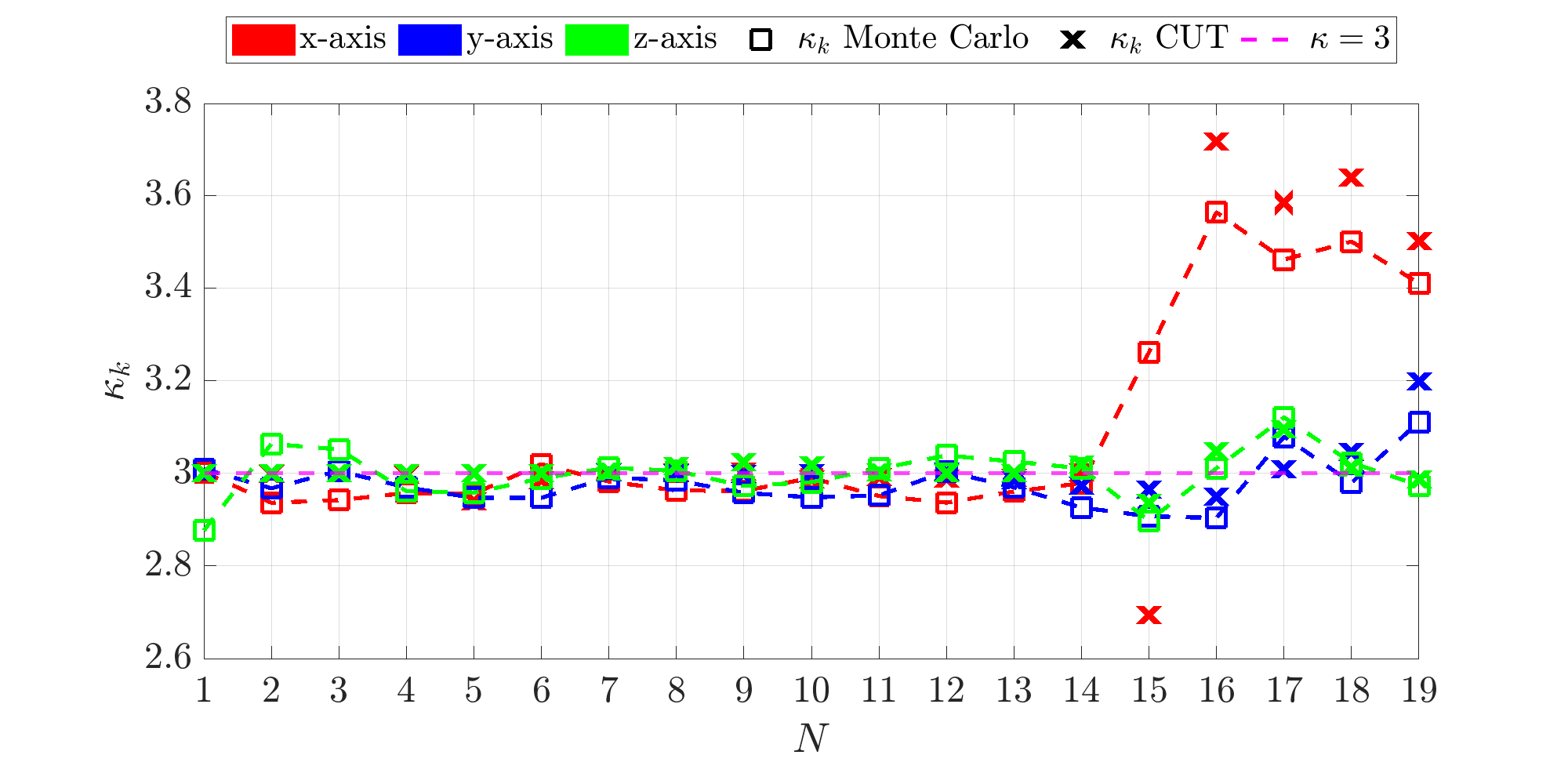}}
    \caption{Monte Carlo ($n_{\text{samples}}=10,000$) for CR3BP example with CUT6 and kurtosis constraint: Time history of higher moments predicted by CUT vs Monte Carlo.}
    \label{fig: CR3BP CUT6, higher moments}
\end{figure}

Figure~\ref{fig: CR3BP CUT6, higher moments} shows the time history of skewness for the CR3BP example with the additional kurtosis constraint. It can be seen that the 6th-order CUT prediction of these moments are much better than that of the 4th-order CUT. Firstly, the prediction of skewness in Figure~\ref{fig: CR3BP CUT6, skewness} shows no divergence in the estimation from the Monte Carlo values. In the kurtosis results shown in Figure~\ref{fig: CR3BP CUT6, kurt}, certain nodes exhibit anomalously poor predictions. But given the scale of the kurtosis, these deviations are relatively modest compared to the more pronounced divergence observed in the 4th-order CUT predictions from Figure~\ref{fig: CR3BP CUT4, kurt}. This exemplifies the power of CUT's prediction on even the higher-ordered moments, and statistical moment steering's ability to control them given a high enough order of CUT. 

\section{Remarks on Statistical Moment Steering}\label{sec: discussion and future works}
\subsection{Convergence Analysis of Monte Carlo}
This paper utilizes Monte Carlo simulations to evaluate the accuracy of statistical moment steering. The Monte Carlo sampling size is chosen to be $n_{\text{samples}}=10,000$ to balance accuracy with computational efficiency. While CUT and statistical moment steering both only approximate the true distribution, it must be made clear that Monte Carlo simulations are also an approximation of the distribution. This section justifies that $10,000$ Monte Carlo samples are sufficiently large to well approximate the true distribution and draw conclusions on the accuracy of CUT. 

One approach to Monte Carlo convergence analysis is to first generate a much larger sample set to serve as a reference for the ``true'' distribution \cite{MC-analysis-Ballio, MC-analysis-Bishop}. A quarter of a million samples are used for this analysis. The statistical moments of the terminal distribution are computed with the first $n_{\text{samples}} \leq 250,000$ samples to get $\mu_{n}$, $3\sigma_{n}$, $\gamma_{n}$, and $\kappa_{n}$. Then these are compared with the corresponding statistical moment computed with $n_{\text{samples}}=250,000$, denoted by $\mu_{\text{250k}}$, $3\sigma_{\text{250k}}$, $\gamma_{\text{250k}}$, and $\kappa_{\text{250k}}$. If the moments appear to converge to $n_{\text{samples}}=250,000$ before $n_{\text{samples}} = 10,000$, then it can be concluded that $n_{\text{samples}} = 10,000$ is sufficiently large to approximate the true distribution.

\begin{figure}[!htb]
  \centering 
    ~
    \subcaptionbox
    {Mean.\label{fig: MC Convergence, mean}}
    {\includegraphics[width=0.48\textwidth]{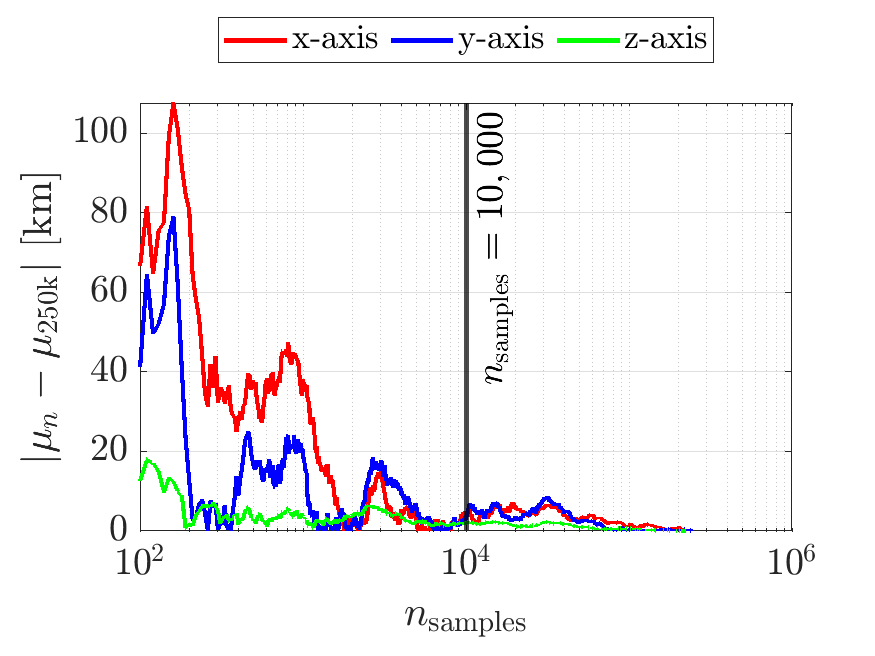}}
    \hskip 0.1truein
    \subcaptionbox
    {$3\sigma$.\label{fig: MC Convergence, 3sigma}}
    {\includegraphics[width=0.48\textwidth]{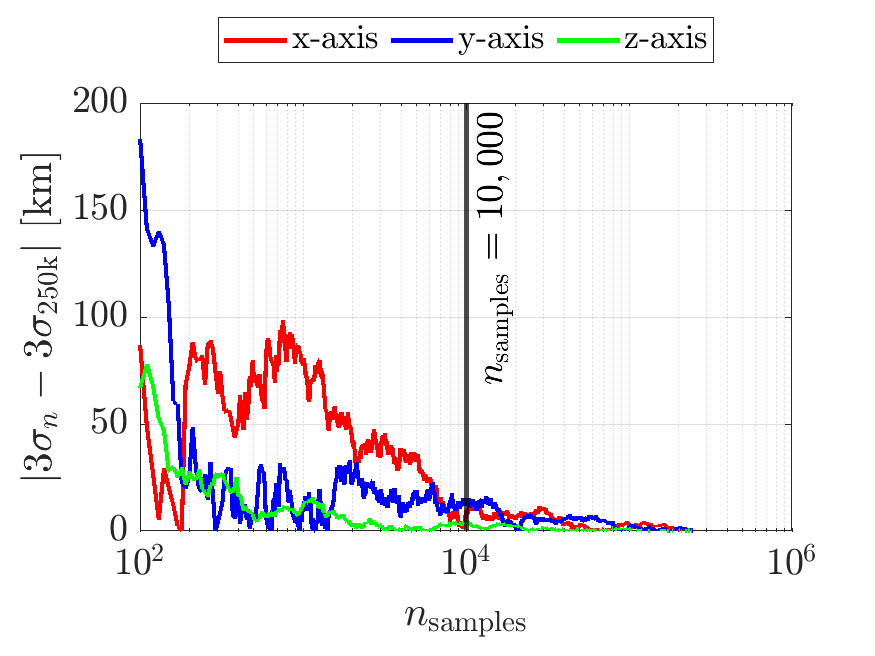}}
    ~
    \subcaptionbox
    {Skewness.\label{fig: MC Convergence, skew}}
    {\includegraphics[width=0.48\textwidth]{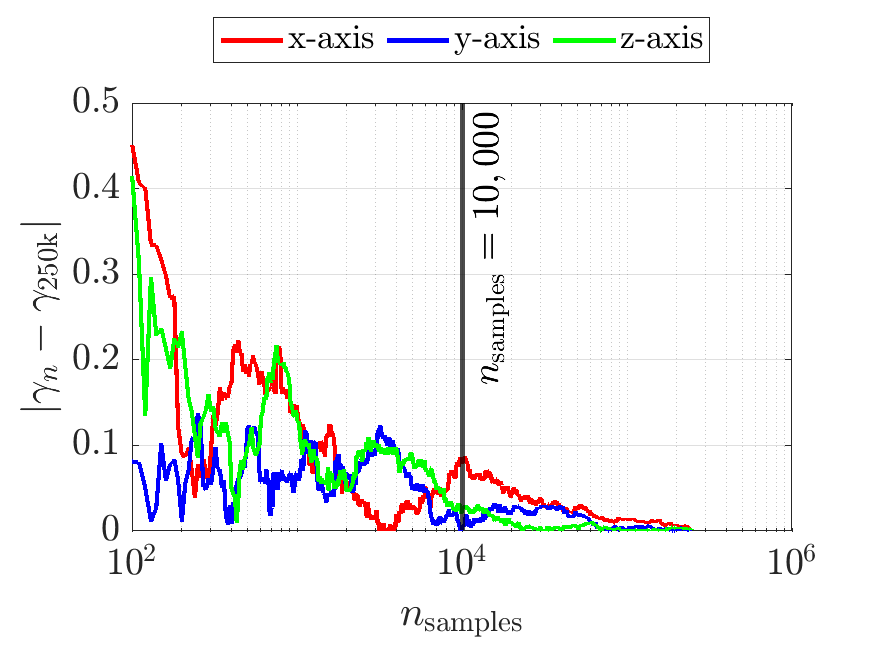}}
    \hskip 0.1truein
    \subcaptionbox
    {Kurtosis.\label{fig: MC Convergence, kurt}}
    {\includegraphics[width=0.48\textwidth]{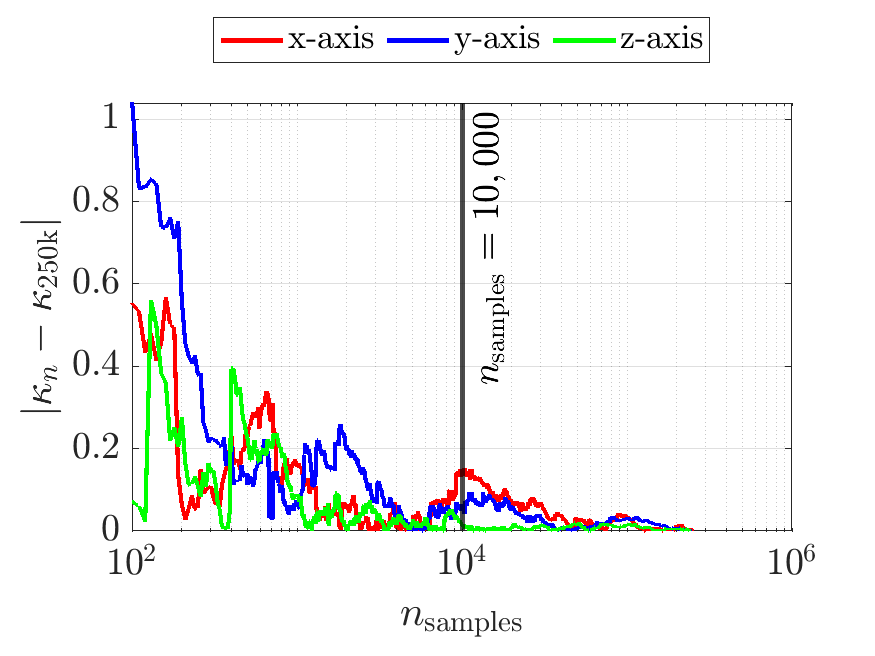}}
    ~
    \caption{Convergence analysis using CR3BP example with CUT6 and kurtosis constraint. \emph{X-axis in log scale.}}
    \label{fig: MC Convergence}
\end{figure}

Figure~\ref{fig: MC Convergence} shows the convergence profile of the Monte Carlo simulation. It can be seen that when $n_{\text{samples}}$ is small, the difference between the statistical parameters and those with the 250,000 sampled Monte Carlo is relatively large. As the number of samples approaches 250,000, the moments also approach a steady state error. It can be seen in Figure~\ref{fig: MC Convergence, mean} and \ref{fig: MC Convergence, 3sigma} that $n_{\text{samples}} = 10,000$ is sufficiently large to get a converged approximation of the mean and covariance. From Figures~\ref{fig: MC Convergence, skew} and \ref{fig: MC Convergence, kurt}, it can be observed that the errors in the higher-order statistical moments are comparatively larger along the x-axis; however, their magnitudes are smaller than the effects induced by statistical moment steering, indicating that the Monte Carlo sampling errors are sufficiently small to justify claims regarding the steering of distributional moments. In summary, the choice of $n_{\text{samples}} = 10,000$ is sufficiently large as a Monte Carlo sampling size.

\subsection{Summary of Discussions for Statistical Moment Steering}
This method is a large-scale nonlinear optimization problem, and thus inherits most of the challenges associated with this class of problem. Firstly, the convergence rate and the local solution will depend on the input parameters of \texttt{SCvx*} and the initial reference solution. Investigating \texttt{SCvx*}'s optimization parameters, as well as better initial reference generation, can improve the robustness of convergence. To add, using other convex solvers such as \texttt{YALMIP} may decrease runtime compared to \texttt{CVX} due to their differences in handling convex programming problems. 

Secondly, because of the inherent limitations of CUT, its prediction of the distribution can become unreliable when the distribution is highly non-Gaussian or overly dispersed. In such cases, the nominal and feedback gains optimized by statistical moment steering may no longer be valid for the rest of the distribution. This motivated the placement of intermediate covariance constraints in the halo stationkeeping example to tighten the distribution and prevent excessive dispersion. Increasing the number of sigma points for CUT will increase the accuracy of the estimation, but will result in more optimization variables. The method can be generalized to any CUT order, but one must be reminded of the computational difficulties with handling the increased number of CUT points.

Next, this paper does not consider the addition of navigational errors or process noise. More realistically, the control of the spacecraft will be based on an estimated state rather than the true state, but the current formulation does not consider the quality of the estimation in its control planning. Previous works in covariance steering papers \cite{Oguri-Chance-Paper, Naoya-Sequential-Cov-Steering} are able to account for this, so a future step in this research is to extend this capability to statistical moment steering. 

Lastly, the versatility of this approach lends itself to a wide range of applications, both within astrodynamics and across other domains where moment-based control or distribution shaping is relevant. For instance, this study primarily employs impulsive control, leaving a potential investigation on how performance might differ under continuous control for low-thrust trajectories. Moreover, other nonlinear environments, such as those encountered in proximity operations around asteroids, may benefit from statistical moment steering as a viable framework for autonomous guidance and control.

\section{Conclusion}\label{sec: conclusion}
This paper presents the idea of statistical moment steering. Statistical moment steering extends previous works on linear covariance steering by developing a feedback control policy for the control of higher-order statistical moments in nonlinear systems. As a result, it eliminates the need for the Gaussian assumption made in linear covariance steering.

The proposed method leverages Conjugate Unscented Transformation (CUT) to quantify the distribution's moments through nonlinear transformations and enforces constraints on these quantified moments by optimizing the control gains. This paper casts the optimal statistical moment steering problem as a nonlinear optimization and develops a sequential convex programming approach to solving it. A large Monte Carlo simulation verifies that the optimized control policy successfully steers the distribution, with its statistical moments remaining consistent with those estimated by CUT.

This paper also presents two nonlinear astrodynamics examples in which non-Gaussian distributions are controlled by statistical moment steering. One example highlights that skewness can be directly controlled with statistical moment steering, and the other example demonstrates the versatility this controller has even in the highly nonlinear three-body problem. One downside of this controller is that the CUT remains an approximation, necessitating validation of its accuracy through Monte Carlo simulations. Nonetheless, statistical moment steering provides a systematic framework for designing a stochastic guidance policy in the inherently challenging problem of controlling non-Gaussian distributions in nonlinear environments.

\section{Appendix}
\subsection{Efficient Computation of \texorpdfstring{$A^{(\gamma)}$}{A\^gamma} and \texorpdfstring{$A^{(^m C)}$}{A\^mC}}\label{appendix: computation of gamma matrix}
The main difficulty is computing $\sum_i w_i (e_j E_i\boldsymbol{z}_k)^m e_j E_i$. Although this summation can still be done in finite time, there is a more efficient way that leverages element-wise operations. It can be identified that the summation of $e_j E_i$ results in a  Boolean matrix,
\begin{equation}
    \begin{bmatrix}
        \sum_i e_{1} E_i \\
        \vdots \\
        \sum_i e_{n_x} E_i
    \end{bmatrix}
    = 
    [\underbrace{I_{n_x}\;\ldots\;I_{n_x}}_{\times n_s}]
    = \bar{I}^\top
\end{equation}

For each $(i,j)$, $w_i (e_{j} E_i\boldsymbol{z}_k)^m$ can be mapped one-to-one to a unique $(a,b)$ in $(\bar{I}^\top)_{a,b}$ wherever $(\bar{I}^\top)_{a,b} = 1$. Both $A^{(\gamma)}$ and $A^{(^m C)}$ can then be computed summation-free with just element-wise operations.

\subsection{Scaled Linear Covariance}\label{appendix: Scaled Linear Covariance}
A linear covariance approach is used to generate an initial reference for SCP. However, in chaotic systems like CR3BP, uncontrolled linear covariance can quickly result in the covariance becoming singular. The goal is to develop a method for generating a good initial reference for SCP while including the rotational information of the natural dynamics as well as not allowing the covariance to grow too large. Consider $\Tilde{P}_k$ to be the scaled covariance at $t_k$. Recall the typically linear covariance propagation to $t_{k+1}$:
\begin{equation}
    P_{k+1} = A_k \Tilde{P}_k A_k^\top
\end{equation}

The covariance matrix for a real-valued random variable, by definition, is a real symmetric matrix. This can be decomposed into an orthogonal matrix $Q$ whose columns are the eigenvectors, and the diagonal matrix $\Lambda$ whose diagonal entries are the eigenvalues.
\begin{equation}
    P_{k+1} = Q_{k+1} \Lambda_{k+1} Q_{k+1}^\top
\end{equation}
Assuming that the eigenvectors are normalized, the entries in $\Lambda$ correspond to the size of the principal axis direction. These values determine the magnitude of the covariance and serve as the basis for scaling. For simplicity, the covariance is scaled to the initial covariance: $P_0 = \Tilde{P}_0$ or $\Lambda_0 = \Tilde{\Lambda}_0$. The scaled linear covariance would be
\begin{equation}
    \Tilde{P}_{k+1} = Q_{k+1} \Tilde{\Lambda}_0 Q_{k+1}^\top
\end{equation}

\section*{Funding Sources}
This material is based upon work supported by the Air Force Office of Scientific Research under award number FA9550-23-1-0512. 

\bibliography{references}

\end{document}